\documentclass[twocolumn]{aastex63}

\usepackage{amsmath}
\usepackage{amssymb}
\usepackage[T1]{fontenc}
\usepackage{gensymb}
\usepackage{enumitem}
\usepackage{url}

\received{June 1, 2019}
\revised{January 10, 2019}
\accepted{\today}

\submitjournal{ApJ}

\shorttitle{A tail of the GC object G2/DSO}
\shortauthors{Pei{\ss}ker et al.}

\graphicspath{{./}{figures/}}

\begin{document}

\title{The apparent tail of the Galactic center object G2/DSO}


\correspondingauthor{Florian Pei{\ss}ker}
\email{peissker@ph1.uni-koeln.de}

\author[0000-0002-9850-2708]{Florian Pei$\beta$ker}
\affil{I.Physikalisches Institut der Universit\"at zu K\"oln, Z\"ulpicher Str. 77, 50937 K\"oln, Germany}

\author[0000-0001-6450-1187]{Michal Zaja\v{c}ek}
\affil{Department of Theoretical Physics and Astrophysics, Faculty of Science, Masaryk University, Kotl\'a\v{r}sk\'a 2, Brno, 611\,37, Czech Republic}
\affil{Center for Theoretical Physics, Al. Lotników 32/46, 02-668 Warsaw, Poland}
\affil{I.Physikalisches Institut der Universit\"at zu K\"oln, Z\"ulpicher Str. 77, 50937 K\"oln, Germany}
\author[0000-0001-6049-3132]{Andreas Eckart}
\affil{I.Physikalisches Institut der Universit\"at zu K\"oln, Z\"ulpicher Str. 77, 50937 K\"oln, Germany}
\affil{Max-Plank-Institut f\"ur Radioastronomie, Auf dem H\"ugel 69, 53121 Bonn, Germany}

\author[0000-0002-5728-4054]{Basel Ali}
\affil{I.Physikalisches Institut der Universit\"at zu K\"oln, Z\"ulpicher Str. 77, 50937 K\"oln, Germany}

\author[0000-0002-5760-0459]{Vladim\'{\i}r Karas}
\affil{Astronomical Institute, Czech Academy of Sciences, Bo\v{c}n\'{\i} II 1401, 141\,00 Prague, Czech Republic}

\author[0000-0001-7134-9005]{Nadeen B. Sabha}
\affil{Institut f\"ur Astro- und Teilchenphysik, Universit\"at Innsbruck, Technikerstr. 25, 6020 Innsbruck, Austria}

\author[0000-0001-5429-2369]{Rebekka Grellmann}
\affil{I.Physikalisches Institut der Universit\"at zu K\"oln, Z\"ulpicher Str. 77, 50937 K\"oln, Germany}


\author[0000-0001-5342-5713]{Lucas Labadie}
\affil{I.Physikalisches Institut der Universit\"at zu K\"oln, Z\"ulpicher Str. 77, 50937 K\"oln, Germany}

\author[0000-0001-6437-6806]{Banafsheh Shahzamanian}
\affil{Instituto de Astrofísica de Andalucía, Glorieta de Astronomía, 18008 Granada, Spain}


\begin{abstract}

The observations of the near-infrared excess object G2/DSO induced an increased attention towards the Galactic center and its vicinity. The predicted flaring event in 2014 and the outcome of the intense monitoring of the supermassive black hole in the center of our Galaxy did not fulfill all predictions about a significantly enhanced accretion event. Subsequent observations furthermore addressed the question concerning the nature of the object because of its compact shape, especially during its periapse in 2014. Theoretical approaches have attempted to answer the contradicting behavior of the object, resisting the expected dissolution of a gaseous cloud due to tidal forces in combination with evaporation and hydrodynamical instabilities. However, assuming that the object is rather a dust-enshrouded young stellar object seems to be in line with the predictions of several groups and observations presented in numerous publications. Here we present a detailed overview and analysis of the observations of the object that have been performed with SINFONI (VLT) and provide a comprehensive approach to clarify the nature of G2/DSO. We show that the tail emission consists of two isolated and compact sources with different orbital elements for each source rather than an extended and stretched component as it appeared in previous representations of the same data. \normalfont{Considering} our recent publications, we propose that the monitored dust-enshrouded objects are remnants of a dissolved young stellar cluster whose formation was initiated in the Circum-nuclear Disk. This indicates a shared history which \normalfont{agrees} with our analysis of the D- and X-sources.

\end{abstract}


\section{Introduction} \label{sec:intro}
Observations of the direct vicinity of the supermassive black hole Sgr~A* in the center of our Galaxy have been bringing unexpected findings. For instance, \normalfont{the S-cluster with a projected diameter of about 40 mpc around Sgr~A* harbors young stars, predominantly of B-type spectral class. The origin of the related S-stars} and the question, where they have formed, is still unresolved \citep{morris1996, Ghez2003}. Several publications aim to answer the two possible scenarios, namely, whether the in-situ star formation (SF) or dynamical-segregation processes lead to the presence of young early-type stars \citep[see, e.g., ][]{Nayakshin2007, Jalali2014, Moser2017} that form a cusp-like surface-brightness density distribution. Another example demonstrating that the Galactic center (GC) is a unique dynamical laboratory is the recent outcome of \cite{Ali2020} where the authors find a non-randomized distribution of the S-stars, quite contrary to the earlier results. The stellar members belong to a multi-disk arrangement that shapes the S-cluster. On even smaller scales \normalfont{(< 6 mpc)}, the observation with GRAVITY of the pericenter passage of S2 \citep{Schoedel2002} in 2018 confirmed the Schwarzschild procession \citep{gravity2018, Do2019S2}, which was tentatively investigated by \cite{Parsa2017}. Recently, \cite{Fragione2020} used the orbital parameters of the newly discovered S-cluster members S4711-S4715 and S62 \citep{Gillessen2009, peissker2020a, Peissker2020d} to derive an upper limit for the spin of Sgr~A*.\newline
However, not only does the analysis of stellar orbits reveal details about the nature of the SMBH and its environment \citep[][]{Zajacek2019, Zajacek2020, Hosseini2020}. Because \normalfont{a part} of the \normalfont{GC data is gathered through the Integral Field Unit} (IFU) of the Spectrograph for INtegral Field Observations in the Near Infrared (SINFONI) mounted at the Very Large Telescope (VLT), it is possible to access several emission \normalfont{lines in} the near-infrared (NIR) \citep{Eisenhauer2003, Bonnet2004}. With this information, it is possible to reveal, for example, large scale structures, such as the prominent Br$\gamma$-bar \citep[][]{Schoedel2010, Peissker2020c}. Furthermore, Doppler-shifted lines, which represent the line-of-sight (LOS) velocity of gas that is dynamically detached from the background and foreground stationary medium, can be detected. Herewith, \cite{Gillessen2012} reported the observation of a $\approx\,3\,M_{\oplus}$ gas cloud (and its tail) that had been detected in the red-shifted Br$\gamma$ regime\footnote{Also Doppler-shifted Pa$\alpha$ and HeI lines were reported.}. The cloud moved on a highly eccentric Keplerian orbit towards Sgr~A*. \normalfont{According to \cite{Gillessen2012}, the cloud} G2 \citep[named by][]{Burkert2012} \normalfont{was expected to be tidally stretched} before and during its pericenter passage. \normalfont{The same authors furthermore predicted that the cloud gets destroyed when it encounters the hot atmosphere surrounding Sgr~A* due to the combined effect of the tidal stretching, evaporation due to heat conduction, and the quick development of hydrodynamical instabilities. Therefore, the material of the destroyed cloud was supposed to enrich} the accretion depot of the SMBH. \normalfont{As a consequence, such a tidal process should have resulted in an enhanced accretion activity of Sgr~A*, which would be manifested by an increased NIR and X-ray flare activity or a ``firework" as underlined by the the authors}. Every attempt to observe the predicted increased flaring activity of Sgr~A*, however, failed. The clockwise stellar disk of massive OB/Wolf-Rayet stars \citep[][]{Paumard2006} was supposed to serve as a birth place for the cloud \citep[][]{Burkert2012}. \normalfont{It was claimed, that shocked winds of the stellar members of the clockwise disk with velocities up to 1000 km/s create hot plasma that is attracted by Sgr~A* (and consequently interpreted as G2).}\newline
Furthermore, \cite{Gillessen2019} reported the detection of a drag force acting on G2 to explain the orbital motion of the compact object on a bound trajectory which was claimed to deviate from the Keplerian orbit. While \cite{Gillessen2012} and \cite{Gillessen2019} pursue a \normalfont{core-less} gaseous cloud model, \normalfont{\cite{MurrayClay2012} interpret the findings using the model of a low-mass star surrounded by a protoplanetary disk. Also,} other authors agree that the observed characteristics of G2 are more consistent with a stellar source that is embedded in a dense gaseous-dusty envelope \citep[see selected references like, e.g., ][]{Eckart2013, Scoville2013, Zajacek2014, Witzel2014, Valencia-S.2015, Prodan2015, Shahzamanian2016, stephan2016, Kohler2017}. Lately, \cite{Peissker2020b} presented a multiwavelength analysis of G2 where the authors use the name Dusty S-cluster Object (DSO) for G2 to underline the dusty nature of the source, in particular its prominent near-infrared excess that is related to the effective blackbody temperature of $\sim 500\,{\rm K}$ \citep[see also][]{Eckart2013}. In contrast, many publications find no K-band counterpart of G2 \citep[e.g.][]{Gillessen2012, Pfuhl2015, Plewa2017, Ciurlo2020}.\normalfont{To reflect the disputed nature of the source, we adapt the name G2/DSO throughout the manuscript.}
The overall spectral energy distribution (SED) that corresponds to the continuum emission of the \normalfont{G2/DSO} in the H-, K-, L-, and M-bands as presented in \cite{Peissker2020b} is based on a two-component fit, which corresponds to the star--envelope system. This analysis emphasizes the nature of the \normalfont{G2/DSO} as a young stellar object \normalfont{(hereafter YSO)} that is embedded in a dense gaseous-dusty envelope.\newline
This interpretation seems to be in agreement with the analysis of G1, a predecessor of \normalfont{G2/DSO} \citep[see][]{Pfuhl2015}. The G1 object, found by \cite{Clenet2005a}, was originally classified as a gas cloud. \cite{Pfuhl2015} connected the orbit of G1 and \normalfont{G2/DSO} to claim that both sources are part of a common gaseous streamer. The authors propose that G1 and \normalfont{G2/DSO} share similar orbits. Hence, \normalfont{G2/DSO} should follow the trajectory of the cloud G1. However, recent observations presented by \cite{Witzel2017} do not support this scenario since the orbits of these objects differ substantially. Furthermore, it seems that several other objects (10+) share a similar dynamical history \citep{Ciurlo2020, Peissker2020b}, potentially forming a unique population within the S cluster. For \normalfont{some} of these sources \normalfont{that are mainly found in the Doppler-shifted Br$\gamma$ regime}, a \normalfont{K-band continuum-}emission counterpart can be detected \normalfont{indicating a stellar nature} \citep[see, for example, X7 or G2/DSO as presented in][]{Peissker2020b}. Also, the sources of the D-complex \citep[][]{Peissker2020b} \normalfont{imply} a common origin which seem to support the in-situ star forming scenario \citep[for simulations, see also][]{Jalali2014}.\newline
In this work, we will investigate the G2/DSO source in detail. \normalfont{Since we use high-pass filter and avoid Gaussian smoothing for the here presented data, we present a new orbital solution based on SINFONI data covering 2005-2019. We will underline the orbital analysis with related Br$\gamma$ line maps that are accompanied by K-band continuum detections of G2/DSO}. Additionally, we compare position-position-velocity maps with the literature. Based on the analysis, we introduce new sources that are following the G2/DSO source that we call Obedient Star 1 and 2 (OS1 and OS2)\footnote{This nomenclature is motivated by the close distance to the \normalfont{G2/DSO} and their potentially shared history.}.\newline
In Section~\ref{sec:data}, we will introduce the used data and explain the applied analyzing tools. In Section~\ref{sec:results}, we will show the \normalfont{G2/DSO} with its Keplerian orbit around Sgr~A* and the detection of the newly discovered sources OS1 and OS2. The discussion part in Section~\ref{sec:discussion} is followed by the final conclusion.\newline
For a comprehensive list of the here used data, please consider Appendix \ref{sec:app_data}.

\section{Data and Analysis} \label{sec:data}

Here we explain the data accumulation, the instrument settings, and the applied analysis tools.  

\subsection{SINFONI and the VLT}

SINFONI was previously mounted at the unit telescope (UT) 4 (VLT), afterwards it was relocated to UT3 (VLT). It is now decommissioned. The instrument uses a slicer to create pseudo longslits. Then, a spectral dispersion creates groups of wavelength-dependent longslits\footnote{From these groups, single line maps can be created.}. After this process, a 3d data cube is reconstructed with 2 spatial dimensions and 1 spectral dimension. For a fraction of the data, a laser guide star (LGS) was used. Since UT3 does not support LGS-guidance, the data that were observed after the relocation used exclusively a natural guide star (NGS). Typically, this NGS is located $15".54$ north and $8".85$ south of Sgr~A*. Since SINFONI uses an optical wavefront sensor, the selection of possible bright (14-15 mag) NGS-sources is limited. The location of the bright radio source Sgr~A*, which can be associated with the SMBH \citep[see][]{Eckart2002,Eckart2017}, can be found at Right Ascension (RA) $17:45:40.05$ and Declination (DEC) $-29^{\circ}\,00'\,28.120"$ (J2000). In the following subsections, we will explain the procedure of deriving the position of Sgr~A* in the individual data cubes. \normalfont{In Appendix \ref{sec:app_data}, we list the investigated SINFONI cubes including the quality, exposure time of the individual observations, the related IDs, and the publications where we already used the data.}

\subsection{Dataset and instrument settings}

For the observations, the smallest available plate scale was used \normalfont{(12.5 mas)} and the wavelength range was set to the H+K-band ($1.45\,\mu m\,-\,2.45\,\mu m$). The exposure time for a single data cube was set between 400 and 600 seconds. \normalfont{We use the standard object-sky-object nod cadence. With that, the sky} corrections \normalfont{can be} applied \normalfont{to the individual data cubes}. Because of the background noise and the small field of view (FOV) of $0.8\,\times\,0.8$ arcsec, the GC/S-cluster observations are dithered around the position of Sgr~A* or S2. After the usual reduction steps that are applied with the ESO pipeline \citep{Modigliani2007}, the single data cubes of several nights (see Appendix \ref{sec:app_data} for the used data) are stacked to create a mosaic \normalfont{for each year between 2005 and 2019} with a higher signal-to-noise (S/N) ratio but also an increased FOV.
Furthermore, the SINFONI pipeline automatically applies a barycentric and heliocentric correction.

\subsection{The position of Sgr~A*}

From the well-observed orbit of the brightest (in K-band) S-cluster member S2 \citep[see, e.g., ][]{Parsa2017, Ali2020}, we derive the position of Sgr~A* with an uncertainty of less than 12.5 mas. This uncertainty already contains a linear transformation but also corrections for a distorted FOV. While this is applied in a satisfying procedure for the data discussed in \cite{Parsa2017} and \cite{Ali2020}, the SINFONI data suffers from image motion as a function of wavelength. While this effect is certainly suppressed in single-band observations, the H+K-band observations with SINFONI do show a non-negligible movement of the stars between the H- and K-band. For example, the position of S2 does change by over 1 pixel (= 12.5 mas) by comparing individual line maps. While \cite{Jia2019} addresses some of the GC observation problems for the KECK telescope \normalfont{that can also be applied to the VLT data (stellar confusion, variable PSF, artifical PSF-wing sources)}, we do not agree with the 1 mas uncertainty given by \citet{Gillessen2017}. \normalfont{This underestimates general crowding problems \citep[for example, blend stars, see][]{Sabha2012} and does not reflect the noise character of the SINFONI data.} As pointed out by the SINFONI manual\footnote{\url{www.eso.org}}, the shape and intensity of the point spread function is a function of the source position on the detector. \normalfont{This results in unaccurate positions of stellar sources.} Furthermore, \cite{Eisenhauer2003} discussed the issue of image motion for high-exposure observations (above several hours). Since this effect is nonlinear and depends on the total integration time as well as the weather conditions, we will adapt a conservative uncertainty of 12.5 mas for the position of Sgr~A* in the SINFONI data. In the following, we will elaborate on this issue in more detail.

\subsubsection{Image motion of SINFONI long time exposures}

As mentioned by \cite{Eisenhauer2003}, short-time exposures do not suffer from image motion \normalfont{(i.e. the apparent movement of a source as a function of wavelength and hence channel)}. Unfortunately, the image motion of SINFONI is not broadly covered by the literature. Hence, we will investigate this effect with the available GC data. We randomly pick 5 single exposures with integration times of $\geq\,400$ sec. Furthermore, we will compare the image motion of single exposures and final mosaics. We note that the image motion only appears in the horizontal direction.
\begin{table}[htb]
\centering
\begin{tabular}{cccccc}\hline \hline
ID  & Date & DIT &x$_1$ & x$_2$ & $\Delta$x  \\
   & & [sec] & [px] & [px] & [px] \\ \hline 
081.B-0568(A)&06.04.2008&600&38.756&35.404&-3.352\\
087.B-0117(I)&02.05.2011&600&22.421&23.518&1.097\\
093.B-0932(A)&03.04.2014&400&45.155&45.143&-0.012\\
594.B-0498(R)&14.04.2016&600&37.742&38.2538&0.511\\
091.B-0183(H)&26.03.2018&600&49.592&47.595&-1.997\\
\hline \hline
\end{tabular}
\caption{Image motion of S2 for randomly picked SINFONI exposures. For every item, we use the channel 385 (x$_1$) and 1985 (x$_2$), which covers a wavelength range between $1.6\,\mu m\,-\,2.4\,\mu m$. We use a Gaussian with a Kernel of 5 pixels to extract the related positions. The usual positioning uncertainty is of the order of about $1\,-\,2\,\%$.}
\label{tab:image_motion_1}
\end{table}
In Table \ref{tab:image_motion_1}, we list a few exemplary exposures. However, this nonlinear image motion can be observed in every single observation. The effect increases with the exposure time. Since a technical discussion is beyond the scope of this work, we will categorize the image motion as a sporadic statistical behavior of the data. Hence, the more single exposures are combined, the more the image motion decreases.
\begin{table}[htb]
\centering
\begin{tabular}{ccccc}\hline \hline
Final mosaic &Total Exp&x$_1$ & x$_2$ & $\Delta$x  \\
             &         &[px] & [px] & [px] \\ \hline 
2008&21&51.125&50.513&0.612\\
2011&43&51.618&51.103&0.515\\
2014&310&60.187&59.765&0.422\\
2016&60&61.822&60.587&1.235\\
2018&114&61.237&60.832&0.405\\
\hline \hline
\end{tabular}
\caption{Image motion of S2 for the final mosaics that are a combination of several single exposures. We use the same approach as for the listed cubes in Tab. \ref{tab:image_motion_1}.}
\label{tab:image_motion_2}
\end{table}
While the image motion of a single exposure is an unavoidable mandatory condition, it can efficiently be decreased by combining many data cubes to a final mosaic (see Table \ref{tab:image_motion_2}). As mentioned before, the effect is the most profound for sources in the center of the FOV (here: S2). For objects at the border of the data cube, the effect differs by $1\,-\,20\,\%$. For example, the S-cluster star S4 (see Fig. \ref{fig:fc}) in 2016 moves by $\Delta x_{S4}\,=\,0.112$ px while S2 shows a difference of $\Delta x_{S2}\,=\,1.235$ px (see Table \ref{tab:image_motion_2}).\newline The outcome of the example is expected since the SINFONI manual\footnote{\url{www.eso.org}} states that the shape of a PSF differs depending on the position on the CCD chip. This may not influence the observation of a single source but does impact the analysis of a crowded field like the GC.

\subsection{\normalfont{Line} maps}

A line-map and the related channel of a data cube represents a specific wavelength and if this wavelength is Doppler-shifted, a LOS velocity $v_z$ with $v_z\,\neq\,0$ \normalfont{can be derived}. A SINFONI 3d data cube consists of 2000+ single line maps \normalfont{that can also be called channels}. To isolate a single line, one has to subtract the underlying continuum. Typically, a polynomial fit \normalfont{(here, 2nd degree)} will help to get rid of the continuum and partially the background emission (see Fig. \ref{fig:fitted_spectrum}\normalfont{, Appendix \ref{sec:conti_fit_app}}).
However, the emission line itself can be heavily influenced by a variable background or atmospheric OH emission lines. For the Doppler-shifted Pa$\alpha$, HeI, and Br$\gamma$ lines that have been used for the analysis of G2/DSO \citep[see e.g.][]{Gillessen2013b}, one has to consider the OH vibrational transition states 7-5, 8-6, and 9-7. As pointed out by \cite{Davies2007}, these OH lines do have a non-negligible influence on the shape, intensity, position, and consequently a velocity of the object of interest.      

\subsection{Position-Position-Velocity maps}

In this work, we study Position-Position-Velocity (PPV) maps instead of Position-Velocity (PV) diagrams because in this way we preserve more accurate information. For this purpose, we transform the Doppler-shifted wavelength information of a single pixel (i.e., spaxel) to a LOS velocity. For the analysis, we will not use a smoothing kernel since this will have an impact on the result \normalfont{(see Table \ref{tab:vg_error_comparison} and Fig. \ref{fig:ppv_2}, Appendix \ref{sec:tail_smoothed_app})}.

\subsection{Orbit analysis and MCMC simulations}

Given the distance of the \normalfont{G2/DSO} from Sgr~A* and its LOS velocity evolution, we apply a Keplerian model fit to reconstruct the trajectory of the object. \normalfont{For the Keplerian fit, we use a SMBH mass of $M_{\rm Sgr~A*}\,=\,4.15\,\times\,10^6\,M_{\odot}$ with a distance of $d\,=\,8.3$ kpc. For the MCMC simulations, we leave the boundaries for the different parameters, except for the pericenter passage, open.} Considering the uncertainty of the position of Sgr~A* as well as the sensitive emission lines of the \normalfont{G2/DSO}, we categorize different \normalfont{non-Keplerian} interpretations of the orbit that involve more parameters (in particular the magnetohydrodynamic drag force analyzed by \citeauthor{Gillessen2019}, \citeyear{Gillessen2019} or the fermionic dark matter dense core-diluted halo model of \citeauthor{2020A&A...641A..34B}, \citeyear{2020A&A...641A..34B}) as challenging and unnecessary given the current data quality.  

\subsection{High-pass filter}

As described in \cite{Peissker2019, peissker2020a, Peissker2020b, Peissker2020d}, a high-pass filter can be used to access information which is suppressed by overlapping PSF wings. Since the natural SINFONI PSF (NPSF) does show irreparable imperfections \normalfont{because no source is} isolated \normalfont{in the small and crowded SINFONI FOV ($\sim\,0.8\,\times\,0.8$ arcsec)}, we construct an artificial PSF (APSF) with comparable parameters \normalfont{(x- and y-FWHM, angle with respect to North)} with respect to the NPSF. For that, we fit a PSF ($\sim\,6$ pixel) sized Gaussian to S2, the brightest source in the FOV. From this, we derive the necessary input parameter to construct the NPSF. The fit uncertainties are in the range of about $1\%$.\\
Then, we place the input file in a $256\,\times\,256$ array. We use 10000 iterations to minimize the chance of a false positive. Furthermore, we do a background subtraction, which is of the order of $15\,-\,30\,\%$. With this procedure, the background noise can be suppressed. Subsequently, we apply the Lucy-Richardson algorithm \citep{Lucy1974} and a delta map is created. By convolving this map with a suitable Gaussian kernel (around $70\%$ of the size of the input PSF), we get the final high-pass filtered image.
We verify the robustness of the resulting image by comparing known stellar positions to it. In every step, the input files are normalized to the peak intensity. In most cases, the peak intensity is associated with S2. In every other case, the peak intensity is related to S35 (see Fig. \ref{fig:fc}).

\subsection{OH emission lines/sky correction}
\label{sec:oh_emission}
The OH emission lines in the H+K band and the related correction are widely discussed by \cite{Davies2007} and more recently by \cite{ulmer-moll2019}. Since the emission lines of G2/DSO are Doppler-shifted, they coincide with several vibrational transitions of OH \citep[][]{Rousselot2000}. Since \cite{Davies2007} provides the sky correction for the SINFONI pipeline but also describes an under- and over-subtraction of various OH/sky lines in the NIR, we want to investigate the influence on the high-exposure observations carried out in the GC. The typical observation scheme for these observations is object (o) - sky (s) - object (o). However, several observational programs that are available in the ESO archive do show a nontypical observational scheme. Instead of o-s-o, these programs use o-o-s-o-o with a long exposure time of 600 seconds. This results in nonmatching sky/OH emission from the science object. Because of the exposure time, the effect is maximized.
\begin{figure}[htbp!]
	\centering
	\includegraphics[width=.5\textwidth]{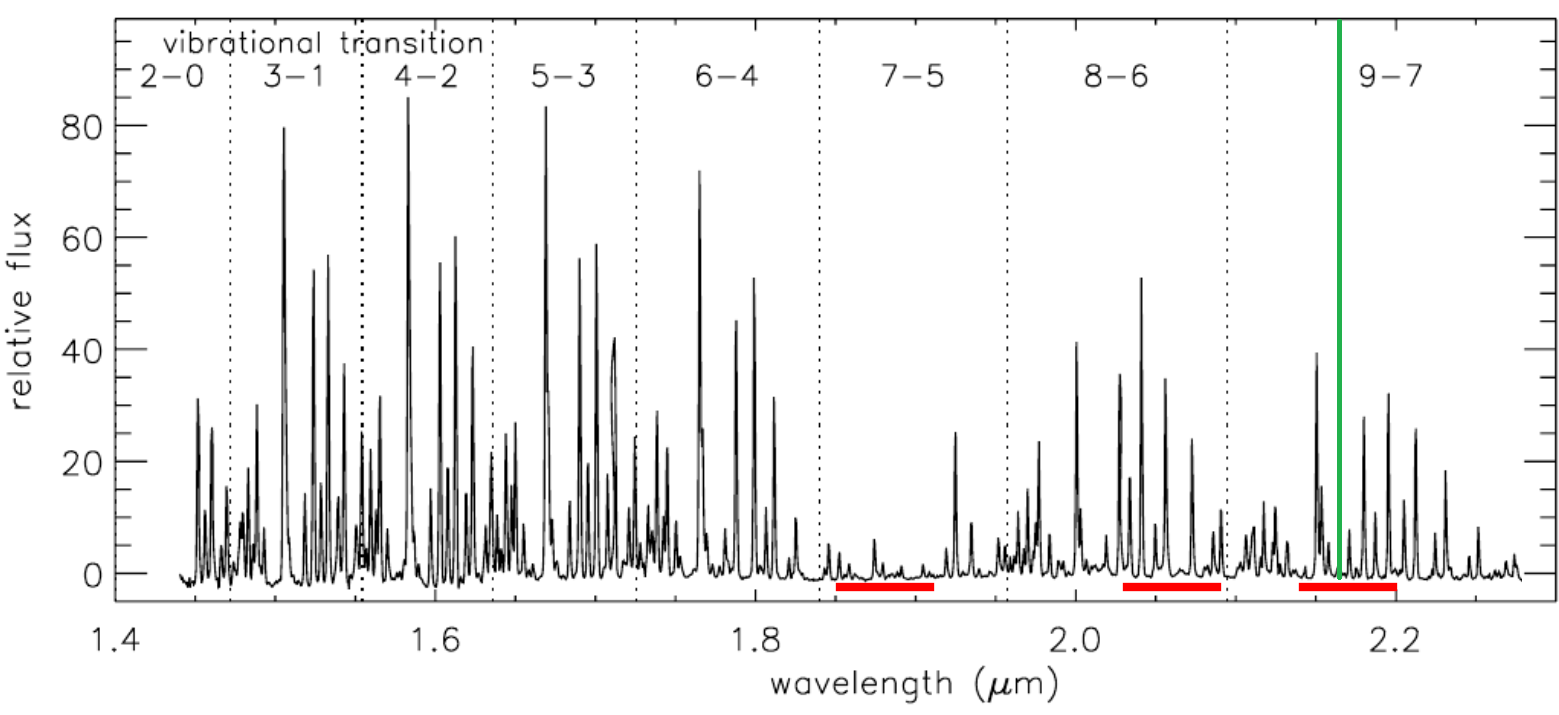}
	\caption{Near-infrared OH emission adapted from \cite{Davies2007}. We use red bars to mark the spectral range of the Doppler-shifted Pa$\alpha$, HeI, and Br$\gamma$ line that coincide with the vibrational transition states 7-5, 8-6, and 9-7 respectively. \normalfont{The Br$\gamma$ rest-wavelength at $2.1661\,\mu m$ is indicated with a green vertical line and embeded in the spectral region of the OH transition states 9-7 (see also Table \ref{tab:oh_line_list}).}}
\label{fig:oh_lines}
\end{figure}
As shown in Fig. \ref{fig:oh_lines}, the Pa$\alpha$ line suffers from strong telluric emission/absorption features. In contrast, the OH emission lines do not influence the spectral range between $1.85-1.95\,\mu m$ at a noticeable level since the relative flux is below $10\%$. This is not valid for the Doppler-shifted HeI and Br$\gamma$ regime. Since the latter line is more prominent than the HeI line by about $40-60\%$, we will emphasize the analysis of the spectral range around the Br$\gamma$ rest wavelength of $2.1661\,\mu m$ in Sec. \ref{sec:results}. This is consistent with the analysis and statements given by \cite{Gillessen2013b}.\newline
In Table \ref{tab:oh_line_list}, we list some prominent OH lines that could impact the line shape of the \normalfont{G2/DSO} (see the following section).
\begin{table}[htb]
\centering
\begin{tabular}{cccc}\hline \hline
OH line & Transition & $\lambda _{vac}$&$\lambda _{air}$  \\
             &         & [$\mu m$]  & [$\mu m$]\\ \hline 
Q2(0.5)  & 9-7 & 21505.044 & 21499.176 \\
Q1(1.5)  & 9-7 & 21507.308 & 21501.440 \\
P1(2.5)  & 9-7 & 21802.312 & 21796.363 \\
P2(2.5)  & 9-7 & 21873.518 & 21867.550 \\
\hline \hline
\end{tabular}
\caption{OH lines as observed in the NIR at Paranal by \cite{Rousselot2000}. Because of the Doppler-shift of the Br$\gamma$ line of the \normalfont{G2/DSO}, these OH lines interfere with the observed source emission.}
\label{tab:oh_line_list}
\end{table}

\section{Results} \label{sec:results}

In this section, we present the main results of the observational analysis. 
First, we will present line maps that show the \normalfont{G2/DSO} approaching Sgr~A*. For comparison, the data between 2014.5-2019.5 cover the approaching post-pericenter part of the \normalfont{G2/DSO} orbit, which will be shown afterwards. Markov-Chain-Monte-Carlo (MCMC) simulation emphasizes the robustness of the orbital elements. Non-smoothed PPV maps investigate a possible velocity gradient of the \normalfont{G2/DSO}. The $K$-band detection of the \normalfont{G2/DSO} emphasizes the stellar nature of the object. This is followed by the analysis of the proposed tail (Sec. \ref{sec:intro}) and the identification of the additional sources OS1 and OS2. 
\begin{figure}[htbp!]
	\centering
	\includegraphics[width=.5\textwidth]{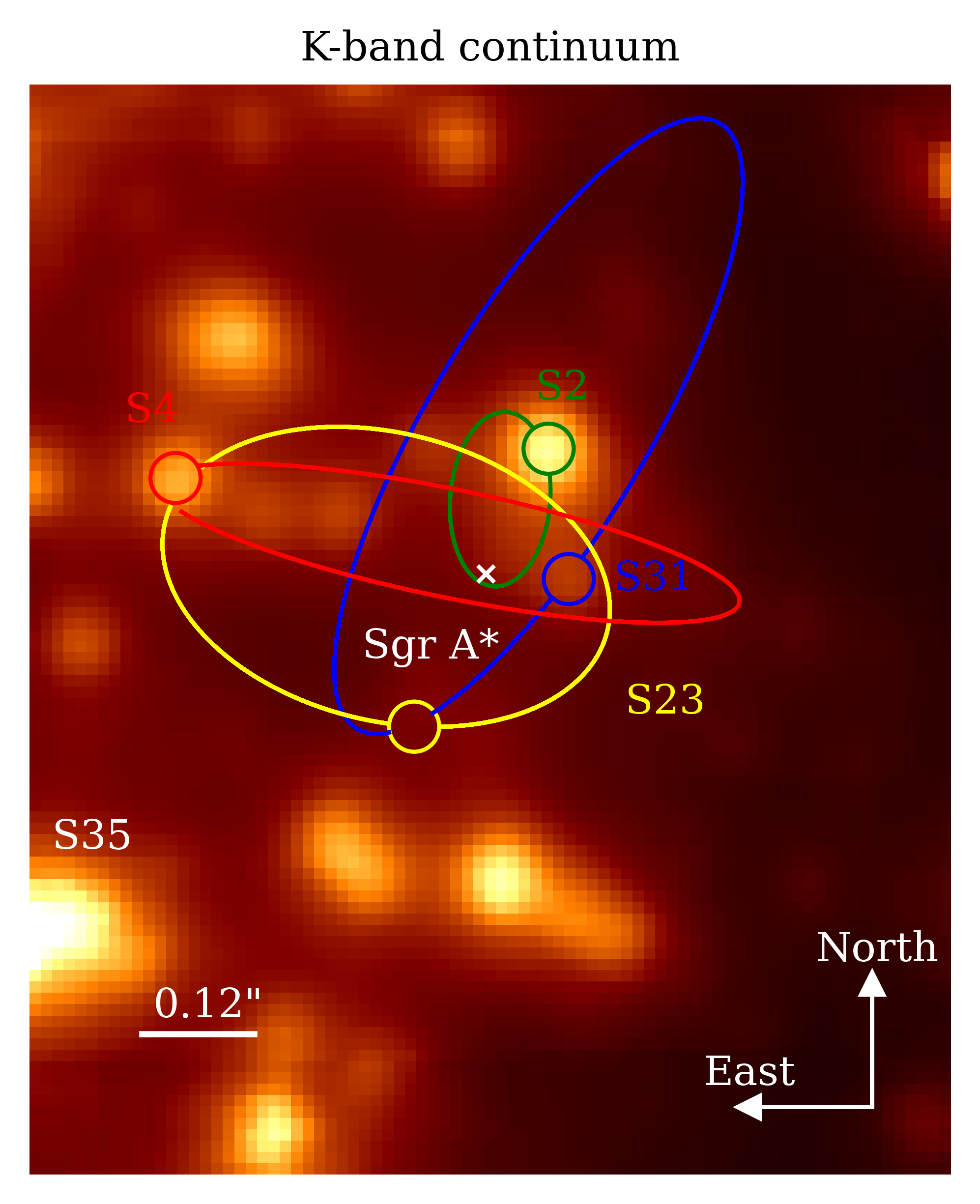}
	\caption{Finding chart of the S-cluster. The x marks the position of Sgr~A*, the green line represents the Keplerian orbit fit of S2. In addition we show the orbits of S4, S23, S31, and S35. The image is extracted from a SINFONI data cube by collapsing the wavelength range between $2.0\,\mu m\,-\,2.2\,\mu m$ (K-band). The observation took place in 2014 and the total on-source integration time is 24 hours (216 single cubes, 400 seconds each).}
\label{fig:fc}
\end{figure}
To provide a confusion-free overview of the S-cluster, please see the finding chart in Fig. \ref{fig:fc} where we mark the most prominent $K$-band source S2\normalfont{, the related orbit, and the in this work discussed stars.}

\subsection{Line map detection}

\begin{figure*}[htbp!]
	\centering
	\includegraphics[width=1.\textwidth]{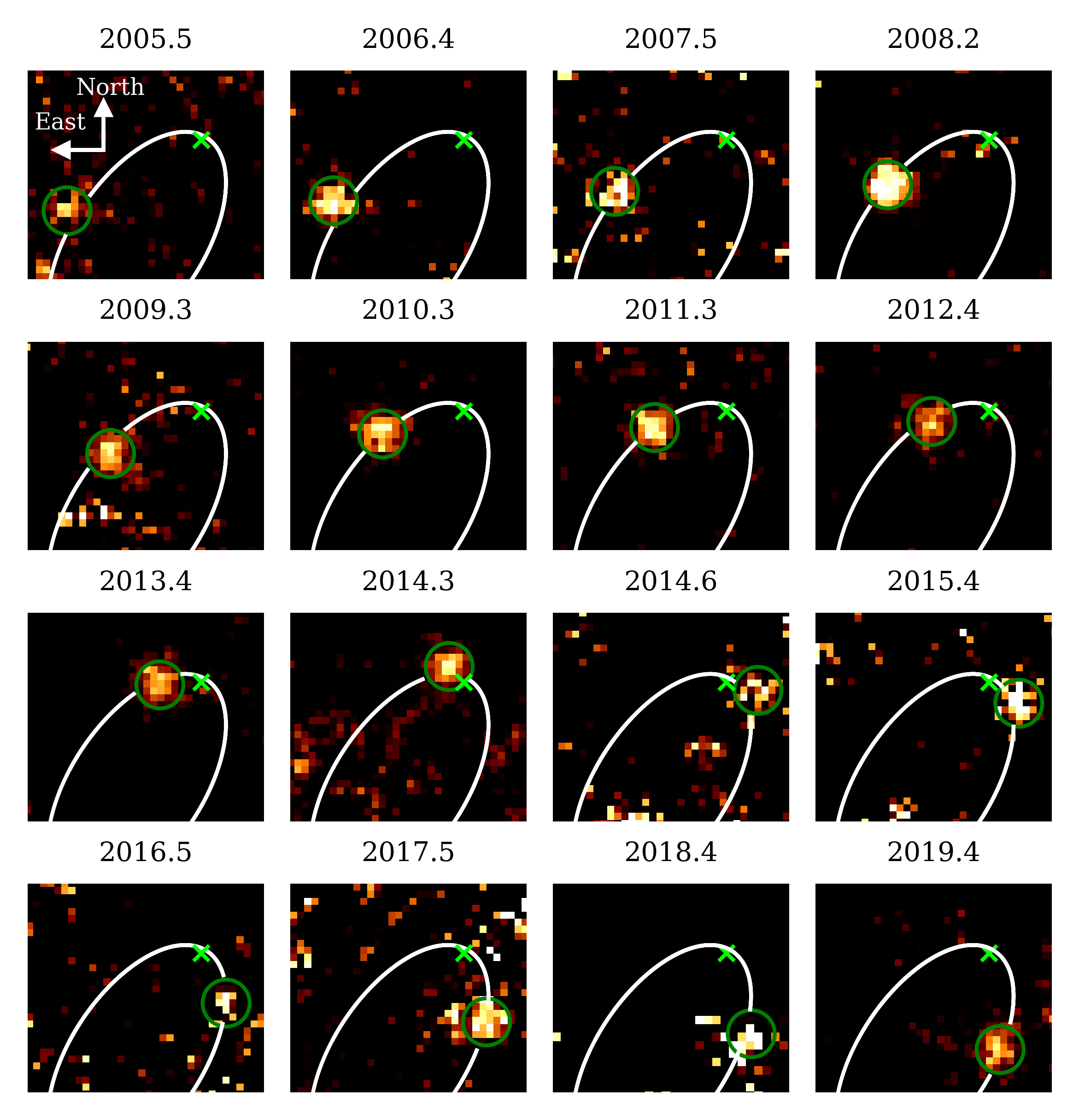}
	\caption{Doppler-shifted Br$\gamma$ line evolution of the \normalfont{G2/DSO} as it approaches and passes around Sgr~A*. The $\times$ marks the position of Sgr~A*, the green circle indicates the position of the \normalfont{G2/DSO}. North is up, east is to the left. As discussed in \cite{Valencia-S.2015}, the forth-shortening factor converges to unity around 2014.3 where we are able to observe the true size of the \normalfont{G2/DSO at the periapse}. The orbital trajectory (in white) is adapted from the fit presented in Sec. \ref{sec:orbit_result}.}
\label{fig:dso_line_evo}
\end{figure*}

The line maps (Fig. \ref{fig:dso_line_evo}) are extracted from the final mosaic data cube of the related year. We emphasize the investigation of the red-shifted (until 2014.3) and blue-shifted (after 2014.4) Br$\gamma$ line. After the Doppler-shifted Br$\gamma$ line is selected, we subtract the underlying continuum. In every dataset, the \normalfont{G2/DSO} can be identified without confusion. The local background, the distance to the surrounding stars, weather conditions, and the on-source integration time do have a major impact on the shape of the source (see Fig. \ref{fig:dso_line_evo}). \normalfont{The} line maps \normalfont{show} the obvious periapse of the \normalfont{G2/DSO} in 2014. Since the source can exclusively be observed in the redshifted Br$\gamma$ regime until 2014.35 and subsequently in the blueshifted domain after 2014.45, the periapse passage must have happened in between.
In the following, we use the identification of the \normalfont{G2/DSO} in the \normalfont{line} maps to derive an exact value for the time of periapse from the Keplerian model fit.

\subsubsection{Br$\gamma$ line evolution for \normalfont{G2/DSO}, OS1, and OS2}

\begin{figure*}[htbp!]
	\centering
	\includegraphics[width=0.9\textwidth]{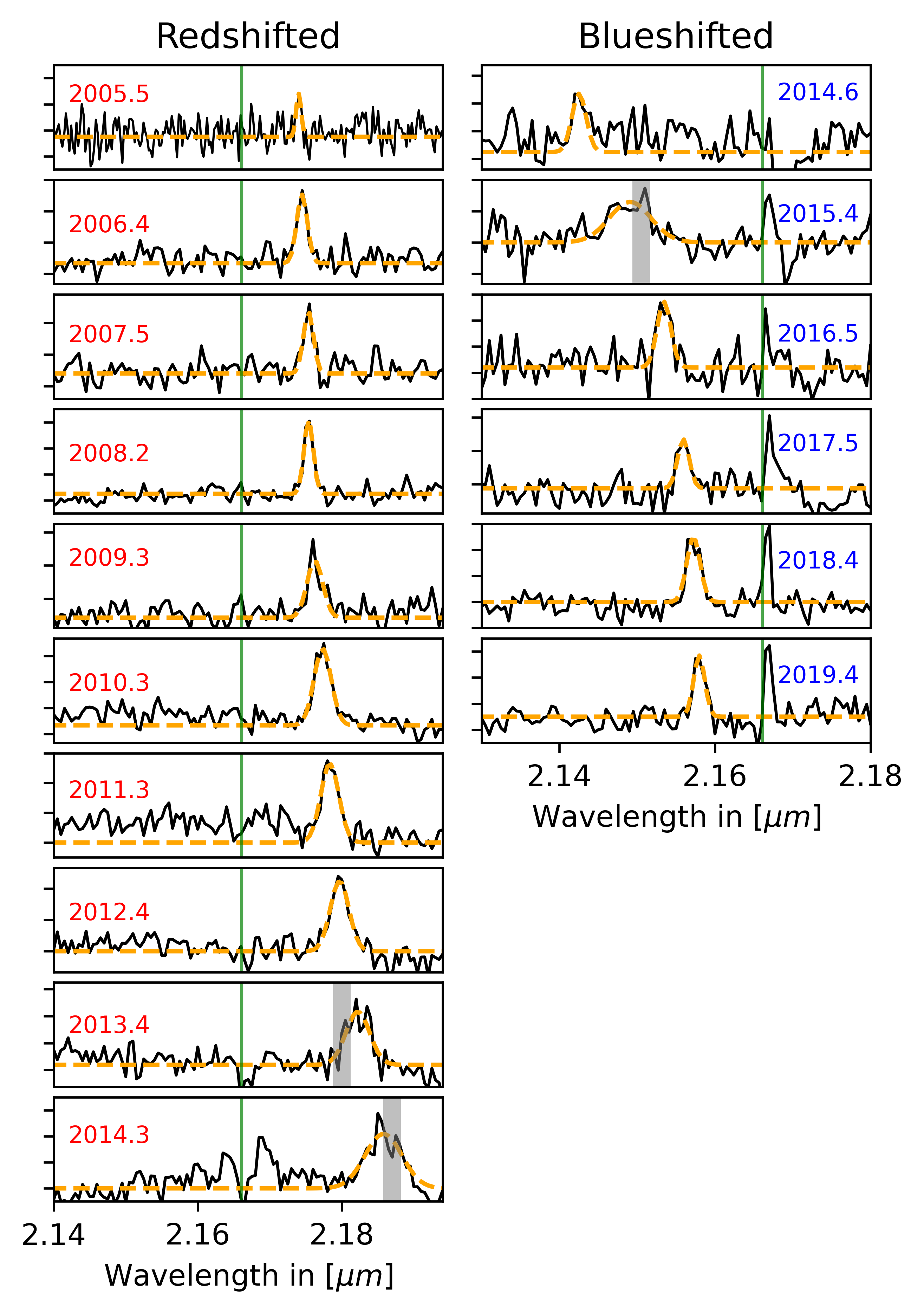}
	\caption{Doppler-shifted Br$\gamma$ line evolution between 2005 and 2019. The green line marks the Br$\gamma$ rest wavelength at $2.1661\mu m$. \normalfont{The grey bars indicate the spectral position of the OH lines Q1, Q2, P1, and P2 (see Sec. \ref{sec:oh_emission} and Table \ref{tab:oh_line_list}) that blend with the Dopper-shifted Br$\gamma$ line.}}
\label{fig:dso_spectral_line_evo}
\end{figure*}

\begin{figure}[htbp!]
	\centering
	\includegraphics[width=.5\textwidth]{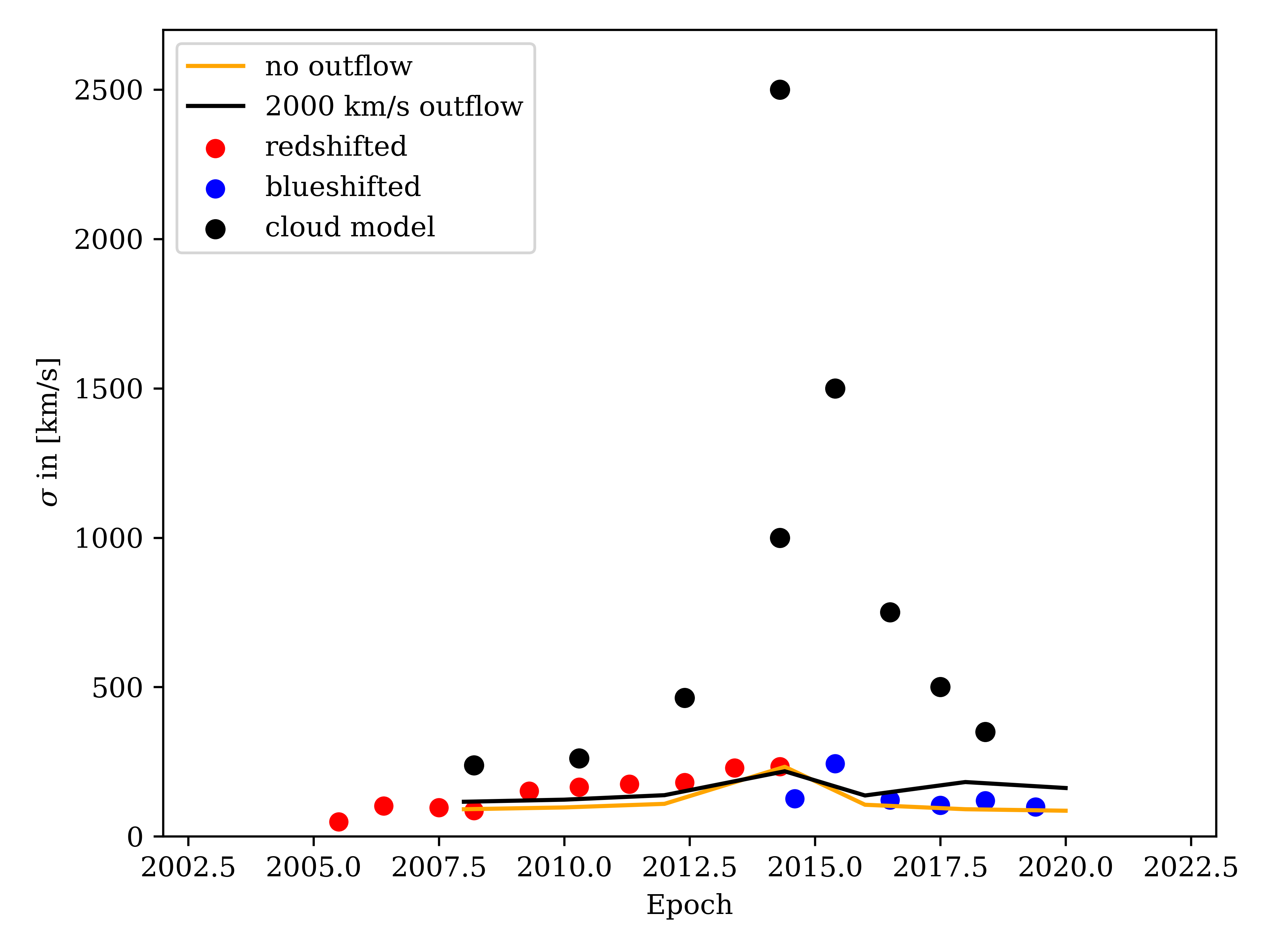}
	\caption{Standard deviation as a measure of the line width $\sigma$ of the Br$\gamma$ line expressed as a function of time. \normalfont{Black data points are based on the cloud model \citep[see, e.g.,][]{Gillessen2019}.} Red datapoints describe the pre-periapse measurements, while blue datapoints \normalfont{represents} the measurements in the post-periapse phase. Orange and black lines are referring to the model calculations of line widths corresponding to the bow-shock velocity field without and with a nuclear outflow, respectively, see \citet{Zajacek2016}. Typical uncertainties for the line width are here $\pm\,30$ km/s. Especially the data points in 2014.3 and 2015.4 suffer from the inefficient OH line correction (vibrational transition 9-7) as shown in Fig. \ref{fig:oh_lines} \normalfont{(see Sec. \ref{sec:oh_emission})} and Table \ref{tab:oh_line_list}. At $2.1505\,\mu m$, $2.1507\,\mu m$, and $2.1873\,\mu m$ the OH lines are infecting the line width of the Br$\gamma$ emission of the G2/DSO. Hence, the data points of 2014.3 and 2015.4 are corrected values (see Sec. \ref{sec:discussion_binary}). For a complete overview of the K-band OH lines and the corresponding relative intensities, see \cite{Rousselot2000}. \normalfont{See Fig. \ref{fig:sigma_time_evo_zoom}, Appendix \ref{sec:brgamma_zoom_app} for a zoomed in view of the same figure.}}
\label{fig:sigma_time_evo}
\end{figure}

Based on the clear line map detection of the \normalfont{G2/DSO} between 2005 and 2019, we extract the source spectrum to derive several properties (Fig. \ref{fig:dso_spectral_line_evo}). To that goal, we use an aperture with a radius of 2 pixel (25 mas)\footnote{In total, the aperture counts 14 pixel.}. We subtract an aperture with \normalfont{an inner radius of 3 pixel and an outer} radius of 10 pixels because of the dominant background/continuum. Then, we fit a polynomial function to the spectrum and fit a Gaussian to the Br$\gamma$ line (see Table \ref{tab:spectral_velocity_properties}). With this procedure, we derive the LOS velocity for the \normalfont{G2/DSO} as well as the line width $\sigma$. By studying $\sigma$ as a function of time, we find a quadratic behavior of the Br$\gamma$ line width (Fig. \ref{fig:sigma_time_evo} \normalfont{and Fig. \ref{fig:sigma_time_evo_zoom}, Appendix \ref{sec:brgamma_zoom_app}}). We note that the gradient is expected to be the largest around the time of periapse because of the viewing angle change and the foreshortening factor close to unity (see the following subsection).
Throughout the years, the PSF size of the source is $\sim 75\,-\,100$ mas in x- and y-direction, therefore the \normalfont{G2/DSO} can be described as compact (Table \ref{tab:spectral_velocity_properties}).
\begin{table*}[htb]
\centering
\begin{tabular}{cccccc}\hline \hline
\normalfont{Epoch} & Wavelength in [$\mu m$]&	Velocity in [km/s]  & Standard deviation $\sigma$ in [km/s]	& \multicolumn{2}{c}{Line map size in [mas]}	\\
& & & & x (R.A) & y (DEC.) \\  

\hline 
2005.5 & 2.1740 & 1094.69  & 48.97  & 65 & 40 \\  
2006.4 & 2.1744 & 1159.96  & 102.51 & 94 & 78 \\  
2007.5 & 2.1753 & 1282.89  & 96.60  & 71 & 60 \\  
2008.2 & 2.1753 & 1287.36  & 87.06  & 74 & 73 \\  
2009.3 & 2.1762 & 1405.98  & 151.64 & 68 & 69 \\  
2010.3 & 2.1773 & 1555.53  & 165.31 & 89 & 74 \\ 
2011.3 & 2.1783 & 1693.99  & 174.92 & 85 & 94 \\  
2012.4 & 2.1796 & 1881.95  & 180.40 & 71 & 76 \\  
2013.4 & 2.1822 & 2231.68  & 229.72 & 90 & 98 \\  
2014.3 & 2.1857 & 2726.20  & 355.02 (234.31) & 51 & 59 \\  
2014.6 & 2.1425 & -3265.23 & 126.74 & 56 & 74 \\  
2015.4 & 2.1490 & -2355.72 & 369.12 (243.61)& 60 & 86 \\  
2016.5 & 2.1534 & -1752.34 & 122.32 & 31 & 55 \\  
2017.5 & 2.1559 & -1407.56 & 104.67 & 69 & 86 \\  
2018.4 & 2.1572 & -1228.13 & 119.71 & 56 & 64 \\  
2019.4 & 2.1579 & -1126.27 & 98.87  & 61 & 85 \\  
\hline \hline
\end{tabular}
\caption{Spectral and line map properties of the \normalfont{G2/DSO}. Please see Fig. \ref{fig:dso_spectral_line_evo} for the related fit. Here, the standard deviation $\sigma$ is a measure of the line width. The size is measured by using a 6 pixel Gaussian fit. Usual uncertainties for the size of the \normalfont{G2/DSO} are in the range of 12.5 mas. For the values in 2014.3 and 2015.4, please see \normalfont{Sec. \ref{sec:oh_emission} and Sec. \ref{sec:discussion}. Here, the format of the time is given in decimal years.}}
\label{tab:spectral_velocity_properties}
\end{table*}

The related spectral line fits and tables for OS1 and OS2 can be found in the Appendix \ref{sec:los_os_app} (see Fig. \ref{fig:os1_velorbit}, Fig. \ref{fig:os2_velorbit}, and Table \ref{tab:spectral_velocity_properties_ossources}).

\subsection{Orbit}
\label{sec:orbit_result}

\begin{figure*}[htbp!]
	\centering
	\includegraphics[width=1.\textwidth]{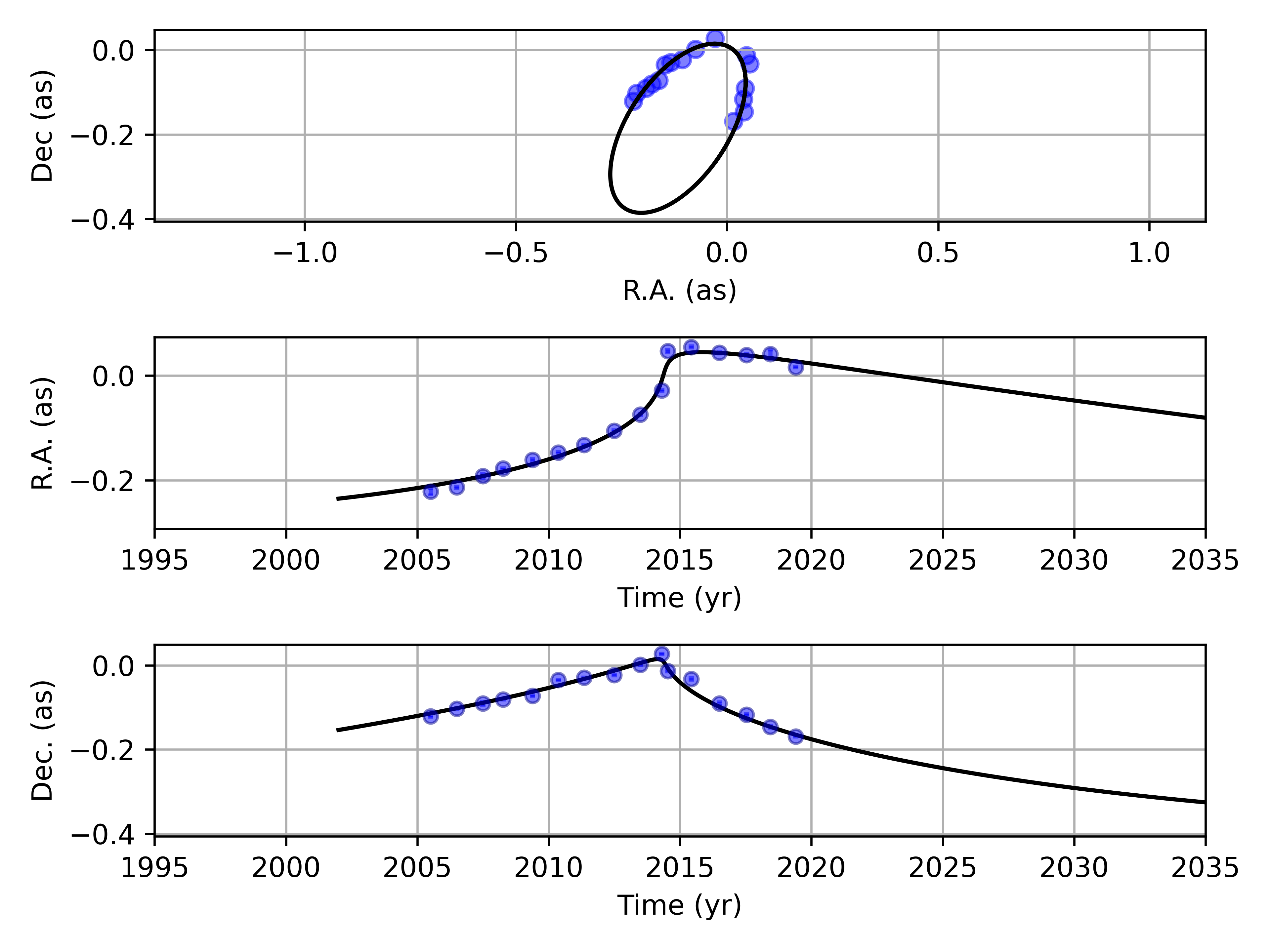}
	\caption{A Keplerian orbit of the \normalfont{G2/DSO} based on the Doppler-shifted Br$\gamma$ emission line. The upper panel shows the on-sky projection of the source. The middle and lower panels show the spatial RA and DEC position as a function of time, respectively. The closest approach to the SMBH can be observed in 2014.38. Given the uncertainties, we find a good agreement of the \normalfont{G2/DSO} trajectory with a Keplerian model fit.}
\label{fig:dso_orbit_1}
\end{figure*}

As mentioned in Sec. \ref{sec:data} and shown in Fig. \ref{fig:fc}, we use the orbit of S2 to derive the position of Sgr~A* with a positional uncertainty of 12.5 mas (= 1 pixel). We list the orbital elements that are adapted from \cite{peissker2020a} in Table \ref{tab:orbit_elements}. For the Keplerian orbital fit, we extract the \normalfont{G2/DSO} position with a Gaussian fit that simultaneously provides a positional uncertainty. Since this value would underestimate the unstable character of the data, we adapt the common uncertainty of 12.5 mas for the spatial position of the \normalfont{G2/DSO} with respect to Sgr~A*. \normalfont{Please consider Table \ref{tab:fit_positions}, Appendix \ref{sec:positions_app} for the derived relative position with respect to Sgr~A*. The related LOS velocities can be found in Table \ref{tab:spectral_velocity_properties}.}\newline
In Figure \ref{fig:dso_orbit_1}, we show the outcome of the Keplerian model fit. This solution underlines the Keplerian trajectory of the \normalfont{G2/DSO}. The best-fit orbital elements are listed in Table \ref{tab:orbit_elements}.
Based on the orbital elements, we find that a periapse passage of the \normalfont{G2/DSO} occurred in $\sim 2014.38$, which is in agreement with the line map detection (see the following subsection and Fig. \ref{fig:dso_line_evo}) as well as with the earlier calculation of the pericenter passage presented in \citet{Valencia-S.2015}, where the authors report $t_{\rm closest}=2014.39 \pm 0.14$.
\begin{figure}[htbp!]
	\centering
	\includegraphics[width=.5\textwidth]{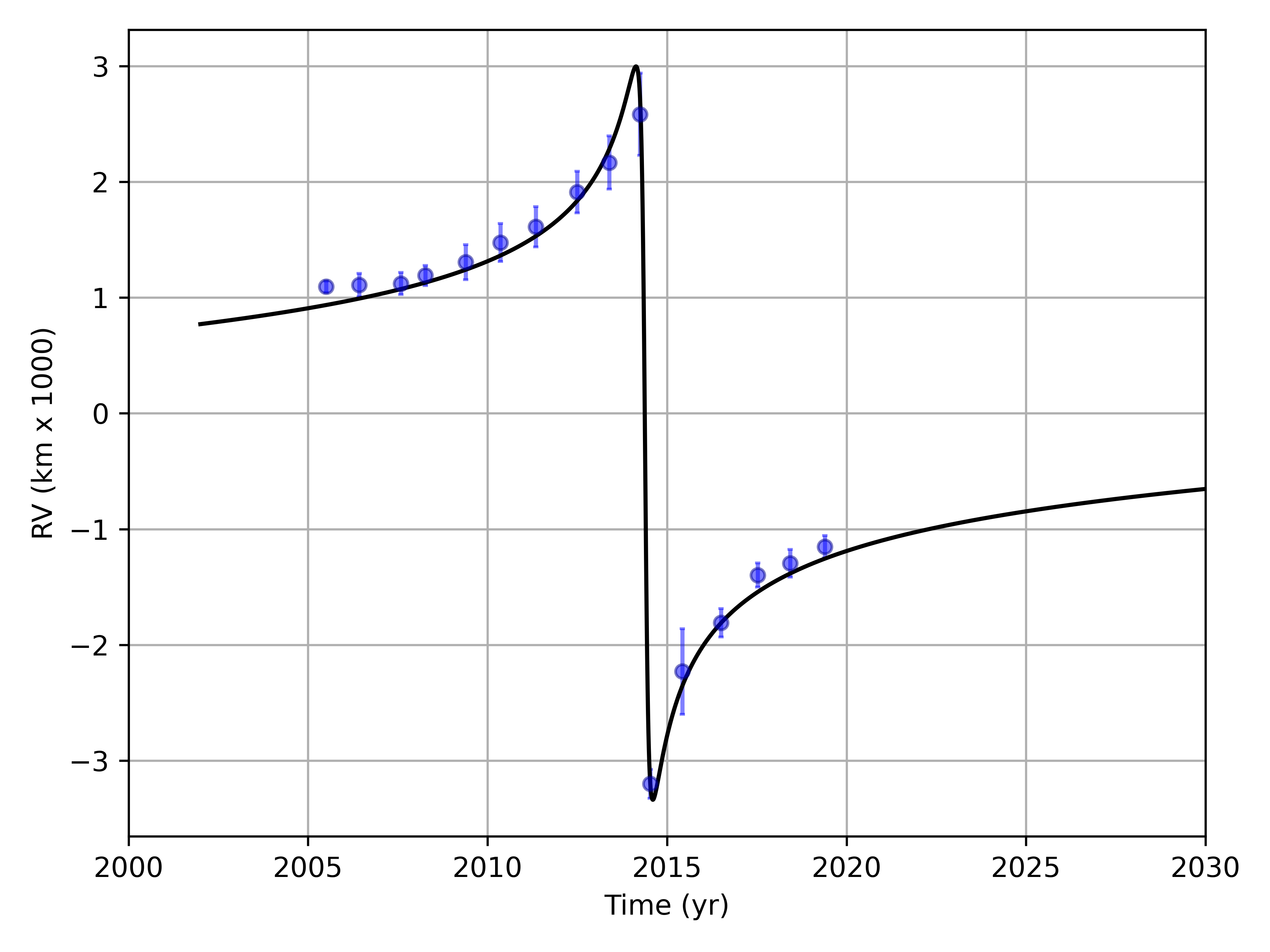}
	\caption{Line-of-sight velocity $v_z$ of the \normalfont{G2/DSO} as a function of time. As displayed, the \normalfont{G2/DSO} accelerates until 2014.3 in the redshifted Br$\gamma$ domain. After 2014.38, the velocity is blueshifted as the \normalfont{G2/DSO} continues its way on a Keplerian orbit. The uncertainties are adapted from Table \ref{tab:spectral_velocity_properties}.}
\label{fig:dso_orbit_2}
\end{figure}
In Fig. \ref{fig:dso_orbit_2}, we furthermore show the LOS velocity evolution of the \normalfont{G2/DSO} as a function of time. The absolute value of the LOS velocity of the \normalfont{G2/DSO} reaches $3338\,{\rm km/s}$. The common uncertainty for the velocities is $\pm\,35\,{\rm km/s}$ \citep[see][]{Peissker2020b}. 
We use the orbital elements given in Table \ref{tab:orbit_elements} to investigate the statistical robustness (Fig. \ref{fig:dso_orbit_3}, \normalfont{Appendix \ref{sec:mcmc_results_app}}). The MCMC simulations agree with the initial input parameters. Since the uncertainty does not reflect the character of the data nor the observations, again we choose a more conservative approach and round up the range of possible values.
\begin{table*}[htb]
\centering
\begin{tabular}{ccccccc}\hline \hline
Source  & $a$ (mpc)&	$e$  &	$i$($^o$)&$\omega$($^o$)&	$\Omega$($^o$)&	$t_{\rm closest}$(yr)\\ \hline 
S2    &5.04 $\pm$ 0.01   &0.884 $\pm$  0.002 &136.88 $\pm$ ~0.40  &~71.33 $\pm$  0.75  &234.51 $\pm$~1.03  &2002.32 $\pm$ 0.02\\
\normalfont{G2/DSO}  &17.45 $\pm$  0.20  &0.962 $\pm$ 0.004&58.72 $\pm$ 2.40&92.81 $\pm$  1.60&295.64 $\pm$ 1.37&2014.38 $\pm$ 0.01\\
\hline \hline
\end{tabular}
\caption{Orbital elements of S2 and \normalfont{G2/DSO} as shown in Fig. \ref{fig:fc}, \ref{fig:dso_line_evo}, \ref{fig:dso_orbit_1}. The orbital parameters for S2 are adapted from \cite{peissker2020a} \normalfont{and are used to determine the position of Sgr~A*}. As estimated in the text, the closest pericentre approach \normalfont{of G2/DSO} is about $0.66\,{\rm mpc}$ or $16.5\,{\rm mas}$ / $137\,{\rm AU}$.}
\label{tab:orbit_elements}
\end{table*}
The final uncertainties for the orbital elements (which are based on the Keplerian fit) are given in Table \ref{tab:orbit_elements}. Additional results of this analysis are the mass $M_{\rm Sgr~A*}$ of Sgr~A* of about $M_{\rm Sgr~A*}\,=\,4.15\,\times\,10^6\,M_{\odot}$ and the distance $d$ to the GC of $d\,=\,8.3$ kpc.
Based on the orbital elements, we find a pericentre distance $r_{\rm p}$ of $r_{\rm p}\,=\,a(1-e)\,\approx\,0.66\,{\rm mpc}$.

\subsection{Periapse of the \normalfont{G2/DSO} in 2014.38}

\begin{figure*}[htbp!]
	\centering
	\includegraphics[width=1.\textwidth]{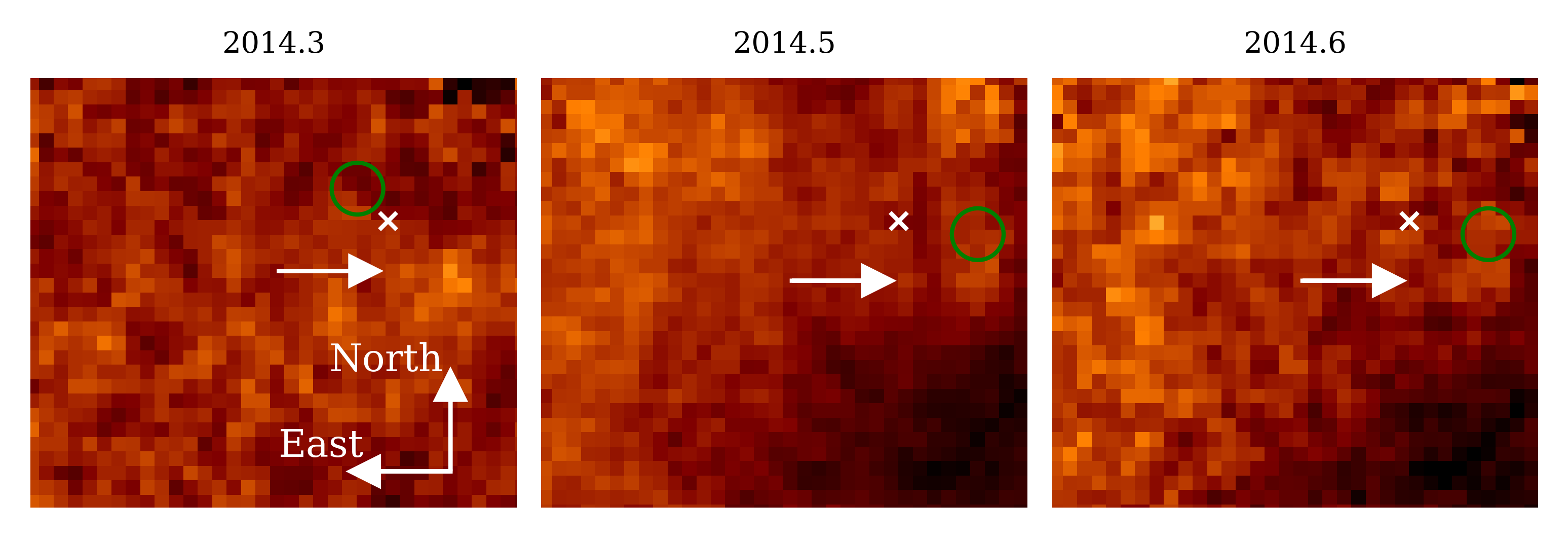}
	\caption{\normalfont{Line} maps that cover a range of about $1500\,{\rm km/s}$. \normalfont{This range is based on the PV diagrams as presented in \cite{Pfuhl2015}.} The white arrow marks the expected position of the blueshifted component of the \normalfont{G2/DSO} based on the analysis presented in \cite{Plewa2017} and \cite{Gillessen2019}. We mark the derived Keplerian position of the \normalfont{G2/DSO} with a green circle (see Fig. \ref{fig:dso_line_evo}). The noisy character of the SINFONI \normalfont{line} maps is here indicated by the intensity variation of the pixels. Because of the wide velocity range that is representative for the spectral range $2.145\,\mu m\,-\,2.155\,\mu m$ (120 channels), the line map is noise dominated which hinders the detection of the \normalfont{G2/DSO}. For each line map, we subtract the averaged 10 neighboring channels of the spectral range. Sgr~A* is located at the white $\times$. The size of every panel is $0.42"\,\times\,0.37"$.}
\label{fig:dso_2014_blue}
\end{figure*}

Since the literature suffers from inconsistencies regarding the nature of the \normalfont{G2/DSO} or the general periapse sequence \citep{Gillessen2012, Witzel2014, Valencia-S.2015, Pfuhl2015}, we focus on the observations that cover the epoch of 2014. Since it was predicted and shown that the gas cloud \normalfont{G2/DSO} can be observed simultaneously before and after its periapse, we will investigate the region of interest \citep[see Fig. 4 shown in ][]{Gillessen2019}. In combination with the velocity \normalfont{(and hence wavelength)} information given in \cite{Pfuhl2015}, we show in Fig. \ref{fig:dso_2014_blue} the reported area of the blueshifted part of \normalfont{G2/DSO} in 2014.3 (216 single exposures), 2014.6 (94 single exposures), and 2014.5 (310 single exposures).
Based on the information of the published data \citep[see Fig. 15 in ][]{Pfuhl2015}, we select the wavelength range $2.145\,\mu m\,-\,2.155\,\mu m$ which corresponds to the blueshifted velocities $1537\,-\,2922\,{\rm km/s}$. As shown in Fig. \ref{fig:dso_2014_blue}, the line maps suffer from noise and artefacts. This is consistent with \cite{Valencia-S.2015} where the authors encounter the same situation. However, we do not agree with the smoothed and artificially enhanced emission (the authors call it `scaling adjustment') that is shown in \cite{Pfuhl2015} for 2014.3. The \normalfont{G2/DSO} in Fig. \ref{fig:dso_2014_blue} cannot be observed since the \normalfont{line} maps are noise dominated (that is due to the large selected spectral range). We advise the interested reader to compare the significance of the Br$\gamma$ emission in Fig. \ref{fig:dso_line_evo} with Fig. \ref{fig:dso_2014_blue}. We will discuss this result in detail in Sec. \ref{sec:discussion}.

\subsection{Position-Position-Velocity maps}

\begin{figure*}[htbp!]
	\centering
	\includegraphics[width=1.\textwidth]{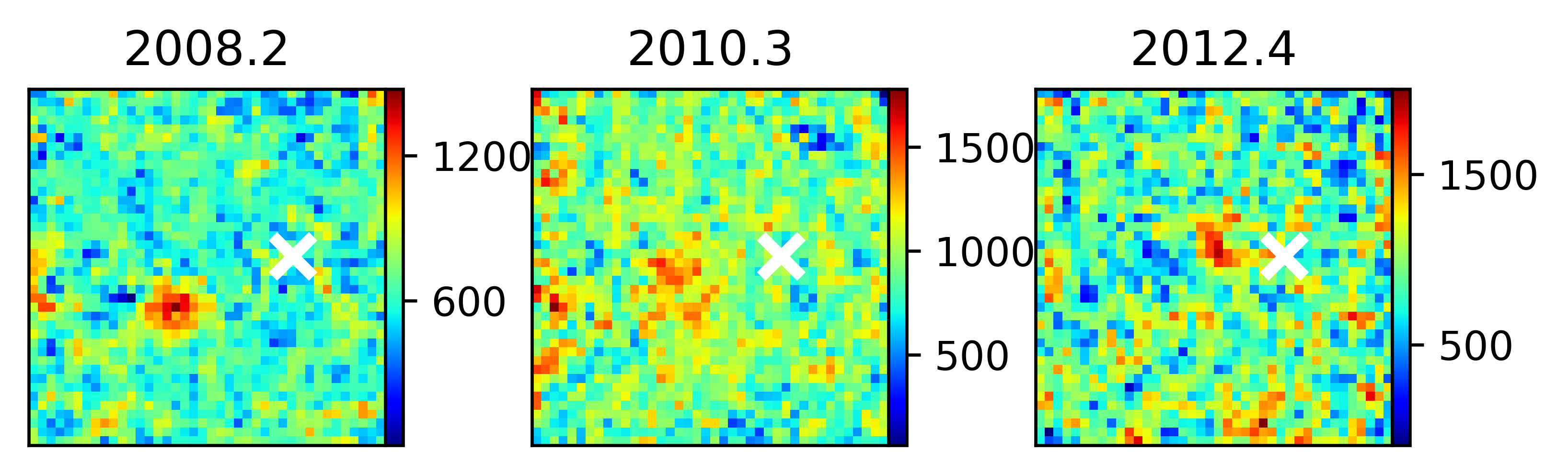}
	\caption{Position-Position-Velocity maps of the \normalfont{G2/DSO} for 2008, 2010, and 2012 epochs. North is up, East is to the left. The size of every panel is $0.5"\,\times\,0.5"$. The position of Sgr~A* is indicated with a white $\times$. We pick these \normalfont{epochs} since the velocity gradient as claimed by \cite{Pfuhl2015}, \cite{Plewa2017}, and \cite{Gillessen2019} should be traceable \normalfont{with reduced confusion compared to other years. In Fig. \ref{fig:ppv_2}, we present a smoothed version of this figure.} A detailed discussion of the result is given in Sec. \ref{sec:discussion}.}
\label{fig:ppv_1}
\end{figure*}

According to \citet{Pfuhl2015}, a clear velocity gradient should be observable, especially in the pre-pericentre phase of the \normalfont{G2/DSO}. To maintain the overall shape of the data and hence provide a certain level of comparability with the pipeline results (i.e., the output mosaic data cube), we use PPV diagrams. \normalfont{We favor this approach since the spatial dimensions of a PV diagram are reduced to one parameter which can result in confused detections. Since the outcome of a PV diagram purely rely on the derived orbit, we question the capability of this technique in the first place. However, for the here presented} PPV plots, the spectral pixel information is transformed to a velocity. In Fig. \ref{fig:ppv_1}, we present a selected overview of the pre-periapse time of the \normalfont{G2/DSO}. We divide the investigated spectral range of each year into equal-sized sections (i.e., spectral slices) and we subtract the neighboring channels. Following this procedure, we preserve the information and avoid a disintegration of the emission.\newline
By comparing the PPV plots, no prominent velocity gradient is present. Furthermore, the data again suffer from artefacts and noise. We underline that we do not use any Gaussian smoothing kernel to enhance nonlinear parts of the \normalfont{G2/DSO}.
Since the background subtraction influences the shape of the Doppler-shifted Br$\gamma$ emission line (see Appendix \ref{sec:tail_smoothed_app}), the velocity distribution may contain artefacts. Please see a smoothed version of Fig. \ref{fig:ppv_1} and a possible tail emission in Appendix \ref{sec:tail_smoothed_app}, Fig. \ref{fig:ppv_2}. In the following, we will investigate the influence of the sky emission on the data.

\subsubsection{Artificial velocity gradient}
As previously discussed in Sec. \ref{sec:data}, we analyse the efficiency of the sky correction in case the \normalfont{object}- and \normalfont{sky}-observations contain irregularities. For that purpose, we manipulate the sky correction with an error of $-10\%$ of the maximum flux of the input files \normalfont{, i.e., we subtract $10\%$ of the peak emission of the sky frames}. Furthermore, we smooth the original position-position-velocity map with a one pixel Gaussian. The results of this analysis are listed in Table \ref{tab:vg_error_comparison}.
\begin{table}[htbp!]
	\centering
	\begin{tabular}{cccc}\hline\hline
		
				 & 2008 & 2010 & 2012  \\
				 & $\left[{\rm km/s}\right]$ & $\left[{\rm km/s}\right]$ & $\left[{\rm km/s}\right]$ \\
	 \hline
		   Expected gradient & $\pm 45.58$ &$\pm 78.77$ & $\pm 101.2$      \\
		   No error & $\pm 19.97$ & $\pm 37.01$ &  $\pm 63.77$     \\
		    $10\%$ error & $\pm 35.99$ & $\pm 88.27$ &  $\pm 138.58$    \\
		    Smoothed, 1 px & $\pm 80.46$ & $\pm 143.94$ &  $\pm 200.5$     \\	
		   \cite{Plewa2017} & $\pm 238$ & $\pm 261$ &  $\pm 464$    \\	
		\hline	\\
	\end{tabular}
	\caption{Influence of a poor sky correction and Gaussian smoothing on the velocity gradient. Details are given in the text.}   
	\label{tab:vg_error_comparison}
\end{table}
While combining the observational data of an object for each year leads to a natural velocity gradient because of the intrinsic motion (``expected gradient"), we find lower values for the Br$\gamma$ emission of the \normalfont{G2/DSO}. Since the observations do not cover a complete year, this result is expected. However, measuring the velocity gradient of the \normalfont{G2/DSO} in the sky manipulated data shows increased values that are close or over the expected values. Smoothing the data increases the velocity gradient by about $30-50\%$. As shown in Fig. \ref{fig:ppv_1}, the noisy character of the data is responsible for this increased velocity gradient. Hence, smoothing data impacts the outcome of the velocity gradient analysis.

\subsection{K-band detection}

To investigate the possibility of a stellar source that is \normalfont{associated with the observed Doppler-shifted Br$\gamma$} line emission, we apply the Lucy-Richardson algorithm to the data \citep[see also ][]{Peissker2020b}. We select the $K$-band in the related data cube and apply a background emission of $15-30\%$ of the peak intensity emission \normalfont{depending on the variable background}. With this approach, we eliminate the chance of a false positive.
\begin{figure*}[htbp!]
	\centering
	\includegraphics[width=1.\textwidth]{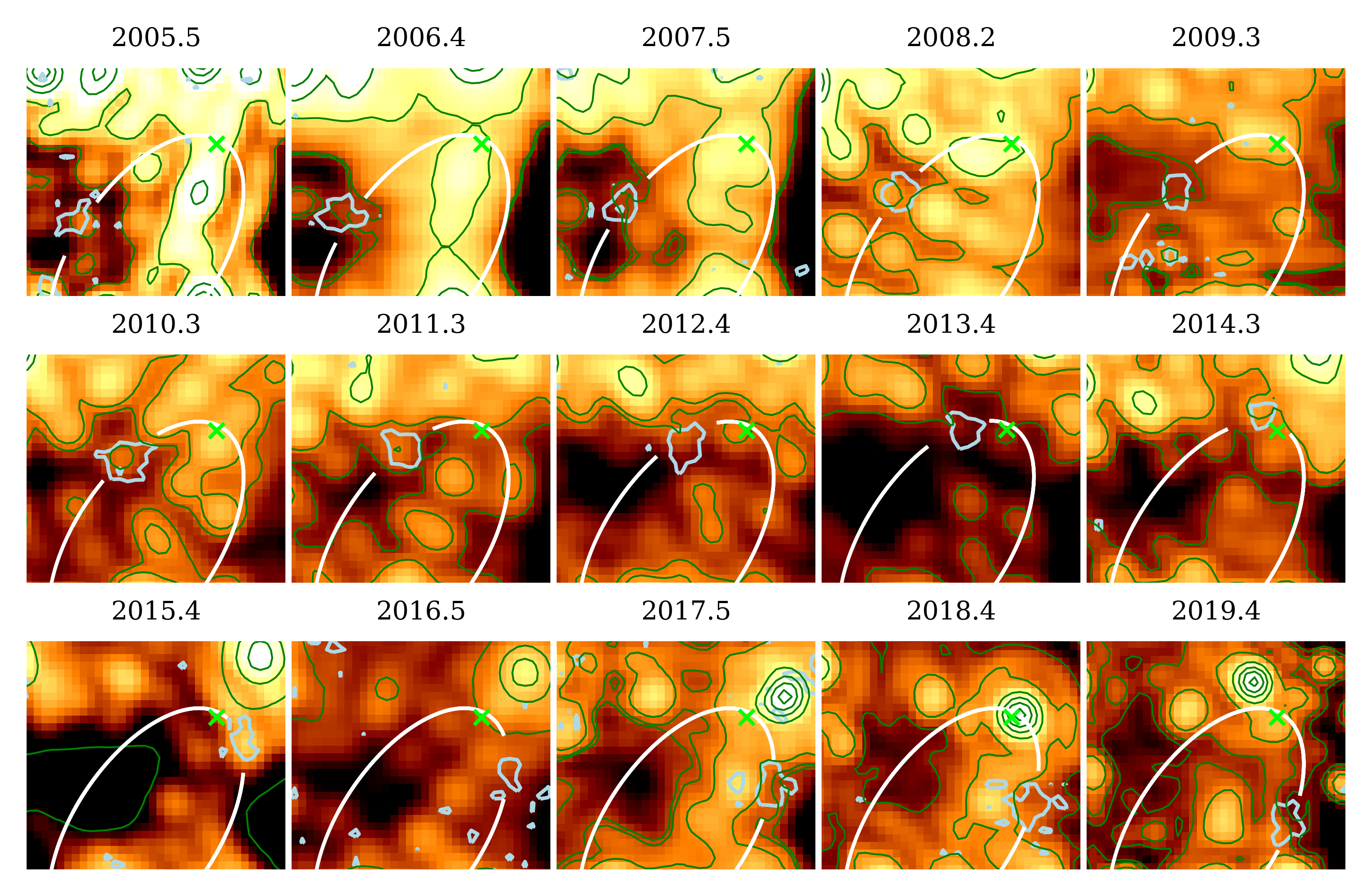}
	\caption{High-pass filtered $K$-band images of the GC region with the focus on the Keplerian \normalfont{G2/DSO} orbit. The location of Sgr~A* is indicated with a lime colored $\times$. The size of each panel is $0.4"\,\times\,0.3"$. North is up, East to the left. The green contour lines are at $10\%,\,30\%,\,50\%,\,70\%,$ and $90\%$ of the normalised S2 peak intensity. The lightblue colored contour lines are adapted from the Doppler-shifted Br$\gamma$ emission shown in Fig. \ref{fig:dso_line_evo}. \normalfont{Because of the high stellar density, the position of the K-band counterpart of G2/DSO may be confused. We refer the reader to the Appendix \ref{sec:positions_app}, Fig. \ref{fig:kband_orbit}, where we show the extracted K-band continuum position of G2/DSO and compare it to the here derived orbit.}}
\label{fig:lr_results}
\end{figure*}
In Fig. \ref{fig:lr_results}, we present the resulting convolved images. In agreement with the position of the Doppler-shifted Br$\gamma$ line emission, we detect a source that is approaching and passing Sgr~A* on a trajectory comparable to the \normalfont{G2/DSO}. \normalfont{By comparing the K-band continuum position with the Br$\gamma$ detection shown in Fig. \ref{fig:dso_line_evo},} we find a small offset of $\leq\,12.5$ mas ($\sim 100$ AU) between the centroid of the stellar source and \normalfont{the line emission. For this comparison, please consider Fig. \ref{fig:kband_orbit}, Appendix \ref{sec:positions_app}.} In 2016 and 2017, S31 and S23 (see Fig. \ref{fig:fc}) coincide with the blueshifted \normalfont{G2/DSO} Br$\gamma$ line position. 

\subsection{Stellar K-band magnitude}

To investigate the $K$-band continuum magnitude and the flux density evolution of the \normalfont{G2/DSO} along the orbit, we use the detection of the high-pass filter analysis shown in Fig. \ref{fig:lr_results}. For the analysis, the zero magnitude flux is set to $653$ Jy and an effective wavelength of $2.180\,\mu m$ is applied. These settings are related to the ESO $K$-band filter \citep[for comparable values, see also][]{Tokunaga2007}.
\begin{figure}[htbp!]
	\centering
	\includegraphics[width=.5\textwidth]{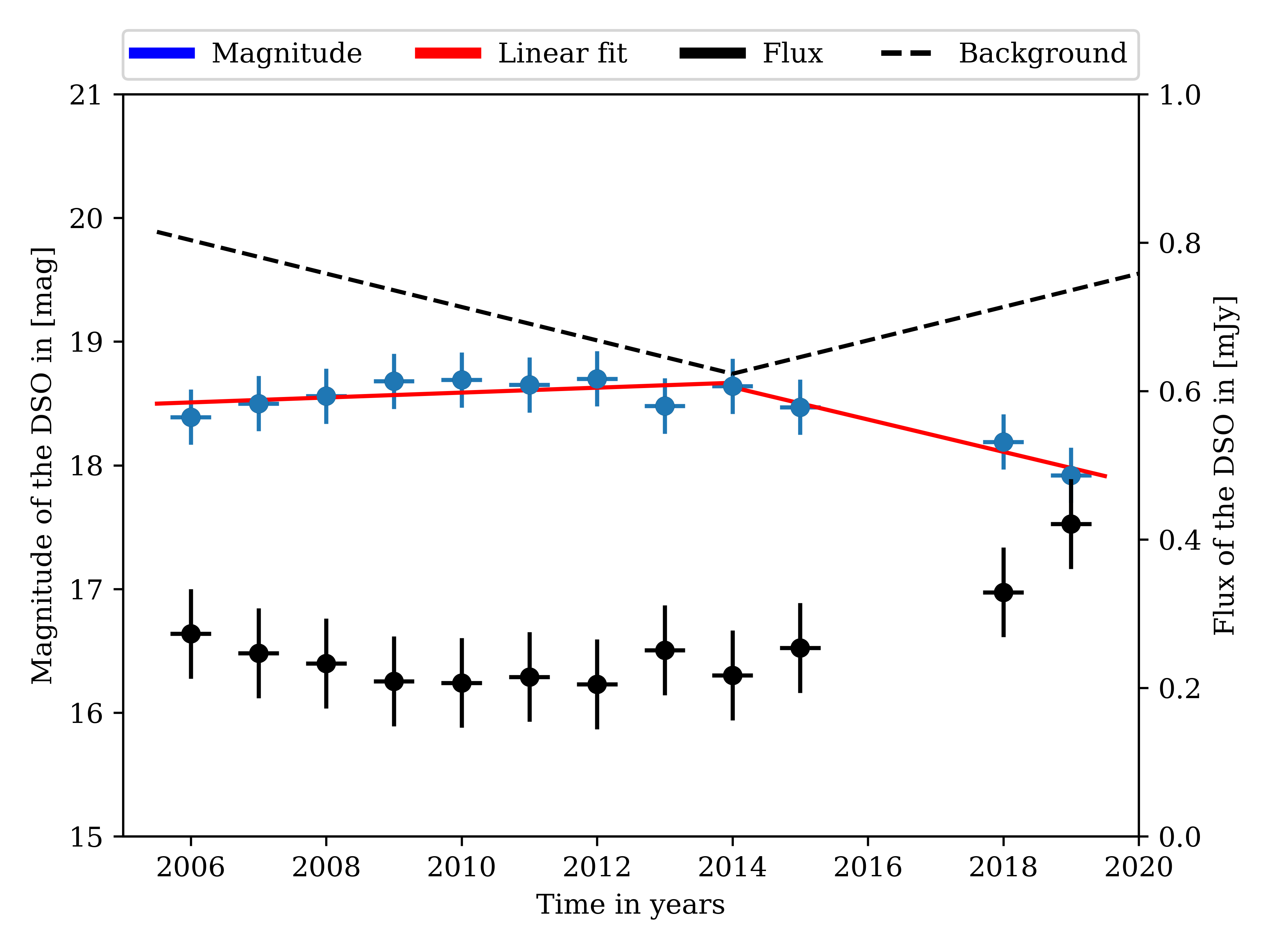}
	\caption{$K$-band flux and magnitude of the \normalfont{G2/DSO} based on the high-pass filter detection. The blue data points are related to the magnitude where lower values correspond to a brighter emission. Black data points represent the magnitude-correlated flux. The error bars represent the standard deviation. \normalfont{The dashed black line corresponds to the observed background magnitude which is measured at a distance of $\sim\,40$ mas to G2/DSO. In 2014, G2/DSO reaches pericenter. Hence, this epoch represents the closest distance towards Sgr~A* which correlates to the increased background light. The derived slope of $0.13$ of the observed background magnitude agrees with the analysis presented in \cite{Sabha2012}.}}
\label{fig:mag}
\end{figure}
For the magnitude, we use the extinction-corrected S2 magnitude of 14.15 mag. In the following, we adopt the bolometric magnitude relation with
\begin{equation}
    m_{\rm \normalfont{G2/DSO}}\,=\,-14.15\,+\,2.5\,\times\,\log(\mathrm{ratio})
\end{equation}
where `ratio' is defined as the counts of the object of interest (here G2/DSO) and the reference star. Since we normalize the data to the flux of S2, `ratio' simplifies to the counts of the G2/DSO. \normalfont{Because of confusion, we exclude the data points of 2005, 2016, and 2017.} While the uncertainty is based on the standard deviation, we find an average magnitude of the \normalfont{G2/DSO} of $18.48\,\pm\,0.22$ mag. For the averaged flux \normalfont{$f$, we adapt}
\begin{equation}
    f_{\rm G2/DSO}\,=\,f_{\rm S2}\,\times\,10^{-0.4(\mathrm{mag}_{\rm G2/DSO}-\mathrm{mag}_{\rm S2})}
\end{equation}
\normalfont{from \cite{Sabha2012} to calculate the related flux density of G2/DSO with $f_{\rm S2}\,=\,14.23\,$mJy and $\mathrm{mag}_{\rm S2}\,=\,14.1$. We get $0.25\,\pm\,0.06\,$mJy which is fully consistent with the literature.}\newline
\normalfont{Comparing the pre- and post-periapse epochs, we find a slightly decreasing magnitude towards Sgr~A*. This is expected since \cite{Sabha2012} shows that the density of (old) faint stars increases towards Sgr~A*. To eliminate the chance that the here presented magnitude is correlated to background fluctuations or nearby stars, we measure its intensity close to G2/DSO with a one-pixel aperture at a distance ($\sim\,40$ mas) larger than the radius of the SINFONI PSF ($31.25-37.5$ mas). Based on Fig. \ref{fig:mag}, we find that neither the pre- nor post-periapse data is correlated to background fluctuations or surrounding stars. It is reasonable to assume, that the K-band magnitude of G2/DSO would increase because of close-by stars or the dominant background light towards Sgr~A*. In Section~\ref{sec:discussion}, we will elaborate on the here mentioned points.}

\subsection{The ``tail"}

\begin{figure}[htbp!]
	\centering
	\includegraphics[width=.5\textwidth]{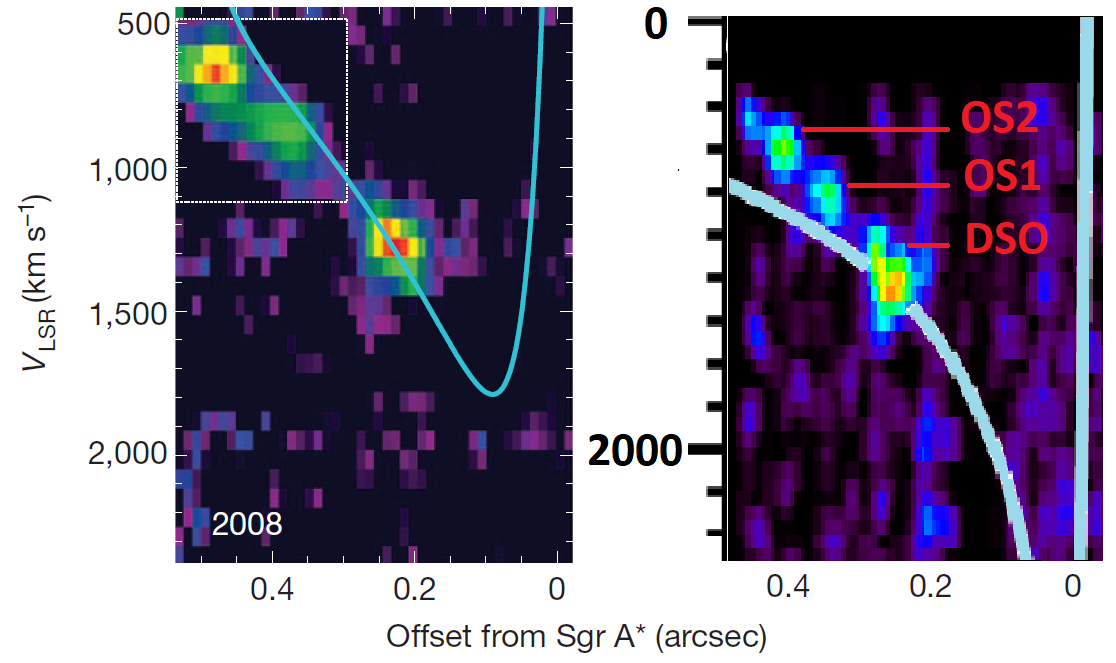}
	\caption{Comparison of Position-Velocity diagrams of 2008 adapted from \cite{Gillessen2012} (left panel) and \cite{Peissker2018} (right panel). The dotted box in the left panel shows smoothed emission. The emission outside the dotted box is not smoothed. The right panel is not treated with an dominant Gaussian kernel and reveals the two isolated sources OS1 and OS2 that are \normalfont{temporarily close to G2/DSO}. In both panels, the Keplerian orbit is shown in lightblue.}
\label{fig:tail_1}
\end{figure}

As it is already shown and described in \cite{Peissker2018}, we created a series of PV diagrams inspired by \cite{Gillessen2012} where the authors show the gas cloud \normalfont{G2/DSO} and its tail component with a Keplerian orbit approaching Sgr~A*. Since \cite{Gillessen2012} smooth parts of the presented image (see the dotted box in the left panel of Fig. \ref{fig:tail_1}), we investigate the non-smoothed emission \normalfont{between $2.1661\,\mu m$ and $2.182\,\mu m$} in contrast.
Without an imperious smoothing kernel that is applied to the data, we confirm emission at the position of the so-called tail that is above the noise level. However, we cannot agree with the interpretation of this emission since we detect isolated sources that show \normalfont{temporary close distance} to the G2/DSO (see the right panel of Fig. \ref{fig:tail_1}). In the following subsections, we will investigate these sources, which we refer to as OS1 and OS2, in detail. Additional material covering the so-called tail can be found in the Appendix \ref{sec:tail_smoothed_app}, see Fig. \ref{fig:art_tail_2006_2008_2012} and Fig. \ref{fig:ppv_2}.

\subsubsection{OS1}

\begin{figure*}[htbp!]
	\centering
	\includegraphics[width=1.\textwidth]{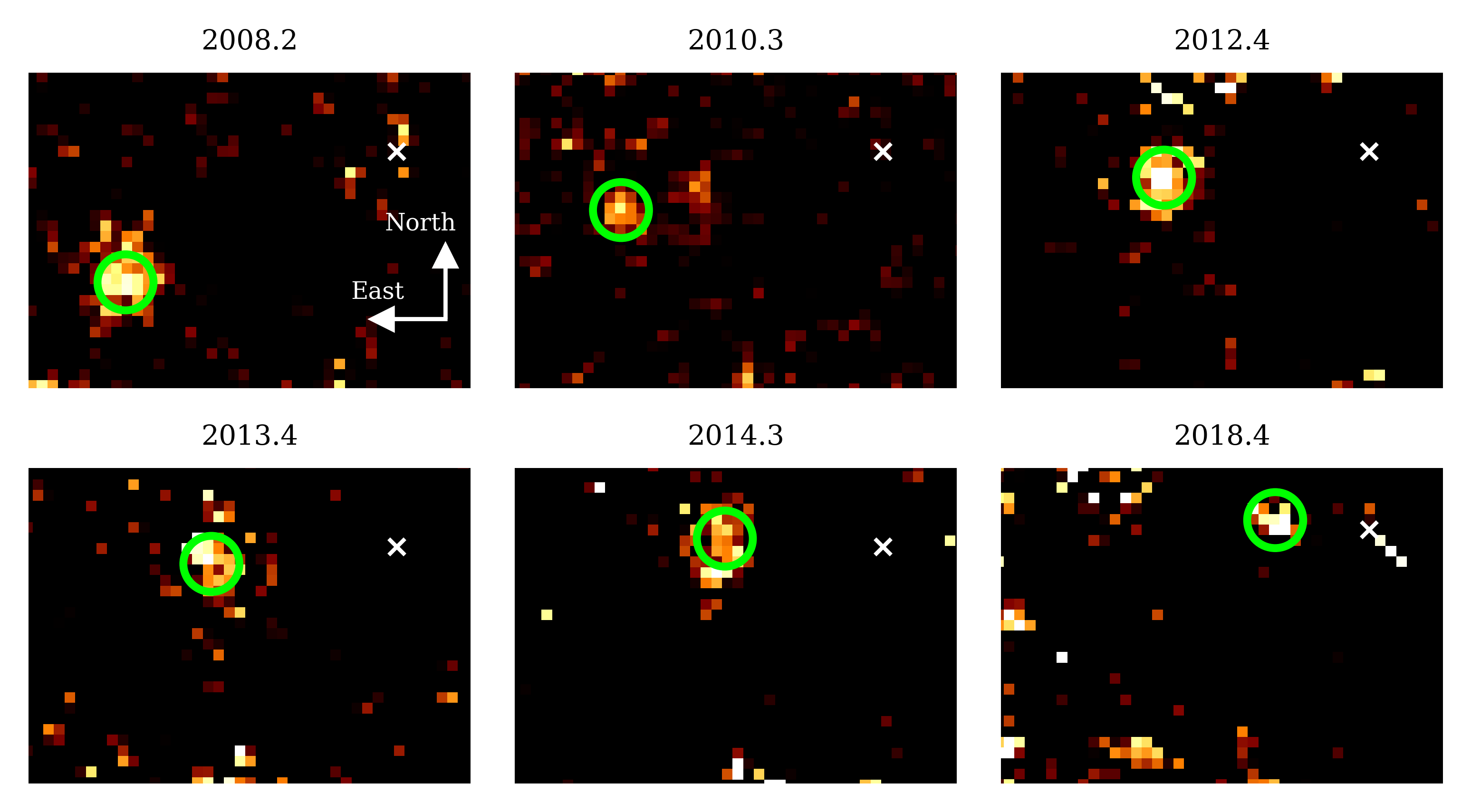}
	\caption{The detection of OS1 in the Doppler-shifted Br$\gamma$ regime (marked with a lime-colored circle). As indicated, North is up, East is to the left. The position of Sgr~A* is marked with a white $\times$.we note that OS1 is consistent with compact object seen at the diffraction limit of the telescope. The size of each panel is $0.52"\,\times\,0.37"$.}
\label{fig:os1_line_1}
\end{figure*}

\begin{figure}[htbp!]
	\centering
	\includegraphics[width=.5\textwidth]{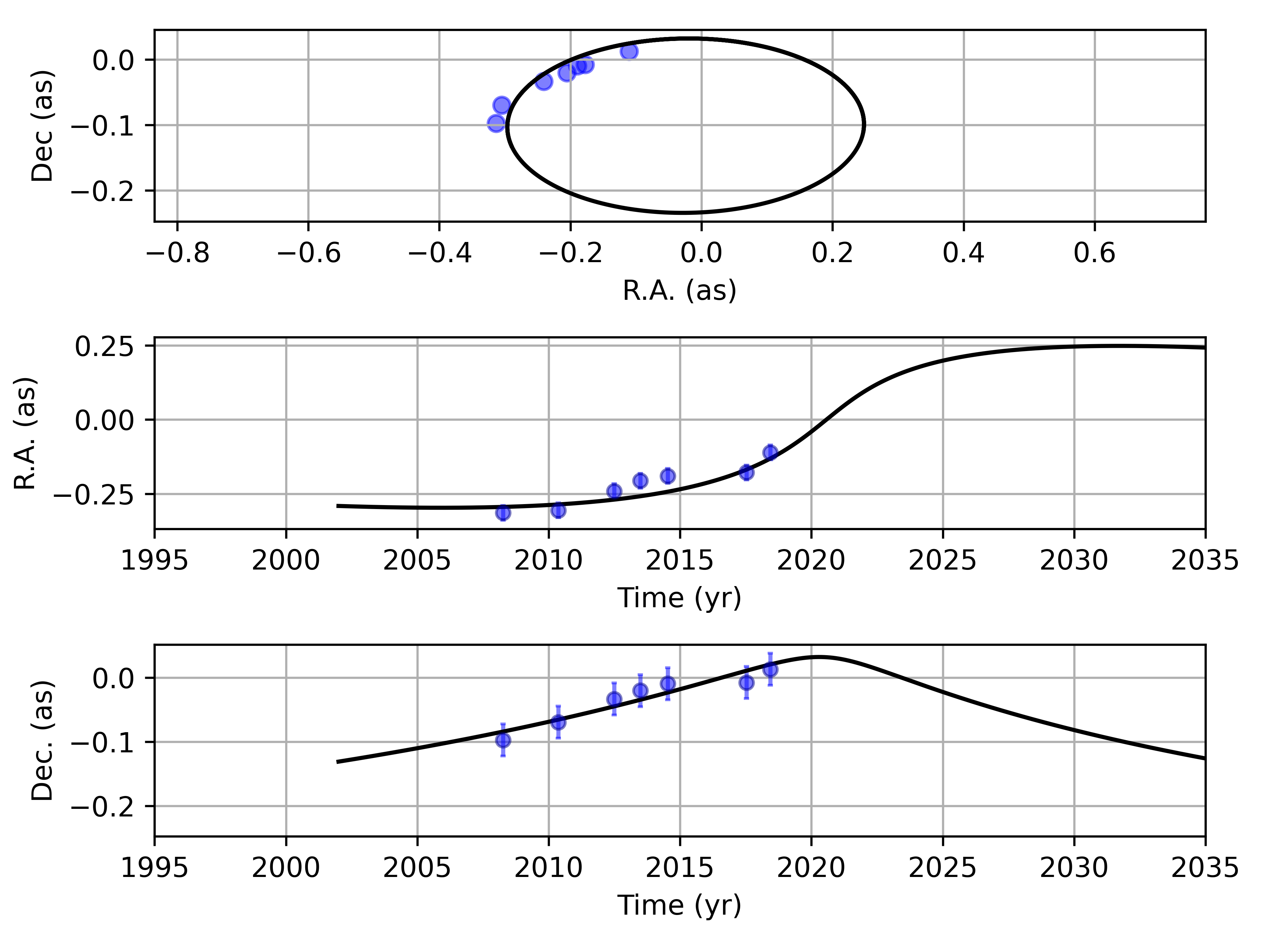}
	\caption{A Keplerian orbit of the line-emission source OS1 based on the Br$\gamma$ analysis. The pericenter passage of OS1 is dated to \normalfont{be around mid-2020.}}
\label{fig:os1_orbit_1}
\end{figure}

As indicated in Fig.~\ref{fig:tail_1} (right panel), we identify OS1 in several epochs following the \normalfont{G2/DSO} on a similar orbit (see Fig. \ref{fig:os1_line_1} and Fig. \ref{fig:combined_orbit}). In several epochs, the identification of OS1 suffers from a decreased data quality and nearby stellar sources.
Using the extracted positions, we derive an orbit for OS1 (see Fig. \ref{fig:os1_orbit_1}). The uncertainties reflect the distance to nearby stellar sources and include the discussed Sgr~A* position range. 
In Table \ref{tab:orbit_elements_os}, we list the related orbital elements. 
\begin{table*}[htb]
\centering
\begin{tabular}{ccccccc}\hline \hline
Source  &$a$ (mpc)&	$e$  &	$i$($^o$)&$\omega$($^o$)&	$\Omega$($^o$)&	t$_{\rm closest}$(yr)\\ \hline 
OS1    &16.75 $\pm$ 0.50   &0.762 $\pm$  0.075 & 71.39$\pm$ 20.68  &93.79 $\pm$  8.42  &271.52 $\pm$ 16.50  &2020.67 $\pm$ 0.02\\
OS2  &12.48 $\pm$  0.19  &0.605 $\pm$ 0.019& 84.62 $\pm$ 4.98& 140.26 $\pm$  4.92&245.28 $\pm$ 2.17&2029.87 $\pm$ 0.05\\
\normalfont{G2/DSO}  &17.45 $\pm$  0.20  &0.962 $\pm$ 0.004&58.72 $\pm$ 2.40&92.81 $\pm$  1.60&295.64 $\pm$ 1.37&2014.38 $\pm$ 0.01\\

\hline \hline
\end{tabular}
\caption{Orbital elements for the sources OS1 and OS2. For comparison, we additionally list the orbital elements of the \normalfont{G2/DSO} (Tab. \ref{tab:orbit_elements}). The pericenter distance of OS1 and OS2 is 3.99 mpc/822 AU and 4.93 mpc/1017 AU, respectively. For comparison, the pericenter distance of the \normalfont{G2/DSO} is 0.66 mpc/137 AU.}
\label{tab:orbit_elements_os}
\end{table*}
With $r_{\rm p}\,=\,a(1-e)$, we derive a pericenter distance of 1201.56 AU (100.08 mas) for OS1.\newline
Please see Fig. \ref{fig:os1_velorbit} in the Appendix \ref{sec:los_os_app} for the spectral line evolution \normalfont{, Table \ref{tab:spectral_velocity_properties_ossources} and Table \ref{tab:fit_positions} for the related positions and LOS velocities.}


\subsubsection{OS2}

\begin{figure}[htbp!]
	\centering
	\includegraphics[width=.5\textwidth]{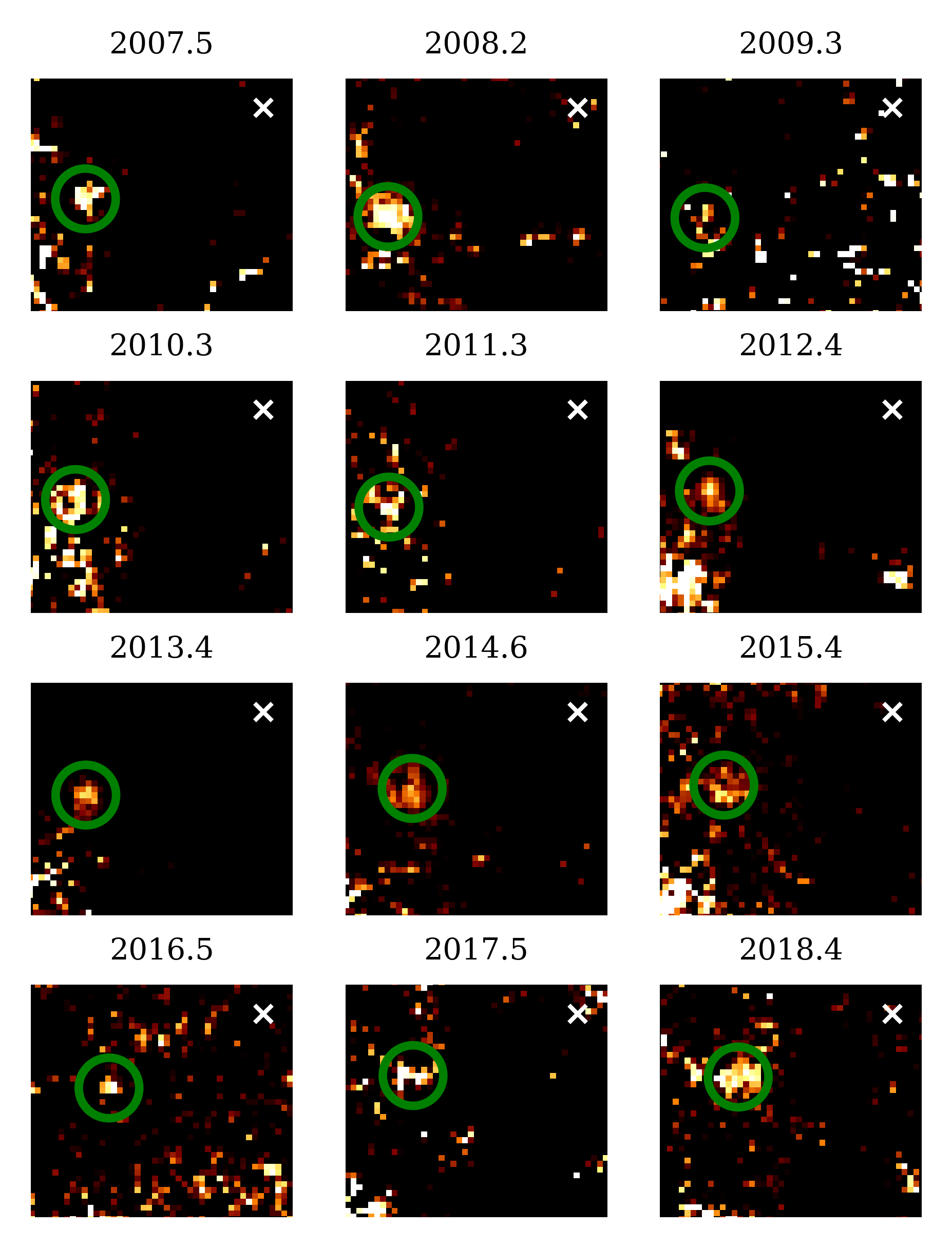}
	\caption{Br$\gamma$ line evolution of OS2 between 2007 and 2018. Because of nearby stars, the Br$\gamma$ line is affected by stellar emission (see especially 2009). In 2012, 2013, and 2014, OS2 is isolated in relation to close-by stars. Sgr~A* is indicated with a white $\times$. Here, North is up, East is to the left. The size of each panel is $0.56"\,\times\,0.50"$.}
\label{fig:os2_line_1}
\end{figure}

\begin{figure}[htbp!]
	\centering
	\includegraphics[width=.5\textwidth]{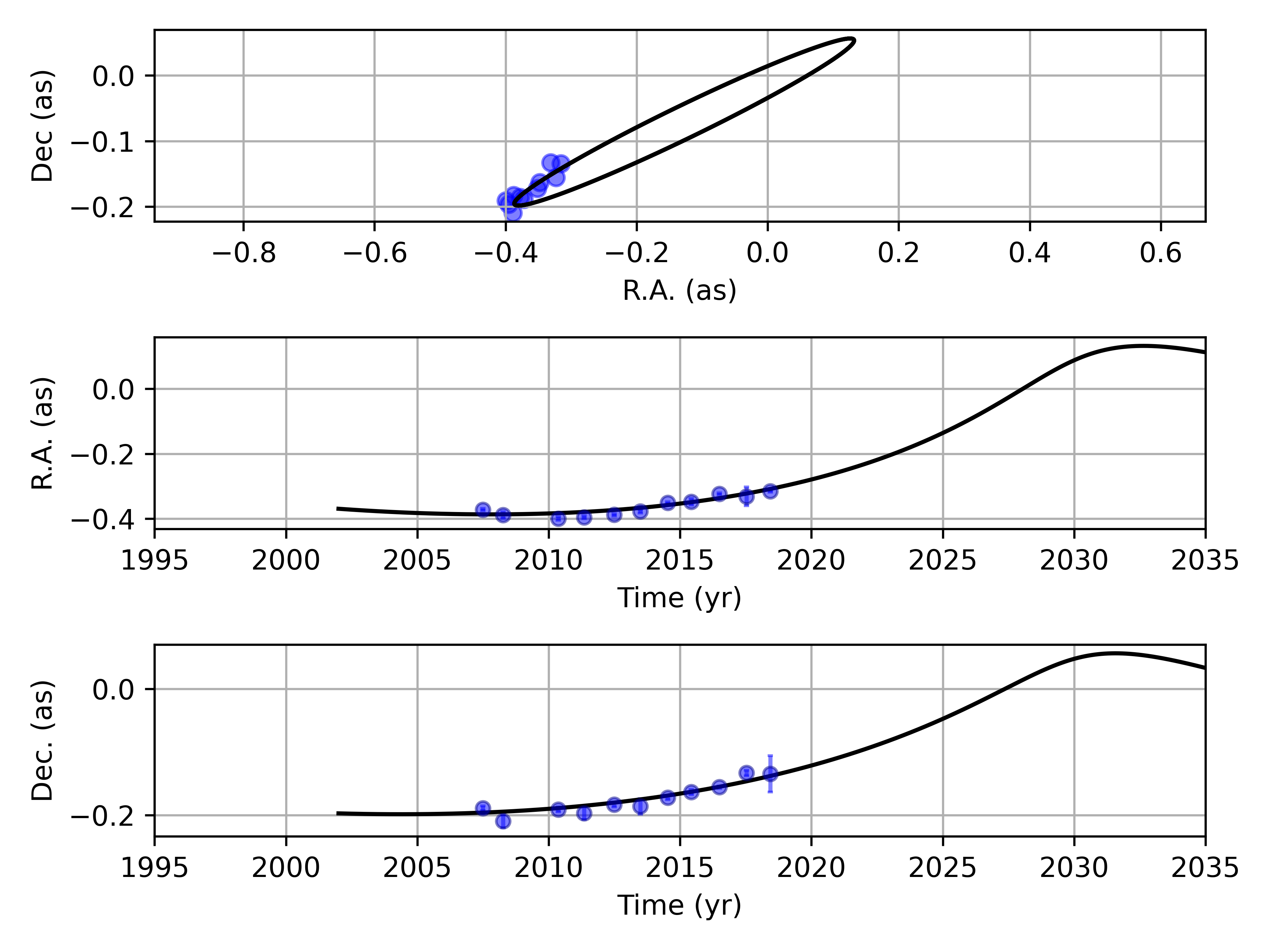}
	\caption{The orbit of OS2 as it approaches Sgr~A*. Because a confusion-free detection of OS2 cannot be guaranteed in 2009, we exclude this data point from the orbital fit. We derive a periapse distance $r_{\rm p}$ of OS2 of about 13 AU in 2029.87 (see the text for details).}
\label{fig:os2_orbit_1}
\end{figure}

\begin{figure}[htbp!]
	\centering
	\includegraphics[width=0.5\textwidth]{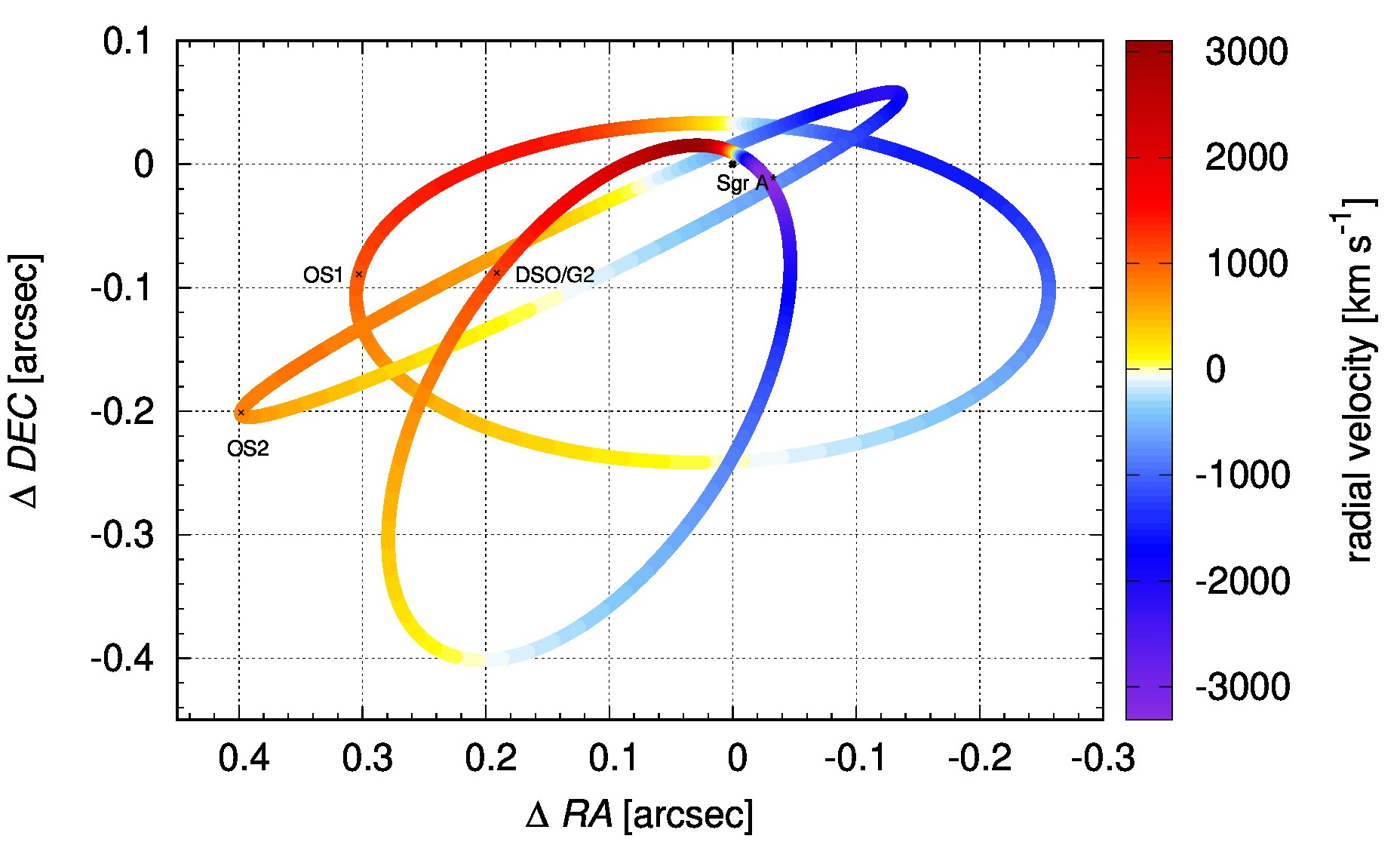}
	\caption{Combined orbit plot of the \normalfont{G2/DSO}, OS1, and OS2 (sources are indicated in the figure). All three compact sources can clearly be separated from each other. The position of Sgr~A* is located at (0,0). The black dots mark the Keplerian derived position of the related source in 2008. The fit is based on the data shown in Fig. \ref{fig:dso_orbit_1}, Fig. \ref{fig:os1_orbit_1}, and Fig. \ref{fig:os2_orbit_1}. We furthermore implement the LOS velocity along the orbit (see also  Fig. \ref{fig:dso_orbit_2}, Fig. \ref{fig:os1_velorbit}, and Fig. \ref{fig:os2_velorbit}).}
\label{fig:combined_orbit}
\end{figure}

\normalfont{Inspecting Fig. \ref{fig:tail_1} (right panel) implies that OS2 can be observed with} an increased intensity count compared to OS1 \normalfont{in 2008}. Hence, the object \normalfont{could be} less prone to confusion due to nearby stellar sources. \normalfont{Given the fluctuating background in the S-cluster and the changing distance to nearby stars, the intensity difference in 2008 may not be true for every other epoch. However, using} the analysing tools that we already applied to the \normalfont{G2/DSO} and OS1, we find OS2 throughout the data approaching Sgr~A* (Fig. \ref{fig:os2_line_1}).
We use a Keplerian orbital fit and find with a satisfying agreement with the data a plausible solution for the trajectory of OS2 (Fig. \ref{fig:os2_orbit_1}).
The related orbital elements can be found in Table \ref{tab:orbit_elements_os}. Using the relation for the pericenter distance, $r_{\rm p}\,=\,a(1-e)$, we calculate a pericenter distance of 1485.82 AU or 123.75 mas for OS2. In the Appendix \ref{sec:los_os_app}, in Fig. \ref{fig:os2_velorbit}, we show the Doppler-shifted LOS velocity evolution (Br$\gamma$ based) with the related fit. \normalfont{Furthermore, we list the positions and LOS velocities in Table \ref{tab:spectral_velocity_properties_ossources} and Table \ref{tab:fit_positions}, respectively.}

\section{Discussion} \label{sec:discussion}
In the analysis, we focused on the profound effects of sky subtraction and the smoothing on the analysis of the \normalfont{G2/DSO}. The source is identified as compact within the measurement uncertainties both before and after the pericenter passage. The identified $K$-band counterpart supports the stellar nature of the source. The previously claimed tail emission can be disentangled into discrete sources in some epochs (OS1 and OS2). \normalfont{We showed that the Gaussian smoothing in noisy line maps enhances the previously claimed tail emission which instead can be disentangled into discrete sources in some epochs (OS1 and OS2) if no smoothing is applied.}
In the following, we will discuss the presented results and provide an answer regarding the nature of the \normalfont{G2/DSO}. 

\subsection{Detection of the \normalfont{G2/DSO} between 2005-2019}

By inspecting and analyzing the SINFONI \normalfont{line} maps, we find the \normalfont{G2/DSO} on its Keplerian trajectory around Sgr~A* between 2005 and 2019 in the Doppler-shifted Br$\gamma$ regime. The Br$\gamma$ line is less affected by tellurics but coincides with OH emission lines with a relative flux between $20\,-\,50\,\%$. As we have shown in Sec. \ref{sec:data} and Sec. \ref{sec:results}, image motion but also the sky emission variability influences the analysis of the \normalfont{G2/DSO}. While the distortions of long-time SINFONI exposure data cubes do show nonlinear inconsistencies regarding object positions (and hence impact positional uncertainties), the shape of the emission lines are affected by an insufficient sky correction. By introducing an error to the sky correction files, we find values for the velocity gradient that are twice as prominent as the observed value. Arguably, our naive approach of the sky variability may not cover every aspect of the topic since this is beyond the purpose of this analysis. Hence, it is safe to assume that this introduced error could be increased or decreased in reality.\newline
Since the final data cubes cover a wide range of data of the related epochs, a ``natural" velocity gradient is expected. However, smoothing the data leads to a velocity gradient that is twice as large as the expected value. This is awaited because of the noisy character of the SINFONI data.\newline
Nevertheless, the confusion-free detection of the Doppler-shifted Br$\gamma$ emission line implies a Keplerian orbital evolution. The periapse of the \normalfont{G2/DSO} can be dated to 2014.38 with a pericenter distance of about 137 AU. Comparing the line emission before and after the pericenter passage reveals a preserved shape of the observed envelope (Fig. \ref{fig:dso_line_evo}), which implies that the gravitational influence of Sgr~A* is in the uncertainty domain \citep[][]{Eckart2013}. The LOS velocity follows the evolution expected for the Keplerian trajectory. Based on the Doppler-shifted \normalfont{G2/DSO} line analysis, we find the maximum values of $v_z\,=\,3000\,{\rm km/s}$ (pre-periapse) and $v_z\,=\,3200\,{\rm km/s}$ (post-periapse) which is about $1\%$ of the speed of light.
We furthermore investigated the blueshifted \normalfont{line} maps in 2014 that should show a prominent structure as shown in \cite{Pfuhl2015}. As it is already indicated in \cite{Valencia-S.2015}, the \normalfont{line} maps do not show comparable structures as it is presented in \cite{Pfuhl2015} and \cite{Plewa2017}. Using a slit along the orbit in combination with a smoothing kernel and enhancing only parts of the presented image may lead to a false interpretation of the data (see Fig. \ref{fig:tail_1} and Tab. \ref{tab:vg_error_comparison}). To underline this point, we presented nonsmoothed position-position-velocity diagrams of the \normalfont{G2/DSO} of 2008, 2010, and 2012. In these years, the source is rather isolated in relation to nearby stars but should also exhibit a noticeable velocity gradient. However, the analysis shows a rather compact source that suffers from the background noise. In the Appendix \ref{sec:tail_smoothed_app}, in Fig. \ref{fig:ppv_2}, we smooth the data with a \normalfont{3} px Gaussian kernel. The results do show indeed some structures that could be interpreted as a possible tail structure (marked in Fig. \ref{fig:ppv_2}, Appendix \ref{sec:tail_smoothed_app}). Unfortunately, the nonsmoothed data do not support this interpretation because of the noise. Hence, placing a slit along the orbit and smoothing the emission will most likely produce artefacts. We will investigate this point separately in detail in an upcoming publication since it would exceed the scope of this work.

\subsection{Stellar counterpart of the gas emission}

\normalfont{As it was first proposed by \cite{MurrayClay2012}, the presence of a stellar counterpart surrounded by a gaseous-dusty photoevaporating proto-planetary disc can explain both the ionized gas of $\sim 10^4\,{\rm K}$ traced by broad Br$\gamma$ emission as well as the dust component revealed by the prominent excess towards longer infrared wavelengths, in particular $L$- and $M$-bands. The} color-color diagram ($H-K$ vs. $K-L$) of \cite{Eckart2013}, the foreshortening factor temporal evolution presented in \cite{Valencia-S.2015}, the detected polarized continuum light by \cite{Shahzamanian2016} as well as the derived SED based on the 3D dusty model by \citet{Zajacek2017} as well as \cite{Peissker2020b} \normalfont{all support the dust-enshrouded star model of G2/DSO. These findings are also consistent with \cite{Scoville2013} who suggested a supersonic low-mass T-Tauri star with a stellar wind that produces a two-layer bow-shock while interacting with the ambient hot X-ray emitting gas. In this scenario, Br$\gamma$ emission line is produced in the colder and denser stellar-wind shock via the collisional ionization at the shock front and/or the cooling X-ray/UV radiation of the post-shock gas. Additionally, \cite{Ciurlo2020} used a stellar model for the findings of the so-called G-sources which is compatible with the discussion of the same objects in \citet{Peissker2020b}. \citet{Witzel2014,Witzel2017}, who also favor a dust-embedded star scenario, came to the same conclusion by analyzing the $L$-band flux density before and after the periapse of G2/DSO and G1, respectively. They found a constant flux density for G2/DSO within uncertainties, which implies the compact dust-enshrouded stellar source, while for G1 they detected a drop by nearly 2 magnitudes that can be interpreted by the tidal truncation of an extended envelope. The constant flux behavior of G2/DSO can be confirmed for its pericenter passage in this work as well. We note a slight flux increase for the data between 2018 and 2019. This could be explained by the partial removal of an envelope material during the pericenter passage where the G2/DSO host star is more revealed and as a result the overall $K$-band flux density increases. \citet{Eckart2013} suggest that the location of the Lagrange point L1 hinders a complete disruption of the envelope, since the denser component that is inside approximately one astronomical unit is bound to the star, see the tidal (Hill) radius estimate given by Eq.~\eqref{eq_Hill_radius}. Detailed numerical models are beyond the scope of this work but should be investigated in future works based on our findings.}\newline 
\normalfont{Since we used} a high-pass filter, we minimized the contaminating influence of overlapping PSF (-wings) and maximize the accessible information. We find a stellar counterpart at the expected position in agreement with the Br$\gamma$ emission line that is moving on the same orbit as the \normalfont{G2/DSO}. \normalfont{We want to note that we already investigated the broad spectrum of results that can be derived by using a high-pass filter, see in particular the results presented in \citet{peissker2020a}. We find that a high number of iterations ($\sim\,10000$ iterations) in combination with a solid PSF can result in robust detections.}\newline 
As discussed in \cite{Eckart2013} and \cite{peissker2021} but also shown in \cite{Sabha2012}, the possibility for a side-by-side flyby becomes negligible after about 3 years. Even though the orbit of S23 and S31 interfered with the trajectory of the \normalfont{G2/DSO} during the years 2015-2017 and 2019, we confirm the \normalfont{robust} detection of a stellar source at the position of the \normalfont{G2/DSO for most of the investigated years}.

\subsection{The tail of G2/DSO}

\begin{figure}[htbp!]
	\centering
	\includegraphics[width=0.5\textwidth]{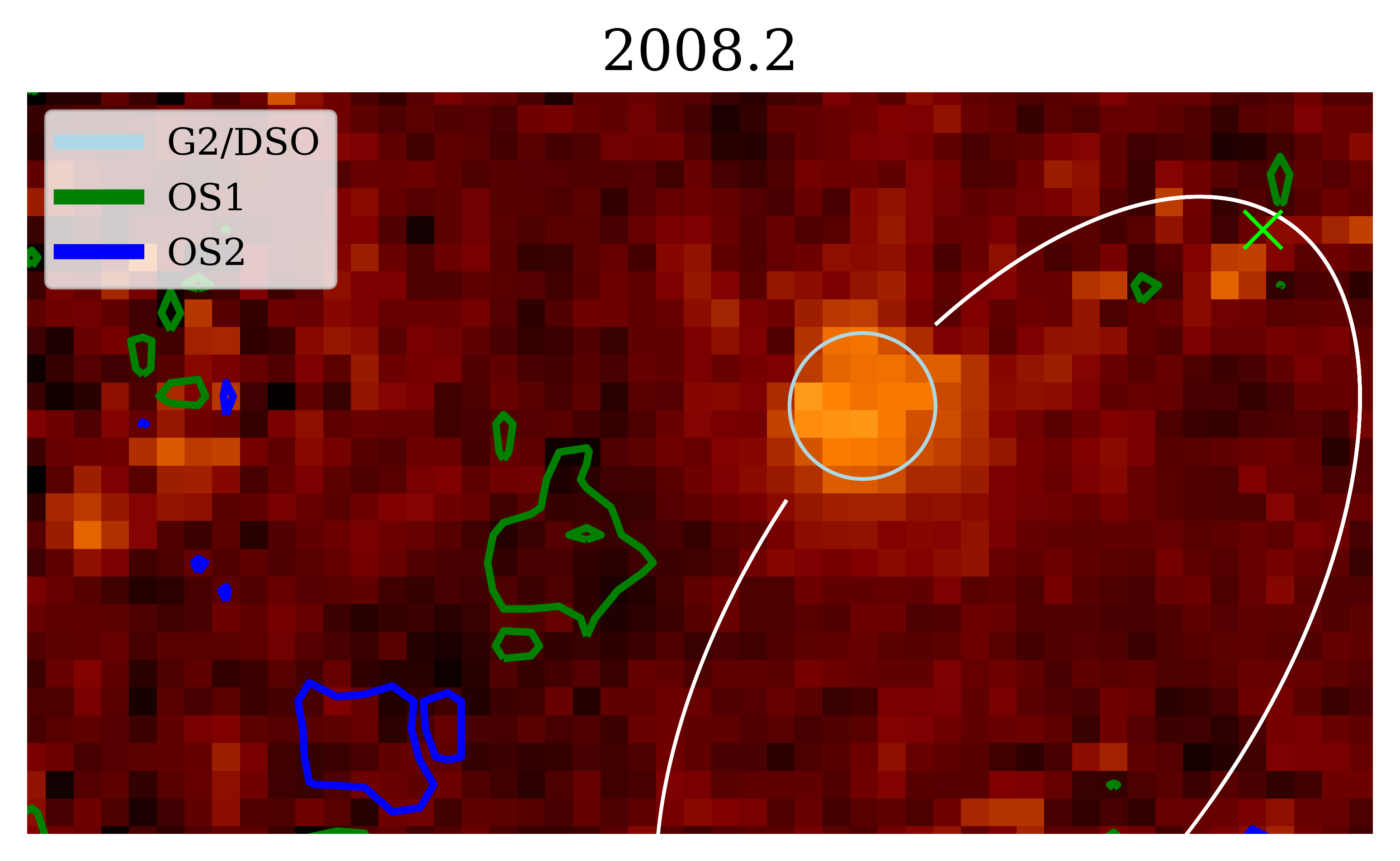}
	\caption{Doppler-shifted Br$\gamma$ line emission of G2/DSO at its velocity of about $1300$km/s and the contours of OS1 (green) and OS2 (blue) based on the data of 2008 presented in this work as shown in Fig. \ref{fig:os1_line_1} and Fig. \ref{fig:os2_line_1}. The orbit solution is based on this work. Placing a slit over all three sources and smoothing the data results in the tail emission as shown in Fig. \ref{fig:tail_1} and Fig. \ref{fig:ppv_2}, Appendix \ref{sec:tail_smoothed_app}. In this figure, North is up, East to the left. Sgr~A* is located at the lime colored $\times$. The size of the panel is $0.6\,\times\,0.3$ arcsec.}
\label{fig:combined_emission_dso_os1_os2}
\end{figure}

Several publications claim the existence of a tail component of G2/DSO that was supposed to be created because of the gravitational and hydrodynamical interaction with the environment of Sgr~A*. Unfortunately, it is not clear why the tail moves on a different orbit than the head component \normalfont{\citep[see Fig. 3 in][]{Pfuhl2015}. If there is a responsible process for this orbit discrepancy, we raise the question why the head is unaffected? Assuming a much higher density for the head compared to the so-called tail, it is controversially reported that G2/DSO was supposed to be destroyed during the periapse. However, \cite{Gillessen2019} reports that G2 is rather compact again after the periapse due to tidal focusing and moves on a drag-force influenced orbit. Assuming material would have been accreted by Sgr~A* during the periapse of G2/DSO, the flare observed by \cite{Do2019} could be a speculative link. However, it is also reasonable to assume, that the periapse of S2 in 2018 \citep[][]{Schoedel2002, Do2019S2} could have created instabilities in the accretion disk of Sgr~A* \citep[][]{Sukova2021} resulting in a bright flare.
In contrast to the drag-force influenced orbit as proposed by \cite{Gillessen2019}, we show in this work that} G2/DSO follows a Keplerian trajectory where the source is not significantly affected by the tidal field of Sgr~A* (see, e.g., Fig. \ref{fig:dso_line_evo}). We furthermore show in Fig. \ref{fig:art_tail_2006_2008_2012}, that the ionized gas, that is associated with the tail \citep[][]{Gillessen2013a} was in the S-cluster before G2/DSO passed by.\newline
By investigating the tail, it consists rather of isolated sources that can be, like the \normalfont{G2/DSO}, detected in the Doppler-shifted Br$\gamma$ line (see Fig. \ref{fig:combined_emission_dso_os1_os2}). 
Considering the observed number of S-stars (about 40), the amount of line-emitting objects is of the same order of magnitude (\normalfont{almost} 20). Hence, the presence of OS1 and OS2 contributes to the overall distribution of line emitting objects \citep[for a complete overview, see][]{Ciurlo2020, Peissker2020b}.

\subsection{Origin and nature of the source G2/DSO}

\begin{figure*}[htbp!]
	\centering
	\includegraphics[width=1.0\textwidth]{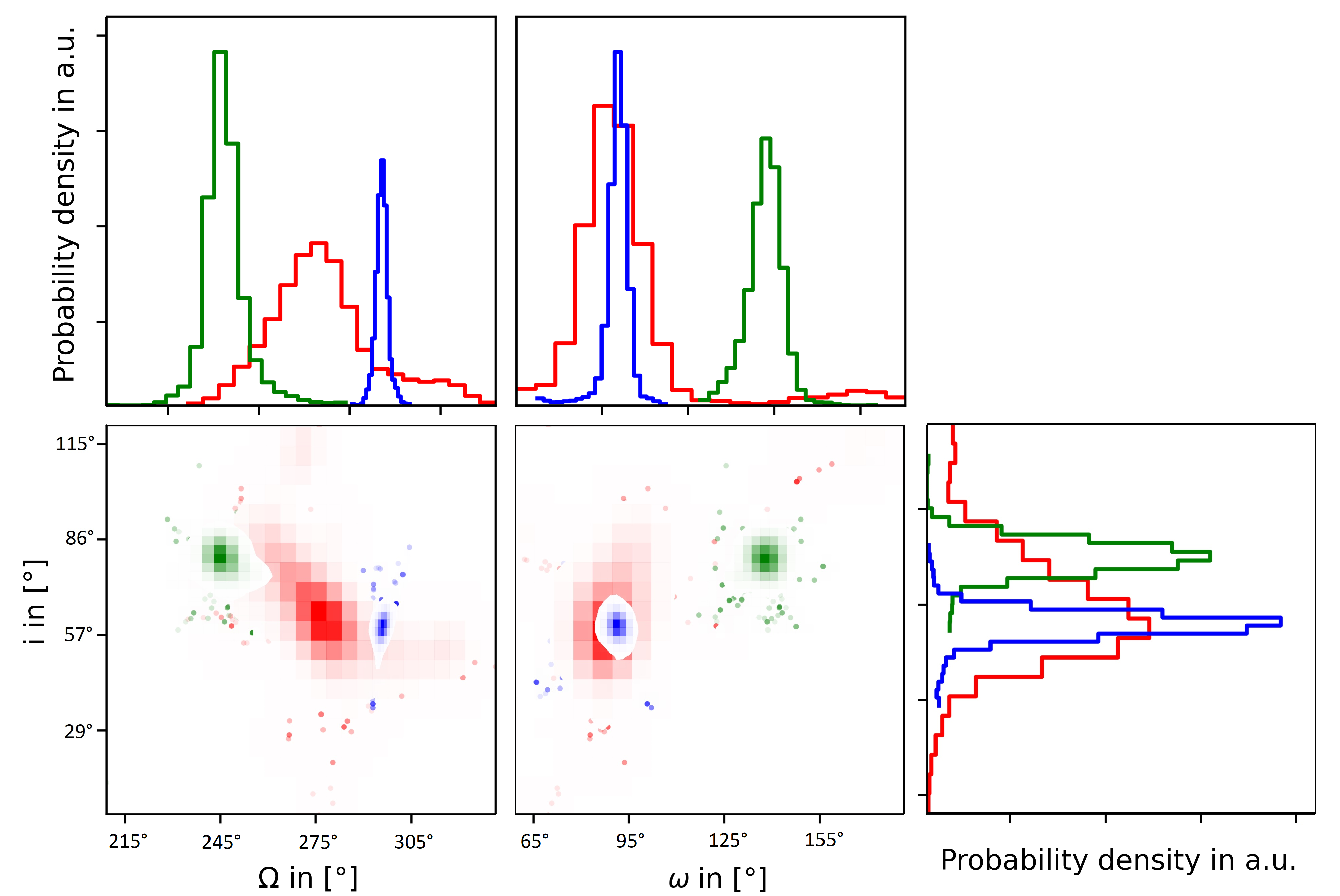}
	\caption{Inclination ($i$) as a function of the longitude of the ascending node ($\Omega$) and the argument of periapse ($\omega$). We adapt the 2d posterior distributions from Fig. \ref{fig:dso_orbit_3}, Fig. \ref{fig:os1_mcmc}, and Fig. \ref{fig:os2_mcmc} \normalfont{(for every mentioned figure, please consider the Appendix \ref{sec:mcmc_results_app})}. \normalfont{Considering the 1-$\sigma$ interval, G2/DSO (blue) and OS1 (red) do show similarities in $\omega$ and $i$ while OS2 (green) differ.} The upper two and the lower right panel (density in arbitrary units a.u.) underline a speculative common origin of the \normalfont{G2/DSO}, OS1, and OS2. The clockwise stellar disk with $i\,=\,(90\,\pm\,10)^{\circ}$ and $\Omega\,=\,(130\,\pm\,10)^{\circ}$ is thus highly unlikely as a birthplace of the \normalfont{G2/DSO} due to a significant offset of its mean orbital elements.}
\label{fig:origin}
\end{figure*}

Due to the Br$\gamma$ compactness of the object in combination with the photometric detection of a $K$-band counterpart at the position of the line-emitting source \normalfont{(Fig. \ref{fig:lr_results} and Fig. \ref{fig:kband_orbit}, Appendix \ref{sec:positions_app})}, we find strong support for a possible young stellar object. As shown in \cite{Lada1987}, a young protostar ($<\,1\,\times\,10^7$ years) consists of a black-body stellar emission as well as a cooler disk/envelope component. Hence, a two-component SED fit as presented in \cite{Peissker2020b} provides a suitable explanation for the observed continuum emission \normalfont{and is in agreement with the predictions by \cite{MurrayClay2012}}. Because of the ongoing accretion processes, the emission lines can show a nonsymmetric profile which is amplified by the interaction with the ambient medium, in particular by the formation of a bow shock layer \citep{Zajacek2016,Shahzamanian2016}. As pointed out by \cite{Zajacek2017}, the sum of the observational results underlines the stellar nature of the object that consists of a non-spherical gaseous-dusty envelope shaped by the bow shock as well as by bipolar cavities. Hence, it is expected that the ionized gas is not centered at the source itself but exhibits an offset. Considering the mentioned bow shock layer \citep[][]{Zajacek2016} in combination with the polarized continuum detection \citep[][]{Shahzamanian2016}, the proposed nature of G2/DSO as a young T-Tauri star by \cite{Scoville2013} and \cite{Eckart2013} seems reasonable. \normalfont{Furthermore, the Br$\gamma$ width variation (Fig. \ref{fig:sigma_time_evo_zoom}, Appendix \ref{sec:brgamma_zoom_app}) of G2/DSO is in line with observations of YSOs by \cite{Stock2020} and emphasizes the proposed classification.}\newline
\normalfont{The pericenter passage of G2/DSO is dated to 2014.38, OS1 and OS2 are following in 2020.67 and 2029.87, respectively. Even though the orbital elements for OS1 and OS2 are different, the former source shows similarities in inclination and the argument of periapse with G2/DSO. Because of the compactness and the Doppler-shifted Br$\gamma$ line emission in combination with the missing [FeIII] detection in contrast to the D-/G-sources \citep[see][]{Ciurlo2020, Peissker2020b}, which are located west of Sgr~A*, we hypothesize if G2/DSO, OS1, and OS2 do share a common history. If so, they should have been formed in the same dynamical process. Following this speculative scenario,}
the combination with the detected reservoir of fast moving molecular cloudlets \citep[][]{Moser2017, Goicoechea2018, Hsieh2021} and the simulation of an infalling cloud presented in \cite{Jalali2014} might provide a suitable explanation for the in-situ star formation event. The authors of \cite{Jalali2014} model a $100\,M_{\odot}$ cloud that crosses the Bondi radius of Sgr~A* \citep[for observations, see also][]{Tsuboi2018}. As an initial setup, \citet{Jalali2014} assume the loss of an angular momentum via the clump-clump collisions within the CND \citep[see also][for the studies where molecular cloud-cloud collisions were considered]{1986ApJ...310L..77S,2000ApJ...536..173T,2020MNRAS.492.2973T,2020MNRAS.492..603T}. Because of the gravitational potential of Sgr~A*, the initial cloud gets stretched and triggers, because of a compression force acting perpendicular to the orbital motion, the creation of YSO associations, similar to IRS13N. We will elaborate on this in more detail in Sec. \ref{sec:discussion_binary}.\newline
It is rather unlikely that the \normalfont{G2/DSO} itself was formed in the clockwise stellar disk located further out (CW disk, see Fig. \ref{fig:origin}) as, for example, claimed by \cite{Burkert2012} because of a large difference of inclinations ($i_{\rm CSD}\,\approx\,115^{\circ}\,\neq\,60^{\circ}\,\approx\,i_{\rm \normalfont{G2/DSO}}$).
The nondetected destruction process of the source \citep[see, e.g., ][]{Burkert2012, Schartmann2012} and the previously derived inclination of about $110^{\circ}\,-\,120^{\circ}$ for \normalfont{G2/DSO} may be explained by the lower data baseline.

\subsection{Properties of the young accreting star}

\begin{figure}[htbp!]
	\centering
	\includegraphics[width=0.5\textwidth]{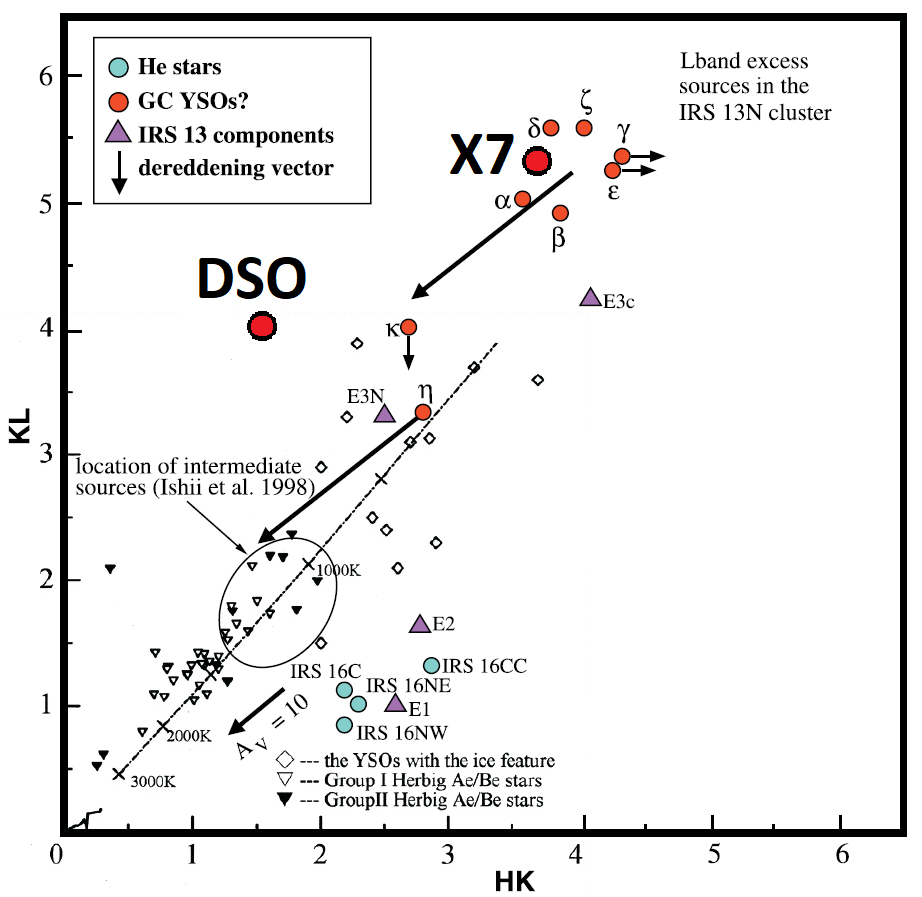}
	\caption{Young stellar objects in the GC. This figure is adapted from \cite{ishii1998} and \cite{Eckart2004a}. We mark X7 \citep[][]{peissker2021} and the \normalfont{G2/DSO} (this work).}
\label{fig:yso_hkl}
\end{figure}

By analysing the $K$-band emission of the continuum counterpart of the Doppler-shifted Br$\gamma$ line source, we find an averaged magnitude of $18.48\,\pm\,0.22$ mag with a correlated flux of \normalfont{$0.25\,\pm\,0.06$ mJy which matches the values presented in \citep{Eckart2013} and \cite{Shahzamanian2016}}.
\normalfont{In agreement with \cite{Sabha2012}, we observe a background magnitude close to G2/DSO and as a function of the distance towards Sgr~A* with a slope of $0.13$. This underlines the robust observation of the K-band magnitude observation of G2/DSO in pre-/post-periapse epochs which is not correlated to the increasing background light towards Sgr~A*.}\newline
The \normalfont{G2/DSO} passed Sgr~A* in 2014.38 at a pericenter distance $r_{\rm p}$ of about 137 AU. From the foreshortening factor, which is maximized during the periapse, the true size of the \normalfont{G2/DSO} can be inferred \citep[][]{Valencia-S.2015}. By applying a Gaussian fit to the Br$\gamma$ line map in 2014.38 (see Fig. \ref{fig:dso_line_evo}), we find a symmetrical shaped FWHM of about 37.5 mas (i.e. 3 px). Over $85\%$ of the total emission of the \normalfont{G2/DSO} is concentrated in a very compact area with a radius of 12.5 mas in 2014.38 (see Fig. \ref{fig:dso_line_evo}). In combination with the foreshortening factor, the character of the \normalfont{G2/DSO} can be classified as compact. The authors of \normalfont{\cite{Scoville2013}, \cite{Eckart2013}, and} \cite{Valencia-S.2015} furthermore derive a mass of $1-2\,M_{\odot}$ for the \normalfont{G2/DSO}. From the observed averaged $K$-band magnitude in this work of $mag_K\,=\,18.48$ in combination with the H- and L'-band magnitude \citep[see][]{Eckart2013, Peissker2020b}, we derive colors of $K-L\,=\,4.08$ mag and $H-K\,=\,1.48$ mag for the \normalfont{G2/DSO} (see Fig. \ref{fig:yso_hkl}). These are common values for Herbig Ae/Be stars with ice features but also the IRS13N sources \citep[see also][]{Eckart2004a, Moultaka2005}. \normalfont{Recently, \cite{Cheng2020} reported matching values for observed YSOs with a circumstellar disk.}
With the upcoming Mid-Infrared Instrument \citep[MIRI, see][]{Bouchet_2015, Rieke_2015, Ressler_2015} for the James Webb Space Telescope (JWST), we will be able to confirm the presence of ice features \citep[][]{moultaka2015}.

The estimated radius of $12.5\,{\rm mas}$ close to the pericenter passage corresponds to the upper limit on the physical length-scale of $\sim 103\,{\rm AU}$. However, the true physical size of the \normalfont{G2/DSO} envelope is plausibly one or two orders of magnitude smaller. Using the derived pericenter distance and the \normalfont{G2/DSO} mass estimate, we obtain the tidal radius at the pericenter of
\begin{align}
    r_{\rm t}^{\rm per}&=a(1-e)\left(\frac{m_{\star}}{3M_{\bullet}}\right)^{1/3}\,\notag\\
    &\sim 0.6\,\left(\frac{m_{\star}}{1\,M_{\odot}}\right)^{1/3}{\rm AU}\,.\label{eq_Hill_radius}
\end{align}
Given the orbital period of the \normalfont{G2/DSO}, $P_{\rm orb}\sim 105\,{\rm yrs}$, the source likely went through several pericenter passages, possibly thousands in case of a stellar nature. Therefore, the length scale of the \normalfont{G2/DSO} is likely small, of the order as expressed by Eq.~\ref{eq_Hill_radius}, which makes the source interesting from the point of view of stellar evolution and extreme star formation. The 3D MCMC radiative transfer simulations performed by \citet{Zajacek2017} showed that the basic continuum properties of the source can be reproduced with the compact gaseous and dusty envelope of the order of an astronomical unit. The constant $L$-band \citep{Witzel2014} as well as $K$-band flux density (this work) implies that the tidal prolongation and truncation of the envelope has rather been small, which is in contrast to the G1 source \citep{Witzel2017} that has exhibited profound SED changes in the post-pericenter phase. The recent increase in the $K$-band flux density after 2017, see Fig.~\ref{fig:mag}, may indicate changes in the envelope morphology, however, this will need to be clarified with more post-pericenter $L$- and $K$-band data. In conclusion, the length-scale of the \normalfont{G2/DSO} has been $\sim 1\,{\rm AU}$ both before and after the pericenter passage.

\subsubsection{Br$\gamma$ line width of the \normalfont{G2/DSO}}
\label{sec:brgamma_line_width}
As proposed by, e.g. \cite{Gillessen2019}, the increasing line width of the Doppler-shifted Br$\gamma$ emission of the \normalfont{G2/DSO} until 2014 could be interpreted via the tidal stretching of the extended cloud. The decreasing line width after 2014 was proposed to be due to the tidal focusing \citep{Gillessen2019}.

\cite{Eckart2013}, \cite{Zajacek2016}, and \cite{Shahzamanian2017} consider a dust-enshrouded star to explain the observations of the \normalfont{G2/DSO}. As shown by \cite{Valencia-S.2015} and this work, a variation of the Br$\gamma$ line width is detectable in the data sets that cover 2005 to 2019. In contrast to the pure cloudy nature of the source, we consider an internal and external explanation for the line width variation. Hence, the observed Br$\gamma$ emission line width could be produced by two proposed mechanisms:
\begin{itemize}
    \item The production within the denser and colder stellar wind bow shock due to the collisional ionization \citep{Scoville2013}. As was explicitly shown by \citet{Zajacek2016}, due to the variable \normalfont{viewing angle of the bow-shock velocity field with respect to the observer}, the line width increases from $\sim 100\,{\rm km\,s^{-1}}$ up to $\sim 200\,{\rm km\,s^{-1}}$ at the pericenter passage ($2014.4$) and then decreases down to $\sim 100\,{\rm km\,s^{-1}}$ in the post-pericenter phase, see in particular the comparison of the model calculations with the observed line width in Fig.~\ref{fig:sigma_time_evo} (orange and black lines for the case with no outflow and an outflow of 2000 km/s, respectively),
    \item a broad Br$\gamma$ line with the line width of a few 100 km/s is commonly detected in young YSOs of class I, and can be produced within the disc winds and/or in the process of magnetospheric accretion \citep[][]{Valencia-S.2015, Stock2020}. 
\end{itemize}

We have shown in Fig. \ref{fig:sigma_time_evo} that the line width between 2006 and 2008 is rather decreasing. \normalfont{Considering also the absence of a tail emission (Fig. \ref{fig:dso_2014_blue}, \ref{fig:ppv_1}, \ref{fig:tail_1}, \ref{fig:art_tail_2006_2008_2012}, \ref{fig:ppv_2}), we strongly question} a tidal stretching scenario. \normalfont{Because of the ongoing accretion from the host star \citep{MurrayClay2012, Scoville2013, Zajacek2016}}, an overall trend can be observed and shows the same behavior as the foreshortening factor \normalfont{\citep{Valencia-S.2015}}. In particular, we pay attention to the outliers in 2014.3 and 2015.4. The width of the Br$\gamma$ line of the \normalfont{G2/DSO} at $2.1490\,{\rm \mu m}$ in 2015.4 and $2.1857\,{\rm \mu m}$ is polluted by the strong OH emission at $2.1505\,{\rm \mu m}$, $2.1507\,{\rm \mu m}$, and $2.1873\,{\rm \mu m}$ \citep[][]{Rousselot2000}. Inspecting the presented spectrum in Fig. \ref{fig:dso_spectral_line_evo} reveals peaks that significantly broaden the Doppler-shifted \normalfont{G2/DSO} Br$\gamma$ line in 2014.3 and 2015.4 at $2.1874\,\mu m$ and $2.1509\,\mu m$, respectively. Effectively, this broadening is due to the OH line emission \normalfont{(see Sec. \ref{sec:oh_emission})} and increases the line width in 2014.3 and 2015.4 by about $30\%$. By applying this correction, we derive a line width of 234 km/s and 243 km/s for the \normalfont{G2/DSO} in 2014.3 and 2015.4, respectively.\newline
Based on the measured line map size (Table \ref{tab:spectral_velocity_properties}), the true size of the \normalfont{G2/DSO} can be observed at the pericenter passage in 2014.38. Since the evolution of the line width coincides with the foreshortening factor trend \citep[see][]{Valencia-S.2015}, we can safely assume that the observed effect is due to the stellar nature and the orientation of the source.

%

\subsection{Magnetohydrodynamic drag force}

\begin{figure}[htbp!]
	\centering
	\includegraphics[width=0.5\textwidth]{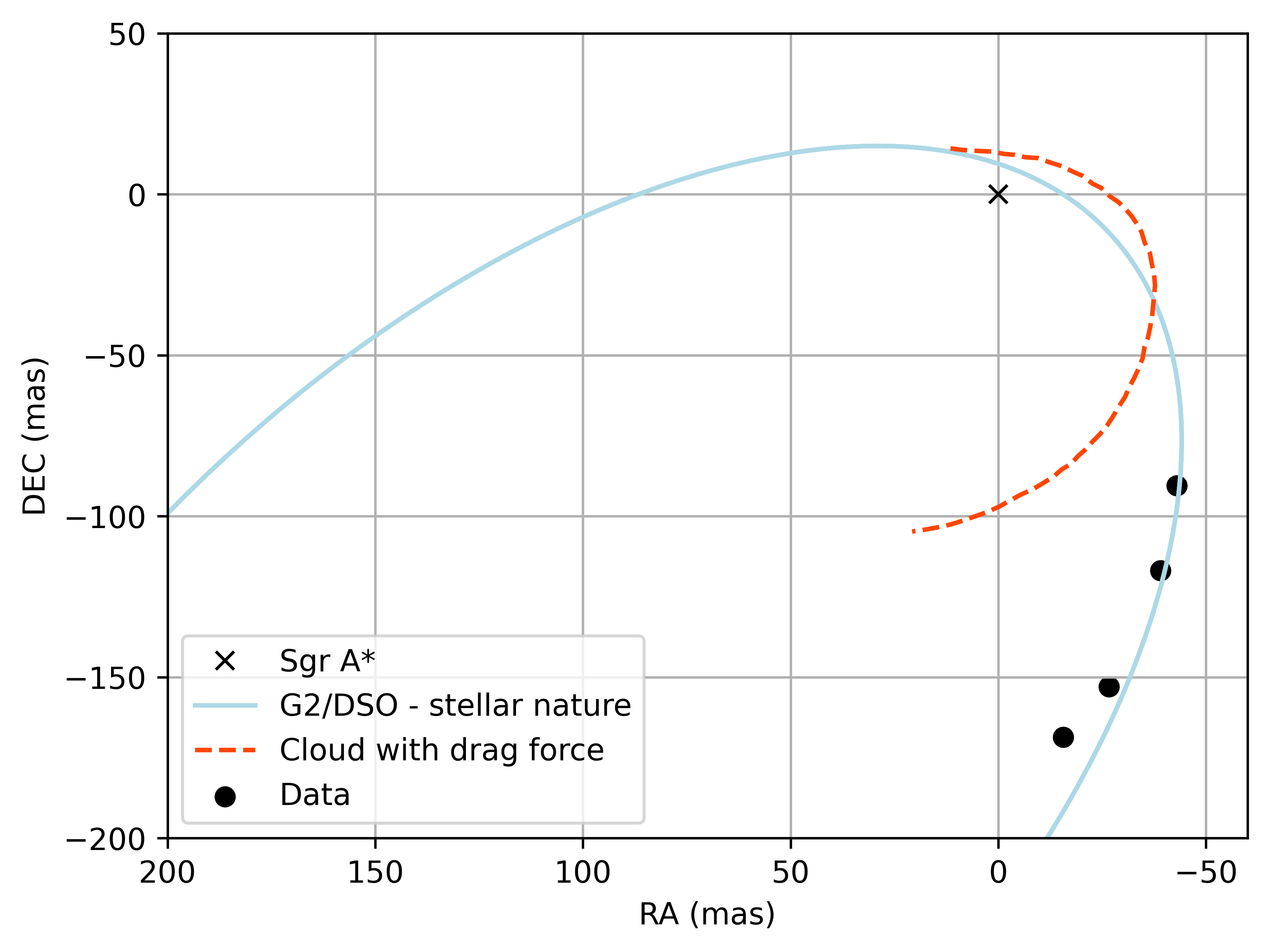}
	\caption{Based on the proposed gas streamer idea of \cite{Pfuhl2015}, the authors connect the orbit of G1 and \normalfont{G2/DSO} to predict the evolution of the gas cloud (red dotted line). Comparing the Br$\gamma$ position of the \normalfont{G2/DSO} after the periapse between 2016-2019 with the gas-streamer idea reveals gaps between the proposed evolution and the observed trajectory as shown by the blue line.}
\label{fig:cloud_model_orbit}
\end{figure}

\citet{Pfuhl2015} discussed and showed a possible drag force acting on the \normalfont{G2/DSO}. Within this model, the authors combined the data of G1 and \normalfont{G2/DSO} to demonstrate the effect of the predicted inspiraling cloud towards Sgr~A* after the pericenter passage \citep[see, e.g., ][]{Gillessen2012}. In Fig. \ref{fig:cloud_model_orbit}, we compare the predicted trajectory of \normalfont{G2/DSO that is following G1 as part of a gas streamr \citep[][]{Pfuhl2015}} with the observed data (black data points) and the derived orbit.
The comparison indicates that no measurable drag force \normalfont{as proposed by \cite{Pfuhl2015}} can be observed and its assumption, \normalfont{that G2/DSO is part of a gas streamer cannot be confirmed. In comparison to Pfuhl et al.,} the magnetohydrodynamic drag force analysis by \cite{Gillessen2019} discusses \normalfont{a smaller drag force}  with respect to the pre-pericenter orbit of G2/DSO. However, the \normalfont{hardly recognizable presented effect is underlined by shortening the observational available pre-pericenter data points in the related publication of \cite{Gillessen2019} (see their Fig. 2)}. When we consider all available SINFONI data up to 2019, the position- and velocity-data can simultaneously be fit well with a simple Keplerian orbital solution, see Fig.~\ref{fig:dso_orbit_1} and Fig.~\ref{fig:dso_orbit_2}. \normalfont{Following Occam's razor \citep[see, e.g.,][]{Ariew1976}, the inclusion of an additional drag term appears redundant.} 
Moreover, we have shown that the influence of image motion (Tab. \ref{tab:image_motion_1}) and sky variability can have a significant impact on the data \normalfont{by up to $40\%$}. Based on the data presented in \cite{Gillessen2019}, we conclude that the arguably small offset \normalfont{(< few percent)} of the Br$\gamma$ source in combination with the presented line shape might not be explained by a drag force \normalfont{and can rather be explained by the effects discussed in this work. By carefully inspecting the provided Kepler- and Drag Force-fits in Gillessen et al. we note, that both models do not perfectly match the shown Br$\gamma$ emission which underlines the noisy character of the SINFONI data.} Based on these points, we do not find a strong need for a drag force to explain the orbital evolution of the \normalfont{G2/DSO} (Fig. \ref{fig:dso_line_evo}). In a similar way, \citet{Witzel2017} found that the evolution of G1 is consistent with the Keplerian motion on a different highly-eccentric orbit to that of the \normalfont{G2/DSO}, which supports the stellar nature of both sources. In this regard, the tests of alternative gravitational theories, such as the fermionic dark matter compact core--diffuse halo \citep{2020A&A...641A..34B}, that has made use of the \normalfont{G2/DSO} orbit should be updated accordingly.

\subsection{Are the \normalfont{G2/DSO}, OS1, and OS2 YSOs or rather related to the stellar binary dynamics?}
\label{sec:discussion_binary}
In \cite{Zajacek2014}, the authors propose the possibility that the \normalfont{G2/DSO} can be associated with a binary or multiple-star dynamics close to the Galactic center. In particular, they showed that in case the \normalfont{G2/DSO} is a binary system, it would lead to the disruption event at the pericenter. This model was soon followed by scenarios that explain the \normalfont{G2/DSO} and related objects as binary merger products \citep[][]{perets2007, stephan2016,stephan2019}. \normalfont{Following the binary merger fraction discussion of \cite{Ciurlo2020}, we adapt}
\begin{equation}
    R\,=\,\frac{1}{2}\frac{N_B}{N_m}
\end{equation}
from this publication and assume, that line emitting and dusty objects are binary merger products. By using this equation, the binary fraction R is calculated with the number of binaries $N_B$ and the number of low-mass stars $N_m$. To provide a certain degree of comparability, we adapt $N_m\,=\,478$ from Ciurlo et al. and consider the additional sources from \cite{Peissker2020b} which results in $N_B\,=\,85$. For R, we derive $\sim\,9\%$ which is almost twice as much as the binary fraction of $\sim\,5\%$ calculated in \cite{Ciurlo2020}. Since we assumed that all objects are binary merger, it is safe to conclude that the assumption is not justified. We obtain a low-mass binary fraction that is almost twice than that derived by Ciurlo et al. and also larger than the overall predicted value of $6-7\%$ \citep[][]{Raghavan2010, stephan2019, Ciurlo2020}. Hence, we conclude that a considerable fraction of the overall population of line and dust emitting sources are not necessarily binary mergers but also YSOs. \normalfont{We would like to emphasize that a binary merger scenario for G2/DSO is not excluded from the overall discussion.}\newline
\normalfont{Assuming a speculative binary} disruption scenario, the pericenter passage \normalfont{at a unspecified time} of the \normalfont{G2/DSO} would have been responsible for the tidal break-up of the binary components that we denote here as \normalfont{G2/DSO}$_1$ and \normalfont{G2/DSO}$_2$. In this regard, \normalfont{G2/DSO}$_2$ is a runaway star, while \normalfont{G2/DSO}$_1$ is a highly eccentric component which is following a slightly modified Keplerian trajectory \citep[][]{Hills1988} with a smaller eccentricity as well as a semi-major axis \citep{Zajacek2014}. \normalfont{Following this,} the post-pericenter trajectory of \normalfont{G2/DSO}$_1$ could mimic the inspiral due to the drag force \citep{Gillessen2019}. If we consider the second star (\normalfont{G2/DSO}$_2$), the magnitude of this component is at most $18.5$ mag. Because of the preserved Br$\gamma$ shape in the post-periapse years of the observed \normalfont{G2/DSO}$_1$, we are allowed to speculate that \normalfont{G2/DSO}$_2$ did not capture most of the surrounding stellar material. This implies a mass estimate of \normalfont{G2/DSO}$_1\,>\,$\normalfont{G2/DSO}$_2$. With the derived mass for the \normalfont{G2/DSO} of about $1-2\,M_{\odot}$, we \normalfont{assume that} an upper limit for the mass of \normalfont{G2/DSO}$_2$ \normalfont{is closer to} $m_{\rm \normalfont{G2/DSO}_2}\,<\,1\,M_{\odot}$.\newline

We note that the disruption scenario applies also to the theory of in-situ star formation, where an infalling cloud that is stretched and compressed forms associations of young stellar objects \citep{Jalali2014}. These associations are first bound systems with a certain velocity dispersion. In case they orbit the SMBH on an eccentric orbit, the tidal radius of the YSO cluster is time-variable depending on the distance from the SMBH. In particular, the tidal (Hill) radius during the pericenter passage is
\begin{align}
    r_{\rm t}^{\rm cluster}&=\overline{a}(1-\overline{e})\left(\frac{m_{\rm cluster}}{3M_{\bullet}} \right)^{1/3}\,\notag\\
    &\sim 14.4\,\left(\frac{\overline{a}}{15.6\,{\rm mpc}} \right)\left(\frac{\overline{e}}{0.78} \right)\left(\frac{m_{\rm cluster}}{100\,M_{\odot}} \right)^{1/3}\,{\rm AU}\,,
\end{align}
where we considered the mean values of the semi-major axis and the eccentricity based on \normalfont{G2/DSO}, OS1, and OS2 (see Table~\ref{tab:orbit_elements_os}). The cluster mass was scaled to the order of magnitude considered in the simulations of \citet{Jalali2014} for an infalling molecular cloud. Since the YSO cluster length-scale approximately given by the Jeans length-scale for a given critical density for self-gravitation is larger than $r_{\rm t}^{\rm cluster}$, the cluster is effectively dissolved at the distances where dusty orbits are seen to orbit now.
Hence, during the pericenter passage, the cluster will tend to dissociate and lose its members that will afterwards orbit around Sgr~A* on independent orbits with similar orbital elements initially (inclination, longitude of the ascending node, argument of the pericenter). Due to the resonant relaxation and the perturbative effects of the S cluster as well as Newtonian (mass) and Schwarzschild precession, the inclination as well as other orbital elements will start to deviate. In this regard, \normalfont{G2/DSO}, OS1, OS2, as well as other dusty objects can share a common origin, despite certain offsets in the orbital elements, see Table~\ref{tab:orbit_elements_os} and Fig.~\ref{fig:origin} for comparison.

To estimate how much individual dynamical processes can alter orbital elements of interest (inclination, longitude of the ascending node, argument of the pericenter), we compare their fundamental timescales with the assumed lifetime of the dusty sources of $\sim 10^5\,{\rm yr}$. In particular, the argument of pericenter shift between mainly G2/DSO and OS2 could be explained by the Schwarzschild precession, which per orbit can be estimated as follows,
\begin{equation}
\Delta \phi=\frac{6\pi G M_{\bullet}}{c^2 a_{\rm G2}(1-e_{\rm G2}^2)}    
\label{eq_Delta_phi}
\end{equation}
which for the G2/DSO yields $\Delta \phi=9.52'$ per orbital period ($P_{\rm G2}\sim 108$ years). The difference for the argument of periapse of $\sim 50^{\circ}$ between OS2 and G2/DSO (see Fig. \ref{fig:origin}) could thus be achieved in $\sim 35\,000$ years due to the much faster prograde relativistic precession of G2/DSO than OS2 ($\sim 1.57'$ per its orbital period). The Newtonian (retrograde) mass precession counter-balances the relativistic precession and acts in an opposite direction. The mass coherence timescale $T_{\rm c}^{M}$, on which the arguments of the periapse would be randomized, can be estimated as,
\begin{align}
    T_{\rm c}^{M}&\sim \frac{M_{\bullet}}{<M_{\star}>}\frac{P_{\rm G2}}{N_{\star}}\,\notag\\
    &\sim 4\times 10^5\left(\frac{<M_{\star}>}{10\,M_{\odot}}\right)^{-1}\left(\frac{P_{\rm G2}}{100\,{\rm yr}}\right) \left(\frac{N_{\star}}{100} \right)^{-1}\,{\rm yr},
    \label{eq_mass_coherence_time}
\end{align}
where $<M_{\star}>$ is the mean stellar mass, $P_{\rm G2}$ is the orbital period of G2/DSO, and $N_{\star}$ is the number of stars inside its orbit. If the lifetime of G2/DSO, OS1, and OS2 is at most $\sim 10^5\,{\rm yr}$, then the mass precession has not significantly altered their orbital orientations.

The resonant relaxation process, which is characteristic of highly symmetric potentials, such as inside the sphere of influence of Sgr~A*, proceeds in two modes: a faster vector resonant relaxation (VRR) and a slower scalar resonant relaxation (SRR). The VRR relaxation timescale is 
\begin{align}
    T_{\rm VRR}&\sim \frac{M_{\bullet}}{<M_{\star}>}\frac{P_{\rm G2}}{\sqrt{N_{\star}}}\,\notag\\
    &\sim 4\,\left(\frac{<M_{\star}>}{10\,M_{\odot}}\right)^{-1}\left(\frac{P_{\rm G2}}{100\,{\rm yr}}\right) \left(\frac{N_{\star}}{100} \right)^{-1/2}{\rm Myr}\,,
    \label{eq_VRR}
\end{align}
while the SRR relaxation, which also changes the magnitude of the angular momentum is about 10 times slower because of the definition of the scalar relaxation time, $T_{\rm SRR}\sim P_{\rm G2}M_{\bullet}/<M_{\star}>$ \citep[see][for details]{2006ApJ...645.1152H,2017ARA&A..55...17A}. In case G2/DSO, OS1, and OS2 were formed approximately in the same orbital plane $\sim 10^5$ years ago, then the VRR has not had enough time to randomize their orbits, but it could have contributed to the $\sim 20^{\circ}$ spread in orbital inclinations over time.

In terms of the deviations in the longitude of the ascending node, the orbital precession can be relevant, especially when the stellar motion is perturbed by the presence of an inclined massive stellar disk. \citet{Ali2020} revealed that the S cluster consists of at least two perpendicular stellar disks, which indicates a non-randomized stellar distribution. Beyond the S cluster, at the radius of $R_{\rm d}\sim 0.1\,{\rm pc}$, there is a stellar disc of about hundred OB stars, with the potential total mass of $M_{\rm d}\sim 10000\,M_{\odot}$ \citep{Paumard2006,2009ApJ...697.1741B}. If the G2/DSO infrared source is inclined by $\beta \sim 60^{\circ}$ with respect to the disc plane, the stellar disk induces torques on the misaligned dust-enshrouded objects, which then precess with a certain period $T_{\rm prec}$ around the symmetry axis, effectively shifting the line of nodes. The rate of this shift depends on the semi-major axis $a_{\star}$ and since $a_{\rm G2}=0.01745\,{\rm pc}<R_{\rm d}\sim 0.1\,{\rm pc}$, one can use the following analytical formula to estimate the stellar precession period with respect to the disc \citep{2005MNRAS.359..545N,2008ApJ...683L.151L},
\begin{equation}
  T_{\rm prec}\simeq \frac{8\pi M_{\bullet}}{3M_{\rm d}\cos{\beta}}\sqrt{\frac{a_{\star}^3}{GM_{\bullet}}}\frac{(a_{\star}^2+R_{\rm d}^2)^{5/2}}{a_{\star}^3R_{\rm d}^2}\,,
  \label{eq_precession_nayakshin05}
\end{equation}
For the G2/DSO, the precession period can be estimated to be $T_{\rm p}\sim 2.34\times 10^{7}\,{\rm yr}$, hence it is a slow process, which leads to the shift of $\sim 1.5$ degrees in terms of the argument of the ascending node during $10^5$ years or $\sim 10.8$ degrees in $700\,000$ years. For OS2 the precession shift is slightly smaller than for G2/DSO and OS1, $\sim 1$ degree per $100\,000$ years, which could have contributed to the ascending node offset of $\sim 6$ degrees during the last million years. Hence, most of the difference in the ascending node is potentially attributable to the intrinsic dispersion during the formation process.

Finally, the presence of a perturbing stellar disk at the scale of $R_{\rm d}\sim 0.1\,{\rm pc}$ with the mass of $M_{\rm d}\sim 10^4\,M_{\odot}$ induces a Kozai-Lidov (KL) oscillations of eccentricity and the inclination. The period of the KL cycle can be estimated as \citep{2005A&A...433..405S},

\begin{equation}
    T_{\rm KL}=2\pi\left(\frac{M_{\bullet}}{M_{\rm d}} \right)\left(\frac{R_{\rm d}}{a_{\rm G2}} \right)^3 P_{\rm G2}\,,
    \label{eq_KL_oscillation}
\end{equation}
which yields $T_{\rm KL}\sim 4.73\times 10^7$ years. Since the timescale is about two orders of magnitude longer than the assumed lifetime of dust-embedded sources, the KL effect can slightly contribute to the above-mentioned effects to account for the overall offset of orbital elements. 

Since all relevant dynamical processes operate on longer timescales than the assumed lifetime of dust-enshrouded stars, most of the offset among orbital elements reflects the way they formed - i.e. from a turbulent fragmenting molecular cloud with an intrinsic offset due to a velocity disperion, or G2/DSO and OS1 formed initially as a binary that disrupted, with OS2 forming separately as a single star. More detailed numerical dynamical studies are beyond the scope of this study.

In the binary merger scenario, the two components of a binary star initially orbit the SMBH, which acts as a more distant perturber. The SMBH as a perturber would induce Kozai-Lidov \normalfont{\citep{Kozai1962, Lidov1962, Naoz2016}} resonances of the two components, where during one cycle the eccentricity growth is exchanged for the inclination decrease and vice versa. Finally, at large orbital eccentricities, the two components are tidally distorted, induce torques on each other, and both stars are finally driven to merge. Such a merger product contracts on a Kelvin-Helmholtz timescale and is often associated with optically thick dusty outflows that give rise to the NIR-excess \citep{stephan2016,stephan2019}. Hence, because of the resemblance to young stellar objects, it is difficult to distinguish between binary mergers and pre-main-sequence stars merely based on the photometry and the spectroscopy of the sources. Moreover, as mentioned before, the SINFONI data suffers from noisy behavior. Hence, the observational indications to verify a merger scenario are rather challenging to determine.\newline 

One distinguishing feature between YSOs that formed in situ in a single star formation event and dust-enshrouded merger products could be their distribution of orbital elements. While for YSOs we expect comparable orbital elements on timescales less than the resonant relaxation timescale within the S cluster, binary mergers should not follow such a condition as they form continuously and on different orbits. Since \normalfont{G2/DSO and OS1 share comparable orbital elements in terms of their inclination, semi-major axis, and} the argument of the pericenter, see also Fig.~\ref{fig:origin}, the common origin in the same star formation event is plausible. \normalfont{Following this argument implies that OS2 might be a binary merger. This would be fully compatible with the discussed infalling cloud scenario since \cite{Jalali2014} predicts that a certain fraction of the resulting sources are binaries and single low- and high-mass YSOs \citep[][]{Yusef-Zadeh2013, Ciurlo2020, Peissker2020b}.}

\subsection{Dust-enshrouded objects as remnants of a disrupted young stellar association}

Combining some of the mechanisms discussed in the previous subsection, we hypothesize that \normalfont{G2/DSO}, OS1, and OS2 as well as other dust-enshrouded objects could be remnant YSOs captured by the SMBH during a nearly parabolic infall of a young star cluster formed further away at the scales of $\sim 1\,{\rm pc}$ and beyond.

The advantage of this scenario is that larger distances from the SMBH put only moderate restrictions on the critical Roche density necessary for the molecular cloud to withstand disruptive tidal forces. The lower limit on the number density of the self-gravitating cloud at the distance $r$ from Sgr~A* is,
\begin{align}
    n_{\rm cloud} &\gtrsim \frac{3M_{\bullet}}{2\pi \mu m_{\rm p} r^3}\,\notag\\
    &\sim 7.7 \times 10^7 \left(\frac{M_{\bullet}}{4 \times 10^6\,M_{\odot}} \right) \left(\frac{r}{1\,{\rm pc}} \right)^{-3}\,{\rm cm^{-3}}\,.\label{eq_roche}
\end{align}
In comparison, the Roche limit according to Eq.~\eqref{eq_roche} within the S cluster ($\sim 0.01$ pc) gives as much as $n_{\rm cloud}>10^{14}\,{\rm cm^{-3}}$.

The critical density for the star-formation to take place could be reached via cloud-cloud collisions \citep{1986ApJ...310L..77S,2000ApJ...536..173T,2004ASPC..322..263T,2009MNRAS.394..191H} within the circum-nuclear disk (CND), where individual clumps have $\sim 10^6-10^{8}\,{\rm cm^{-3}}$ \citep{Hsieh2021}. A further enhancement in the density can be provided by UV radiation pressure originating in NSC OB stars at the inner rim of the CND \citep{2013ApJ...767L..32Y}. Moreover, fast stellar winds and occasional supernova explosions can be another source of a star-formation trigger. External pressure is necessary for individual clumps to overcome the turbulent pressure, which prevents them from collapsing \citep{Hsieh2021}. Clump-clump collisions can partially remove the angular momentum, which helps to set the resulting self-gravitating cloud on the infalling trajectory towards Sgr~A* with a small impact parameter \citep{Jalali2014,2020MNRAS.492.2973T,2020MNRAS.492..603T}. However, the overall hydrodynamics of clump-clump interactions is rather complex and only a fraction of such collisions may end up with a self-gravitating and fragmenting cloud complex falling radially towards Sgr~A*. First, this is due to the complex internal structure of molecular clumps, in particular their turbulent field \citep{2021AJ....161..243S}, and hence the ``hard ball'' approximation does not apply to them. Second, the clumps at larger distances from Sgr~A*, beyond $\sim 2$ pc, follow the Galactic rotation and thus have a larger angular momentum with respect to Sgr~A*, which needs to be removed for the cloud to fall in with a sufficiently small impact parameter. These two points imply that the shearing likely takes place, which can result in a formation of a new cloud without a sufficient loss of the angular momentum or a sheared gaseous streamer. The formation of shearing gaseous streamers from a set of clumps was demonstrated in the 3D N-body/smoothed particle hydrodynamics (SPH) simulations by \citet{2021AJ....161..243S}. In their work, the turbulence was continually injected to mimick the effect of supernovae and stellar winds. In this way, the high dispersion of the gas within the Central Molecular Zone (CMZ) can effectively be reproduced. As shown by \citet{2021AJ....161..243S}, the injected turbulence results in the accretion to smaller scales down to Sgr~A* in the form of turbulent accretion flows \citep{2020arXiv201004170S} or high-density spiral streamers \citep{2021arXiv210209569D}. Regardless of this complex behaviour within the CMZ, for simplicity here we assume that at least once in every $\sim 10^6\,{\rm yr}$ \citep{2008ApJ...683L..37W,Jalali2014} a fragmenting star-forming cloud can reach the vicinity of Sgr~A* with the impact parameter at the length-scale of the S cluster, where it is expected to tidally disintegrate, with a fraction of YSOs being captured by Sgr~A* \citep{2003ApJ...592..935G}, while the remaining fraction being unbound on hyperbolic orbits \citep{2017MNRAS.467..451F}. If the molecular cloud size is comparable or larger than its impact parameter, it can completely engulf Sgr~A* and leave behind a compact star-forming disk \citep{2008ApJ...683L..37W}, which may help explain the multi-disk configuration of the S-cluster \citep{Ali2020}.

The crucial point of the ``infalling-cloud'' model is that YSOs of the age of $\sim 10^5$ years can already be formed during the infall phase. Thus, we require that the infall timescale of the cloud towards Sgr~A*, which is half of the orbital timescale, $t_{\rm infall}=P_{\rm orb}/2$, is longer than the free-fall timescale of the clump with the critical density $n_{\rm cloud}$, $t_{\rm ff}=[3\pi/(32G\mu m_{\rm p} n_{\rm cloud})]^{1/2}$. Hence, we obtain the lower limit on the initial distance of an infalling cloud,
\begin{align}
    d_{\rm cloud} &\gtrsim \left(\frac{3M_{\bullet}}{32\pi \mu m_{\rm p} n_{\rm cloud}} \right)^{1/3}\,\notag\\
    &\sim 0.4\,\left(\frac{M_{\bullet}}{4\times 10^6\,M_{\odot}} \right)^{1/3}\left(\frac{n_{\rm cloud}}{10^8\,{\rm cm^{-3}}} \right)^{-1/3}{\rm pc}\,,\label{eq_cloud}
\end{align}
for the free-fall timescale of $t_{\rm ff}\sim 5000$ years, which can, however, get smaller as the density within the fragmenting clumps increases beyond the limit given by Eq.~\eqref{eq_roche}. The outer distance range can be inferred from the \normalfont{G2/DSO} lifetime of $t_{\rm \normalfont{G2/DSO}}\sim 10^5\,{\rm yr}$ and from the condition $t_{\rm infall}\lesssim t_{\rm \normalfont{G2/DSO}}$,
\begin{align}
    d_{\rm out} & \lesssim (GM_{\bullet})^{1/3} \left(\frac{t_{\rm \normalfont{G2/DSO}}}{\pi}\right)^{2/3}\,\notag\\
    & \sim 2.6 \left(\frac{M_{\bullet}}{4\times 10^6\,M_{\odot}} \right)^{1/3} \left(\frac{t_{\rm \normalfont{G2/DSO}}}{10^5\,{\rm yr}} \right)^{2/3}\,{\rm pc}\,.\label{eq_dout}
\end{align}
Given the distance range between $0.4\,{\rm pc}$ and $2.6\,{\rm pc}$, it is quite plausible that the self-gravitating cloud was formed or rather triggered towards star-formation by external pressure within the CND, which is located between $\sim 1.5\,{\rm pc}$ and $\sim 3-4\,{\rm pc}$ \citep{2005ApJ...622..346C}.

In the further discussion, we analyze the scenario where the self-gravitating cloud from the CND fragmented into a cluster of pre-main-sequence stars on its way towards Sgr~A*. We assume the mass of this minicluster or rather a young stellar association of $m_{\star}=100\,M_{\odot}$, which is in the range of masses of $0.05\,{\rm pc}-0.2\,{\rm pc}$ clumps within the CND \citep{Hsieh2021}. The stellar velocity dispersion is adopted from the IRS 13N association of young stars, $\sigma_{\star}\sim 50\,{\rm km\,s^{-1}}$ \citep{2008A&A...482..173M}. Then from the virial theorem, we obtain the stellar association radius of
\begin{equation}
    r_{\star}=0.2\,\left(\frac{m_{\star}}{100\,M_{\odot}}\right)\left(\frac{\sigma_{\star}}{50\,{\rm km\,s^{-1}}} \right)^{-2}\,{\rm mpc}\,.
\end{equation}
The young stellar cluster on a highly eccentric orbit will dissociate at the tidal disruption radius that can be expressed as
\begin{align}
    r_{\rm dis} &=r_{\star}\left(\frac{M_{\bullet}}{m_{\star}} \right)^{1/3}=\,\notag\\
    &=\frac{GM_{\bullet}^{1/3}m_{\star}^{2/3}}{\sigma_{\star}^2}=\,\notag\\
    &\sim 1214\,\left(\frac{m_{\star}}{100\,M_{\odot}}\right)^{2/3}\left(\frac{\sigma_{\star}}{50\,{\rm km\,s^{-1}}} \right)^{-2}\,{\rm AU}\,,
\end{align}
which is close to the pericenter distances of OS1 and OS2, as well as other dusty sources \citep{Peissker2020b}. A few stars from an infalling cluster will remain bound to the SMBH \citep{2003ApJ...592..935G}, while others will escape the NSC altogether as hypervelocity stars \citep{2017MNRAS.467..451F}. The high eccentricity of the \normalfont{G2/DSO} as well as G1 implies that they might have been constituents of binary systems within the cluster, while the other components escaped \citep{Zajacek2014}. However, given a nearly radial infall of the cluster and its velocity dispersion, a fraction of YSOs should anyway reach the pericenter distance of the \normalfont{G2/DSO} ($\sim 137\,{\rm AU}$) during the disruption event \citep[see also][for statistical estimates of a similar scenario for S2 star]{2003ApJ...592..935G}. The detailed numerical modelling of such a cluster dissociation is beyond the scope of the paper.

Hence, once two clumps within the CND collide or there is another external trigger (intense UV radiation, supernova blast wave, collision with a jet), the clump of several $100\,M_{\odot}$ will start collapsing and fragmenting into protostars on the scale of several thousand years, which is a comparable timescale or shorter than the infall time from the CND. The collisional timescale of clumps within the CND was estimated by \citet{Jalali2014} to be $\tau_{\rm col}\sim 10^5$ years, which is comparable to the lifetime of \normalfont{G2/DSO}-like objects. In this sense, \normalfont{G2/DSO}, OS1, OS2, and similar dusty objects in the Galactic center could be remnants of the previous infall of a fragmented star-forming cloud from the CND. 

\section{Conclusion} \label{sec:conclusion}

We conclude that data smoothing significantly affects the kinematical analysis of \normalfont{G2/DSO}-like sources. By using a smoothing kernel, the impact on a velocity gradient is not negligible and increases its overall range. From the presented analysis, we draw the following final remarks:
\begin{enumerate}
\item
The \normalfont{G2/DSO} can be detected in the SINFONI data as a compact source between 2005 and 2019. No shearing, elongation, or dissolving, which would be expected in case of a core-less, non-self-gravitating cloud, can be observed.
\item
A stellar counterpart can be traced at the position of the \normalfont{G2/DSO} in the K-band. It closely follows the Br$\gamma$ emission throughout the data.  
\item
Since we do not use overlaying smoothing tools for the data, we find that the claimed tail consists of individual sources. 
\item
Smoothing noisy data leads to interpretations that are not supported by the observations. We have shown that a natural velocity gradient gets artificially enhanced by using blurring filters. Hence, we propose that for the analysis of \normalfont{G2/DSO}-like sources, smoothing should be applied with an increased caution. 
\item
The ionized gas, which was supposed to be associated with the \normalfont{G2/DSO} tail, is not connected to the source and it is rather located within the S-cluster. It can be disentangled from the \normalfont{G2/DSO} emission.
\item
The sky emission/absorption variability of NIR data should be kept in mind during GC observations. Using sky frames for long-time exposures that do not match the commonly used observation scheme o-s-o tend to artificially influence the analysis of \normalfont{G2/DSO}-like sources.
\item
We question the need for a magnetohydrodynamic drag force for the orbital solution of the \normalfont{G2/DSO} trajectory based on the robust, purely Keplerian description of the orbit.
\item
The magnitude of the K-band counterpart of the \normalfont{G2/DSO} is constant till 2015. After 2017, the data implies a slightly decreased magnitude, i.e. an increase in the K-band flux density. 
\item The detected increase in the Br$\gamma$ line width up to the pericenter passage and a subsequent decrease can naturally be interpreted as an effect of \normalfont{accretion variations of the young star. As an additional interpretation, the evolution of} a velocity field of an unresolved stellar bow shock associated with the G2/DSO is consistent with the overall trend both quantitatively and qualitatively \citep{Zajacek2016}.
\item
The clockwise disk as a possible birth place for the \normalfont{G2/DSO} is rather unlikely. Because of the long-term data baseline, the confidence level of the derived orbit is increased compared to earlier orbital solutions that indicated a potential connection. However, the current \normalfont{G2/DSO} inclination of $\sim 60^{\circ}$ is significantly offset from the clockwise disk mean inclination of $\sim 115^{\circ}$ \citep[see e.g.][]{2009ApJ...690.1463L}.
\end{enumerate}

\acknowledgements
\normalfont{We highly appreciate the comprehensive comments of the anonymous referees that helped to improve this paper. Furthermore, we would like to thank M. Valencia-S. for her contributions to the analysis.}
This work was supported in part by the
Deutsche Forschungsgemeinschaft (DFG) via the Cologne
Bonn Graduate School (BCGS), the Max Planck Society
through the International Max Planck Research School
(IMPRS) for Astronomy and Astrophysics as well as special
funds through the University of Cologne. Conditions and Impact of Star Formation is carried out within the Collaborative Research Centre 956, sub-project [A01], funded by the Deutsche Forschungsgemeinschaft (DFG) – project ID 184018867. B.Sh. acknowledges financial support from the State Agency for Research of
the Spanish MCIU through the “Center of Excellence Severo Ochoa” award for the Instituto de Astrofisica de Andalucia (SEV-2017- 0709). MZ acknowledges the financial support by the National Science Center, Poland, grant No. 2017/26/A/ST9/00756 (Maestro 9) and the NAWA financial support under the agreement PPN/WYM/2019/1/00064 to perform a three-month exchange stay at the Charles University in Prague and the Astronomical Institute of the Czech Academy of Sciences. \normalfont{MZ also acknowledges the GA\v{C}R EXPRO grant 21-13491X (``Exploring the Hot Universe and Understanding Cosmic Feedback") for financial support.} Part of this
work was supported by fruitful discussions with members of
the European Union funded COST Action MP0905: Black
Holes in a Violent Universe and the Czech Science Foundation
-- DFG collaboration (VK, No.\ 19-01137J). AP, JC, SE, and GB contributed useful points to the discussion. We also would like to 
thank the members of the SINFONI/NACO/VISIR and ESO's Paranal/Chile team for their support and collaboration.


\bibliographystyle{aasjournal}
\bibliography{bib.bib}

\appendix
\label{sec:appendix}
\section{Used data}
\label{sec:app_data}
In Table \ref{tab:data_sinfo1}, \ref{tab:data_sinfo2}, \ref{tab:data_sinfo3}, \ref{tab:data_sinfo4}, \ref{tab:data_sinfo5}, and \ref{tab:data_sinfo6} we list the data used in this work. This data set was already used in \cite{Peissker2018}, \cite{Peissker2019}, \cite{Peissker2020b}, \cite{Peissker2020c}, \cite{Peissker2020d}, \cite{peissker2021}, and \cite{peisser2021b}. 

\begin{table*}[htbp!]
        \centering
        \begin{tabular}{cccccc}
        \hline\hline
        \\      Date & Observation ID  & \multicolumn{3}{c}{Amount of on source exposures} & Exp. Time \\  \cline{3-5} &  & Total & Medium & High &  \\
        (YYYY:MM:DD) &  &  &  &  & (s) \\ \hline\hline 
        
        2005.06.16 & 075.B-0547(B) &  20  &  12   &  8  &    300  \\
        2005.06.18 & 075.B-0547(B) &  21  &  2   &  19  &    60  \\
        2006.03.17 & 076.B-0259(B) &  5  &  0   &  3  &    600  \\
        2006.03.20 & 076.B-0259(B) &   1 &   1  &   0 &    600  \\
        2006.03.21 & 076.B-0259(B) &  2  &   2  &  0  &    600  \\
        2006.04.22 & 077.B-0503(B) &  1  &   0  &  0  &    600  \\
        2006.08.17 & 077.B-0503(C) &  1  &   0  &  1  &    600  \\
        2006.08.18 & 077.B-0503(C) &  5  &   0  &  5  &    600  \\
        2006.09.15 & 077.B-0503(C) &  3  &   0  &  3  &    600  \\
        2007.03.26 & 078.B-0520(A) &  8  &   1  &  2  &    600  \\
        2007.04.22 & 179.B-0261(F) &  7  &   2  &  1  &    600 \\
        2007.04.23 & 179.B-0261(F) &  10 &   0  &  0  &    600 \\
        2007.07.22 & 179.B-0261(F) &  3  &   0  &  2  &    600  \\
        2007.07.24 & 179.B-0261(Z) &  7  &   0  &  7  &    600  \\
        2007.09.03 & 179.B-0261(K) & 11  &   1  &  5  &    600  \\
        2007.09.04 & 179.B-0261(K) &  9  &   0  &  0  &    600  \\
        2008.04.06 & 081.B-0568(A) &  16 &   0  &  15   &    600  \\
        2008.04.07 & 081.B-0568(A) &   4 &   0  &   4 &    600  \\
        2009.05.21 & 183.B-0100(B) &  7 &   0  &  7   &    600  \\
        2009.05.22 & 183.B-0100(B) &   4 &   0  &   4 &    400  \\
        2009.05.23 & 183.B-0100(B) &  2 &   0  &  2  &    400  \\
        2009.05.24 & 183.B-0100(B) &  3 &   0  &  3  &    600  \\
        
        \hline  \\
        \end{tabular}   
        \caption{SINFONI data of 2005, 2006, 2007, 2008, and 2009. The total amount of data is listed.}
        \label{tab:data_sinfo1}
        \end{table*}
\begin{table*}[htbp!]
        \centering
        \begin{tabular}{cccccc}
        \hline\hline
        \\      Date & Observation ID  & \multicolumn{3}{c}{Amount of on source exposures} & Exp. Time \\  \cline{3-5} &  & Total & Medium & High &  \\
        (YYYY:MM:DD) &  &  &  &  & (s) \\ \hline\hline 
        
        2010.05.10 & 183.B-0100(O) & 3 &   0  &  3   &    600  \\
        2010.05.11 & 183.B-0100(O) &  5 &   0  &   5 &    600  \\
        2010.05.12 & 183.B-0100(O) & 13 &   0  &  13  &    600  \\
        2011.04.11 & 087.B-0117(I) &  3 &   0  &   3 &    600  \\
        2011.04.27 & 087.B-0117(I) & 10 &   1  &  9  &    600  \\
        2011.05.02 & 087.B-0117(I) & 6 &   0  &  6  &    600  \\
        2011.05.14 & 087.B-0117(I) & 2 &   0  &  2  &    600  \\
        2011.07.27 & 087.B-0117(J)/087.A-0081(B) & 2 &   1  &  1  &    600  \\     
        2012.03.18 & 288.B-5040(A) &  2 &   0  &   2 &    600  \\
        2012.05.05 & 087.B-0117(J) & 3 &   0  &  3  &    600  \\
        2012.05.20 & 087.B-0117(J) & 1 &   0  &  1  &    600  \\
        2012.06.30 & 288.B-5040(A) & 12 &   0  &  10  &    600  \\
        2012.07.01 & 288.B-5040(A) & 4 &   0  &  4  &    600  \\
        2012.07.08 & 288.B-5040(A)/089.B-0162(I)& 13 &   3  &  8  &    600  \\
        2012.09.08 & 087.B-0117(J)  & 2 &   1  &  1  &    600  \\
        2012.09.14 & 087.B-0117(J)  & 2 &   0  &  2  &    600  \\   
        2013.04.05 & 091.B-0088(A)  &  2 &   0  &   2 &    600  \\
        2013.04.06 & 091.B-0088(A)  & 8 &   0  &  8  &    600  \\
        2013.04.07 & 091.B-0088(A)  & 3 &   0  &  3  &    600  \\
        2013.04.08 & 091.B-0088(A)  & 9 &   0  &  6  &    600  \\
        2013.04.09 & 091.B-0088(A)  & 8 &   1  &  7  &    600  \\
        2013.04.10 & 091.B-0088(A)  & 3 &   0  &  3  &    600  \\
        2013.08.28 & 091.B-0088(B)  & 10 &  1  &  6  &    600 \\
        2013.08.29 & 091.B-0088(B)  & 7 &  2   &  4  &    600 \\
        2013.08.30 & 091.B-0088(B)  & 4 &  2   &  0  &    600 \\
        2013.08.31 & 091.B-0088(B)  & 6 &  0   &  4  &    600 \\
        2013.09.23 & 091.B-0086(A)  & 6 &   0  &  0  &    600  \\
        2013.09.25 & 091.B-0086(A)  & 2 &   1  &  0  &    600  \\
        2013.09.26 & 091.B-0086(A)  & 3 &   1  &  1  &    600  \\   
        
        \hline  \\
        \end{tabular}
        
        \caption{SINFONI data of 2010, 2011, 2012, and 2013.}
        \label{tab:data_sinfo2}
        \end{table*}

\begin{table*}[htbp!]
        \centering
        \begin{tabular}{cccccc}
        \hline\hline
        \\      Date & Observation ID  & \multicolumn{3}{c}{Amount of on source exposures} & Exp. Time \\  \cline{3-5} &  & Total & Medium & High &  \\
        (YYYY:MM:DD) &  &  &  &  & (s) \\ \hline\hline 
        
        2014.02.27 & 092.B-0920(A) &  4 &   1  &  3  &    600   \\
        2014.02.28 & 091.B-0183(H) &  7 &   3  &  1  &    400   \\
        2014.03.01 & 091.B-0183(H) & 11 &   2  &  4  &    400  \\
        2014.03.02 & 091.B-0183(H) &  3 &   0  &  0  &    400   \\
        2014.03.11 & 092.B-0920(A) & 11 &   2  &  9  &    400   \\
        2014.03.12 & 092.B-0920(A) & 13 &   8  &  5  &    400   \\
        2014.03.26 & 092.B-0009(C) & 9  &   3  &  5  &    400   \\
        2014.03.27 & 092.B-0009(C) & 18 &   7  &  5  &    400   \\
        2014.04.02 & 093.B-0932(A) & 18 &   6  &  1  &    400    \\
        2014.04.03 & 093.B-0932(A) & 18 &   1  &  17 &    400    \\
        2014.04.04 & 093.B-0932(B) & 21 &   1  &  20 &    400       \\
        2014.04.06 & 093.B-0092(A) &  5 &   2  &  3  &    400      \\
        2014.04.08 & 093.B-0218(A) & 5  &   1  &  0  &    600   \\
        2014.04.09 & 093.B-0218(A) &  6 &   0  &  6  &    600   \\
        2014.04.10 & 093.B-0218(A) & 14 &   4  &  10  &    600   \\
        2014.05.08 & 093.B-0217(F) & 14 &   0  &  14  &    600   \\
        2014.05.09 & 093.B-0218(D) & 18 &   3  &  13  &    600   \\
        2014.06.09 & 093.B-0092(E) &  14 &   3  &  0  &    400   \\
        2014.06.10 & 092.B-0398(A)/093.B-0092(E) & 5 &   4  &  0   & 400/600 \\
        2014.07.08 & 092.B-0398(A)  & 6 &   1  &  3   &    600 \\
        2014.07.13 & 092.B-0398(A)  & 4 &   0  &  2   &    600 \\
        2014.07.18 & 092.B-0398(A)/093.B-0218(D)  & 1 &   0  &  0   &    600 \\
        2014.08.18 & 093.B-0218(D)  & 2 &   0  &  1   &    600 \\ 
        2014.08.26 & 093.B-0092(G)  &  4 &   3  &   0 &    400   \\
        2014.08.31 & 093.B-0218(B)  & 6 &   3   &   1 &    600 \\
        2014.09.07 & 093.B-0092(F)  & 2 &   0  &  0  &    400   \\
        2015.04.12 & 095.B-0036(A)  & 18 &  2 & 0 & 400 \\
        2015.04.13 & 095.B-0036(A)  & 13 &  7 & 0 & 400 \\
        2015.04.14 & 095.B-0036(A)  & 5  &  1 & 0 & 400 \\
        2015.04.15 & 095.B-0036(A)  & 23 &  13  & 10 & 400 \\
        2015.08.01 & 095.B-0036(C)  & 23 &   7  & 8  & 400 \\
        2015.09.05 & 095.B-0036(D)  & 17 &  11  & 4  & 400 \\

        \hline  \\
        \end{tabular}
        
        \caption{SINFONI data of 2014 and 2015.}
        \label{tab:data_sinfo3}
        \end{table*}
        
\begin{table*}[htbp!]
        \centering
        \begin{tabular}{cccccc}
        \hline\hline
        \\      Date & Observation ID  & \multicolumn{3}{c}{Amount of on source exposures} & Exp. Time \\  \cline{3-5} &  & Total & Medium & High &  \\
        (YYYY:MM:DD) &  &  &  &  & (s) \\ \hline\hline 
        
        2016.03.15 & 096.B-0157(B) & 15 &   0  &  15  &    400   \\
        2016.03.16 & 096.B-0157(B) & 17 &   0  &  17  &    400   \\
        2016.04.14 & 594.B-0498(R) & 12 &   0  &  12  &    600  \\
        2016.04.16 & 594.B-0498(R) & 10 &   0  &  8  &    600   \\
        2016.07.09 & 097.B-0050(A) & 15 &   0  &  2  &    600   \\
        2016.07.11 & 097.B-0050(A) & 38 &   0  &  3  &    600   \\
        2016.07.12 & 097.B-0050(A) & 27 &   0  &  13  &    600   \\
        2017.03.15 & 598.B-0043(D) &  5 &   2  &  0  &    600   \\
        2017.03.19 & 598.B-0043(D) & 11 &   0  &  5  &    600   \\
        2017.03.20 & 598.B-0043(D) & 15 &   4  &  11 &    600   \\
        2017.03.21 & 598.B-0043(D) & 1  &   0  &  0  &    600  \\
        2017.05.20 & 0101.B-0195(B) & 8 &   2  &  6  &    600   \\
        2017.06.01 & 598.B-0043(E) & 5  &   0  &  3  &    600   \\
        2017.06.02 & 598.B-0043(E) & 8  &   0  &  8  &    600  \\
        2017.06.29 & 598.B-0043(E) & 4  &   2  &  17 &    600   \\
        2017.07.20 & 0101.B-0195(C) & 8 &   5  &  0  &    600   \\
        2017.07.28 & 0101.B-0195(C) & 6 &   0  &  0  &    600   \\
        2017.07.29 & 0101.B-0195(D) & 9 &   0  &  0  &    600   \\
        2017.08.01 & 0101.B-0195(E) & 4 &   0  &  0  &    600   \\
        2017.08.19 & 598.B-0043(F) &  8 &   0  &  2  &    600   \\
        2017.09.13 & 598.B-0043(F) &  8 &   0  &  0  &    600   \\
        2017.09.15 & 598.B-0043(F) & 10  &  1  &  1  &    600   \\
        2017.09.29 & 598.B-0043(F) &  2  &  0  &  0  &    600   \\
        2017.10.15 & 0101.B-0195(F) & 2  &  0  &  0  &    600   \\
        2017.10.17 & 0101.B-0195(F) & 4  &  0  &  0  &    600   \\
        2017.10.23 & 598.B-0043(G) &  3  &  0  &  0  &    600   \\
        \hline  \\
        \end{tabular}
        
        \caption{SINFONI data of 2016 and 2017.}
        \label{tab:data_sinfo4}
        \end{table*}

\begin{table*}[htbp!]
        \centering
        \begin{tabular}{cccccc}
        \hline\hline
        \\      Date & Observation ID  & \multicolumn{3}{c}{Amount of on source exposures} & Exp. Time \\  \cline{3-5} &  & Total & Medium & High &  \\
        (YYYY:MM:DD) &  &  &  &  & (s) \\ \hline\hline 
        
        2018.02.13 & 299.B-5056(B) &  3 &   0  &  0  &    600   \\
        2018.02.14 & 299.B-5056(B) &  5 &   0  &  0  &    600   \\
        2018.02.15 & 299.B-5056(B) &  5 &   0  &  0  &    600   \\
        2018.02.16 & 299.B-5056(B) &  5 &   0  &  0  &    600   \\
        2018.03.23 & 598.B-0043(D) &  8 &   0  &  8  &    600   \\
        2018.03.24 & 598.B-0043(D) &  7 &   0  &  0  &    600   \\
        2018.03.25 & 598.B-0043(D) &  9 &   0  &  1  &    600   \\
        2018.03.26 & 598.B-0043(D) & 12 &   1  &  9  &    600  \\
        2018.04.09 & 0101.B-0195(B) &  8 &   0  &  4  &    600   \\
        2018.04.28 & 598.B-0043(E) & 10 &   1  &  1  &    600   \\
        2018.04.30 & 598.B-0043(E) & 11 &   1  &  4  &    600  \\
        2018.05.04 & 598.B-0043(E) & 17 &   0  &  17  &    600   \\
        2018.05.15 & 0101.B-0195(C) &  8 &   0  &  0  &    600   \\
        2018.05.17 & 0101.B-0195(C) &  8 &   0  &  4  &    600   \\
        2018.05.20 & 0101.B-0195(D) &  8 &   0  &  4  &    600   \\
        2018.05.28 & 0101.B-0195(E) &  8 &   3  &  1  &    600   \\
        2018.05.28 & 598.B-0043(F) &  4 &   0  &  4  &    600   \\
        2018.05.30 & 598.B-0043(F) &  8 &   5  &  3  &    600   \\
        2018.06.03 & 598.B-0043(F) &  8 &   0  &  8  &    600   \\
        2018.06.07 & 598.B-0043(F) & 14 &   1  &  7  &    600   \\
        2018.06.14 & 0101.B-0195(F) &  4 &   0  &  0  &    600   \\
        2018.06.23 & 0101.B-0195(F) &  8 &   1  &  1  &    600   \\
        2018.06.23 & 598.B-0043(G) &  7 &   2  &  1  &    600   \\
        2018.06.25 & 598.B-0043(G) & 22 &   5  &  7  &    600   \\
        2018.07.02 & 598.B-0043(G) &  3 &   0  &  0  &    600   \\
        2018.07.03 & 598.B-0043(G) & 22 &  12  & 10  &    600   \\
        2018.07.09 & 0101.B-0195(G) &  8 &   3  &  1  &    600   \\
        2018.07.24 & 598.B-0043(H) &  3 &   0  &  0  &    600   \\
        2018.07.28 & 598.B-0043(H) &  8 &   0  &  3  &    600   \\
        2018.08.03 & 598.B-0043(H) &  8 &   0  &  1  &    600   \\
        2018.08.06 & 598.B-0043(H) &  8 &   1  &  1  &    600   \\
        2018.08.19 & 598.B-0043(I) & 12 &   2  & 10  &    600   \\
        2018.08.20 & 598.B-0043(I) & 12 &   0  & 12  &    600   \\
        2018.09.03 & 598.B-0043(I) &  1 &   0  &  0  &    600   \\
        2018.09.27 & 598.B-0043(J) & 10 &   0  &  0  &    600   \\
        2018.09.28 & 598.B-0043(J) & 10 &   0  &  0  &    600   \\
        2018.09.29 & 598.B-0043(J) &  8 &   0  &  0  &    600   \\
        2018.10.16 & 2102.B-5003(A) &  3 &   0  &  0  &    600   \\

        \hline  \\
        \end{tabular}
        
        \caption{SINFONI data of 2018.}
        \label{tab:data_sinfo5}
        \end{table*}

\begin{table*}[htbp!]
        \centering
        \begin{tabular}{cccccc}
        \hline\hline
        \\      Date & Observation ID  & \multicolumn{3}{c}{Amount of on source exposures} & Exp. Time \\  \cline{3-5} &  & Total & Medium & High &  \\
        (YYYY:MM:DD) &  &  &  &  & (s) \\ \hline\hline 
        
        2019.04.20 & 0103.B-0026(B)  & 9 &   0  &  8  &    600   \\
        2019.04.28 & 0103.B-0026(B)  &  4 &   0  &  2  &    600   \\
        2019.04.29 & 0103.B-0026(B)  &  8 &   0  &  4  &    600   \\
        2019.05.02 & 0103.B-0026(F)  &  8 &   0  &  8  &    600   \\
        2019.05.23 & 5102.B-0086(Q)  &  13 &   0  &  6  &    600   \\
        2019.05.24 & 5102.B-0086(Q)  &  4 &   0  &  2  &    600   \\

        2019.06.01 & 0103.B-0026(F)  &  9 &   0  &  2  &    600   \\
        2019.06.03 & 0103.B-0026(D)  & 8 &   0  &  2  &    600   \\
        2019.06.04 & 5102.B-0086(Q)  &  6 &   0  &  3  &    600   \\
        2019.06.06 & 594.B-0498(Q)   &  11 &   0  &  10  &    600   \\
        2019.06.09 & 5102.B-0086(Q) &  14 &   0  &  10  &    600   \\
        2019.06.14 & 0103.B-0026(D)   &  4 &   0  &  0  &    600   \\
        2019.06.19 & 0103.B-0026(D) &  2 &   0  &  0  &    600   \\

        \hline  \\
        \end{tabular}
        
        \caption{SINFONI data of 2019.}
        \label{tab:data_sinfo6}
        \end{table*} 
        
\section{Continuum fitting}
\label{sec:conti_fit_app}

In this section, we present an example for the continuum fitting. The related and resulting spectrum of 2008 is displayed in Fig. \ref{fig:dso_spectral_line_evo}. 

\begin{figure}[htbp!]
	\centering
	\includegraphics[width=.5\textwidth]{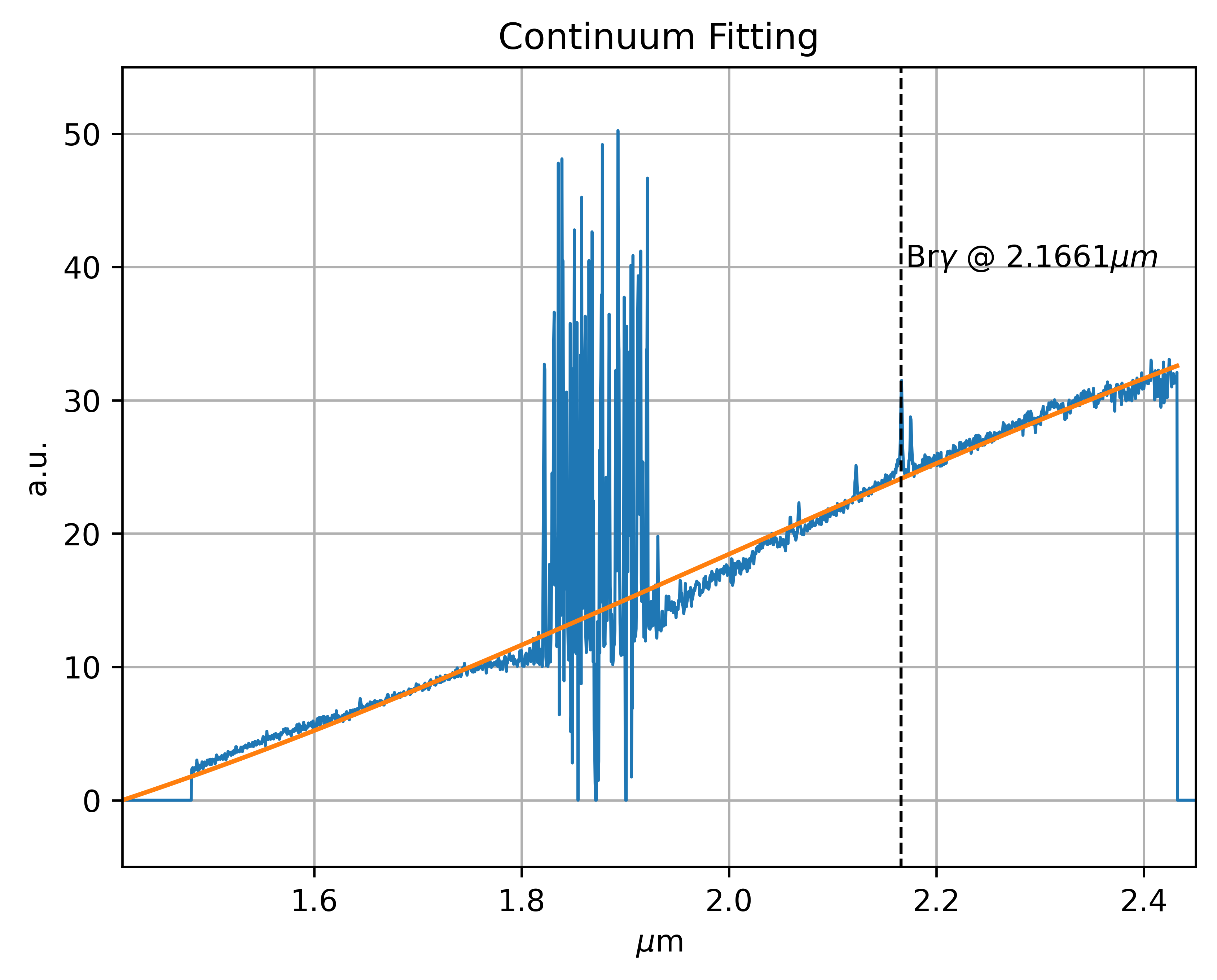}
	\caption{Example of a fit of the underlying H+K continuum \normalfont{of G2/DSO in 2008}. We use a 2nd degree polynomial for the extracted spectrum. \normalfont{The Br$\gamma$ rest wavelength is marked with a dashed line. The prominent telluric absorption lines are located between $1.8\,\mu m$ and $1.95\,\mu m$. Because this spectral region is not free of confusion, we are following the argumentation of \cite{Gillessen2013b} and focus on the Br$\gamma$ analysis.}}
\label{fig:fitted_spectrum}
\end{figure}        

\section{The tail of the \normalfont{G2/DSO} cloud}
\label{sec:tail_smoothed_app}

Here, we compare the results of the literature with the analysis and conclusions from this work. We furthermore apply a Gaussian smoothing beam to the PPV diagrams shown in Sec. \ref{sec:results}.
\begin{figure*}[htbp!]
	\centering
	\includegraphics[width=1.\textwidth]{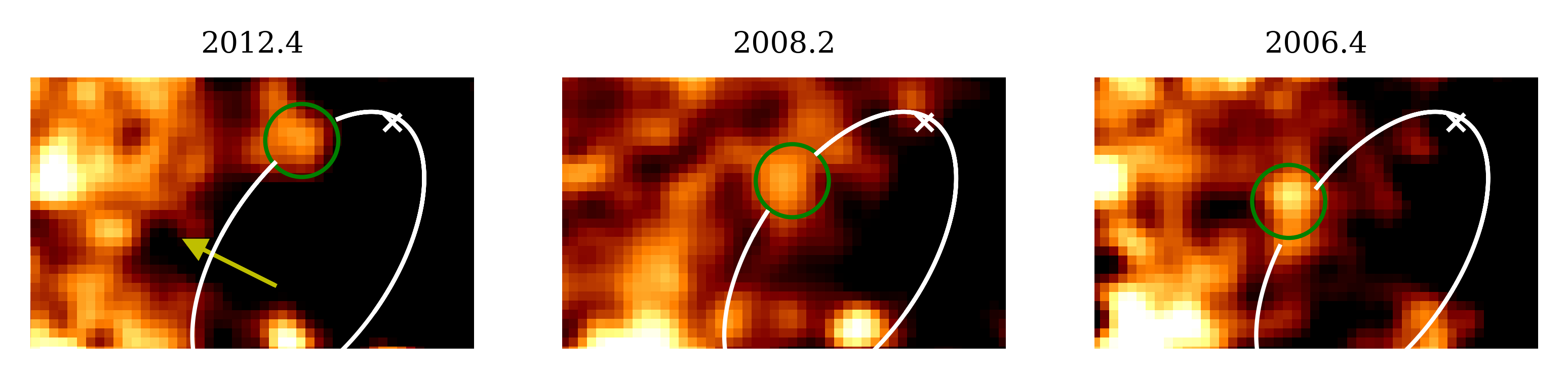}
	\caption{Disputed tail of the assumed cloud. This figure is inspired by \cite{Gillessen2013a} (see Fig. 9 in the related work). The yellow arrow marks the position of the possible tail emission. Sgr~A* is located at the $\times$. The circle indicates the position of the \normalfont{G2/DSO}. North is up, East is to the left. The size of every panel is $0.61"\,\times\,0.37"$.}
\label{fig:art_tail_2006_2008_2012}
\end{figure*}
In Fig. \ref{fig:art_tail_2006_2008_2012}, we show the collapsed line maps where we adapt the settings given in \cite{Gillessen2013a}. We select 25 channels between $2.1695\,\mu m\,-\,2.1815\,\mu m$. Furthermore, we subtract the averaged neighboring channels and smooth the resulting images with a 3 px Gaussian beam. Since the Br$\gamma$ emission, which is associated with the tail, can be found in 2006, 2008, and 2012, we conclude that the indicated feature in \cite{Gillessen2013a} is rather background related.
\normalfont{We advice the reader to compare the orbit and the line emission position of the tail in Fig. 1 and Fig. 3 of \cite{Pfuhl2015}. Please consider the mismatching emission of the tail and the position of the slit as presented in Fig. 10 and Fig. 12 in \cite{Plewa2017} that povides inconsistent information. While the tail seems to be on the same orbit as G2/DSO \citep[Fig. 1 in][]{Pfuhl2015, Gillessen2019}, it is clearly misplaced as shown in the contour plot of Fig. 10 in \cite{Plewa2017}.}

\begin{figure*}[htbp!]
	\centering
	\includegraphics[width=1.\textwidth]{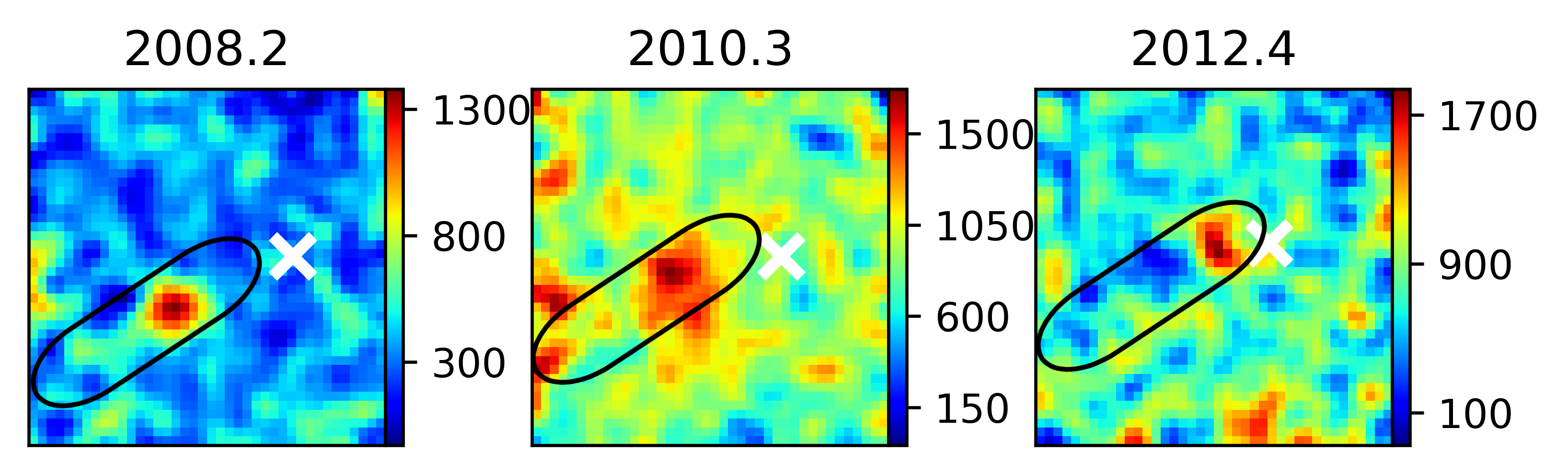}
	\caption{Smoothed PPV images. We smooth the results shown in Fig. \ref{fig:ppv_1} with a Gaussian smoothing kernel. Here, the surrounding noise is enhanced by the Gaussian kernel. Sgr~A* is located at the position of the white $\times$. We mark the assumed position of the disputed tail emission. North is up, East is to the left. As in Fig. \ref{fig:ppv_1}, the size of every panel is $0.5"\,\times\,0.5"$.}
\label{fig:ppv_2}
\end{figure*}

\section{Brackett-$\gamma$ line-width as a function of time}
\label{sec:brgamma_zoom_app}

\begin{figure}[htbp!]
	\centering
	\includegraphics[width=.5\textwidth]{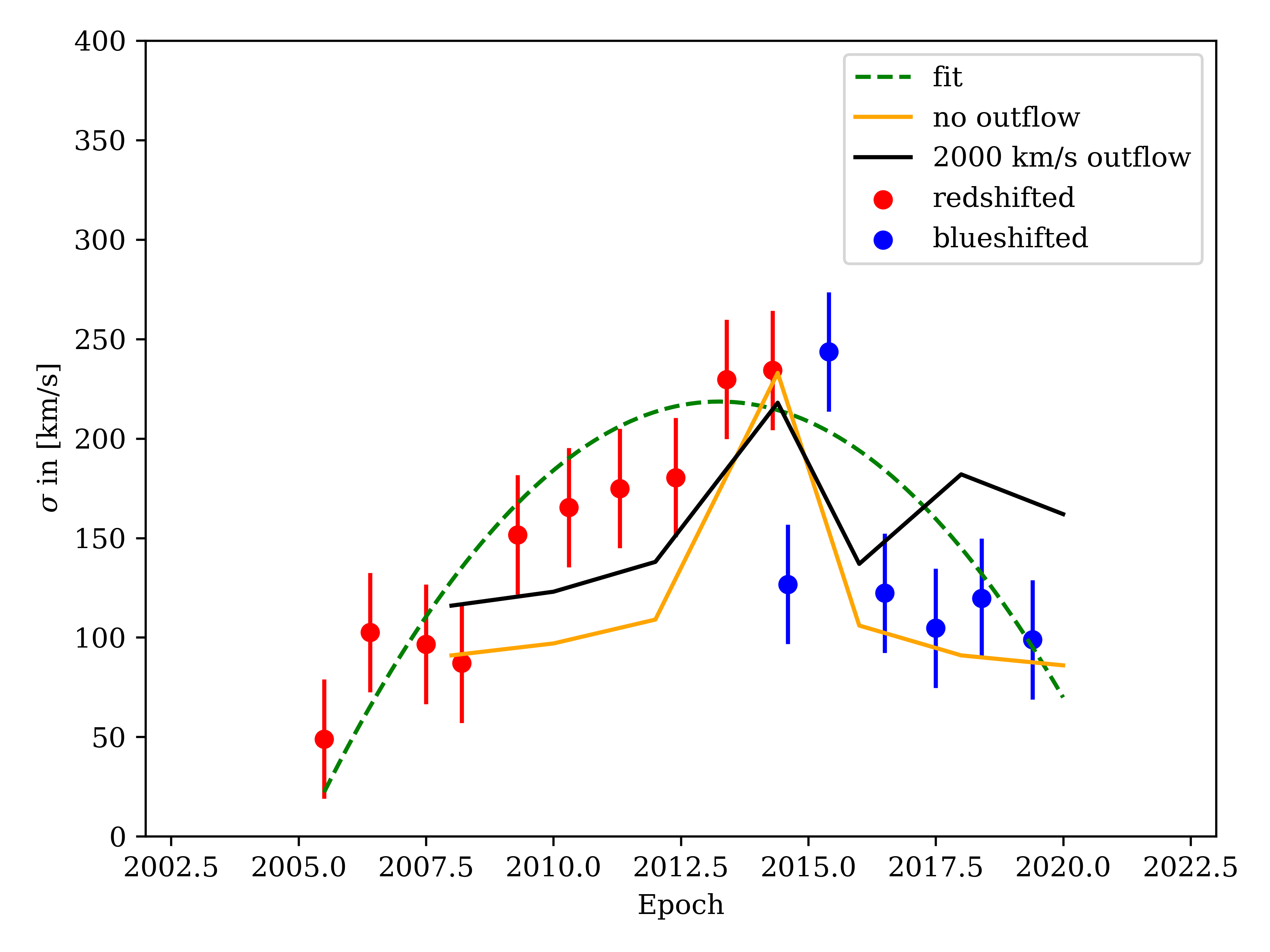}
	\caption{Zoomed-in view of Figure  \ref{fig:sigma_time_evo}. We fit a second-degree polynomial function to the data (green dashed line). Based on  accretion effects in combination of the foreshortening factor (see Sec. \ref{sec:brgamma_line_width} for details), a variation of the line-width is expected. The prominent line-width that is related to a cloud-model (see Fig. \ref{fig:sigma_time_evo}) cannot be confirmed.}
\label{fig:sigma_time_evo_zoom}
\end{figure}

Here we present a zoomed-in view of Figure \ref{fig:sigma_time_evo} with about $15\%$ of the initial boundary range. As mentioned before, the black and orange line are related to the bow-shock model discussed in \cite{Zajacek2016}. This bow-shock model shows a strong correlation to the observed line-width. Due to the OH lines Q1 and Q2 (see Table \ref{tab:oh_line_list}), the pollution of the line-width in 2015 is most probably higher than measured. Hence, the data point in 2015 is an upper limit. Averaging the two post-pericenter data point in 2014.6 and 2015.4 results in an line-width of $247.5\,km/s$ which is in agreement with the presented bow-shock model. The green dashed line in Fig. \ref{fig:sigma_time_evo_zoom} agrees with the foreshortening factor discussed by \cite{Valencia-S.2015}.

\section{Line of sight velocity of the OS sources}
\label{sec:los_os_app}

\begin{figure}[htbp!]
	\centering
	\includegraphics[width=1.\textwidth]{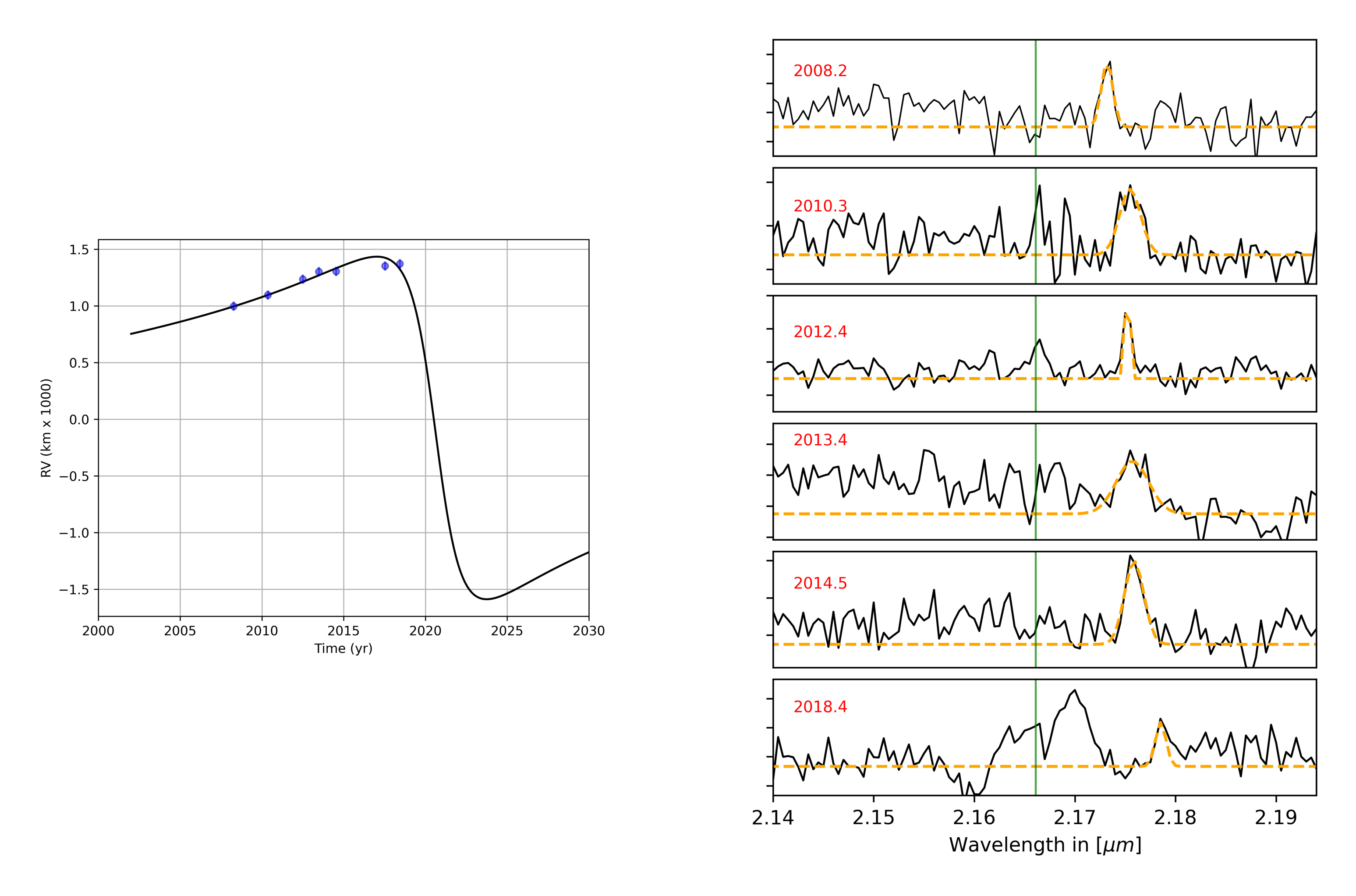}
	\caption{Evolution of the redshifted Br$\gamma$ line of OS1.}
\label{fig:os1_velorbit}
\end{figure}

\begin{figure}[htbp!]
	\centering
	\includegraphics[width=1.\textwidth]{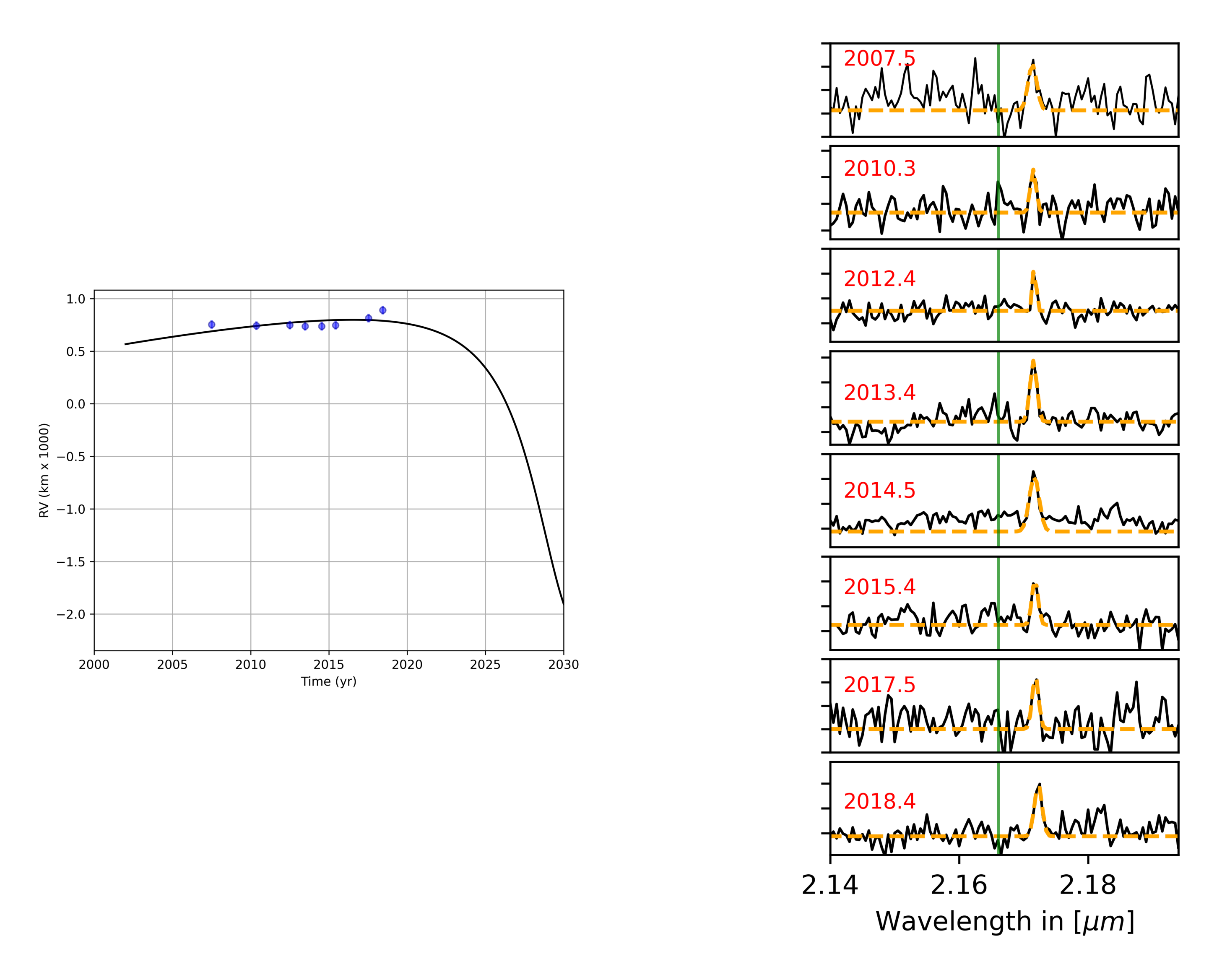}
	\caption{Evolution of the redshifted Br$\gamma$ line of OS2.}
\label{fig:os2_velorbit}
\end{figure}

As for the \normalfont{G2/DSO}, we use the Doppler-shifted Br$\gamma$ line to derive a LOS velocity v$_z$. The line evolution of OS1 and OS2 both are following a Keplerian approach (see Fig. \ref{fig:os1_velorbit} and Fig. \ref{fig:os2_velorbit}). We use the same tools and techniques for the analysis of OS1 and OS2 as for the \normalfont{G2/DSO}.


\begin{table*}[htb]
\centering
\begin{tabular}{|c|ccc|ccc|}\hline \hline 
      & \multicolumn{3}{c}{OS1} & \multicolumn{3}{c|}{OS2} 	\\ 
      \hline
Epoch      & Wavelength &	Velocity & Standard deviation $\sigma$ & Wavelength & Velocity & Standard deviation $\sigma$ in [km/s]	\\
  &in [$\mu m$]  & in     [km/s] & in    [km/s]  & in [$\mu m$]  & in [km/s]  & in [km/s] \\  
\hline 
2007.5 & - & - & - & 2.1713 & 728.94 & 93.37 \\  
2008.2 & 2.1731 & 982.76 & 78.51 & - & - & - \\  
2010.3 & 2.1754 & 1300.48 & 145.25 & 2.1714 & 747.22 & 61.18 \\ 
2012.4 & 2.1752 & 1266.56 & 17.93 & 2.1716 & 773.58 & 33.78 \\  
2013.4 & 2.1757 & 1332.03 & 223.71 & 2.1715 & 750.70 & 72.00 \\  
2014.5 & 2.1758 & 1356.56 & & 2.1716 & 767.65 & 105.85 \\  
2015.4 & - & - & - & 2.1717 & 783.34 & 65.47 \\  
2017.5 & - & - & - & 2.1718 & 793.48 & 68.66 \\  
2018.4 & 2.1785 & 1721.30 & 72.12 & 2.1722 & 853.79 & 80.74 \\  
\hline \hline
\end{tabular}
\caption{Measured Doppler-shifted wavelengths and velocities for the OS sources. In some years, close-by or overlapping stellar sources hindered a confusion-free detection of the emission line. \normalfont{Epoch is given in decimal years.}}
\label{tab:spectral_velocity_properties_ossources}
\end{table*}

\section{Positions of G2/DSO, OS1, and OS2}
\label{sec:positions_app}
In this section, we list the used values for the Keplerian fit that are presented in this work for G2/DSO, OS1, and OS2. The fitted velocity information are listed in Table \ref{tab:spectral_velocity_properties} and Table \ref{tab:spectral_velocity_properties_ossources}. We furthermore show the K-band continuum position of G2/DSO based on the high-pass filtered images presented in Fig. \ref{fig:lr_results}. We use a Gaussian to derive the K-band positions and combine them with the orbital solution (Table \ref{tab:orbit_elements}) of the source. 

\begin{table*}[htb]
\centering
\begin{tabular}{|c|cccc|cccc|cccc|}\hline \hline 
      & \multicolumn{4}{c|}{G2/DSO} & \multicolumn{4}{c|}{OS1} & \multicolumn{4}{c|}{OS2}	\\ 
      \hline
Epoch      & \multicolumn{4}{c|}{Position in [mas]}& \multicolumn{4}{c|}{Position in [mas]} & \multicolumn{4}{c|}{Position in [mas]}	\\
  & x  & y & $\Delta$x & $\Delta$y & x  & y &$\Delta$x & $\Delta$y & x  & y & $\Delta$x & $\Delta$y\\  
\hline 
2005.5 & -212.91 & -116.55  &  5.61 & 2.23 & - & - &  - &  - & - & - &  - &  - \\
2006.4 & -205.05 &  -98.55  &  2.20 & 2.26 & - & - &  - &  - & - & - &  - &  -\\
2007.5 & -183.97 &  -87.08  &  4.59 & 4.69 & - & - &  - &  - & -372.22 & -188.87 & 3.56 & 3.62\\
2008.2 & -170.58  &  -77.62  &  1.22 & 1.28 & -313.43 & -97.25 & 12.5 & 12.5 &-388.52 & -209.72 &10.10 & 10.25\\
2009.3 & -154.71 &  -69.36   &  2.41 & 2.10   & - & - &  - &  - & - & - &  - &  -\\
2010.3 & -141.01 &  -33.39  &  1.60 & 1.28 & -304.86 & -69.6 &12.5 & 12.5   &-399.42 & -191.00   &5.47 & 4.05\\
2011.3 & -127.53 &  -28.15  &  1.59 & 1.74  & - & - &  - &  -                 &-395.33 & -196.80   &4.43 & 9.96\\                                
2012.4 & -101.59 &  -21.80  &  1.78 & 2.11 & -240.65 & -33.46 &12.5 & 12.5  &-387.88 & -183.125 &3.512 & 3.77\\
2013.4 &  -71.59 &    1.42  &  1.88 & 1.53 & -205.47 & -20.37 &12.5 & 12.5 &-377.67 & -186.00  &5.15  & 12.27\\  
2014.3 &  -27.42  &   26.19  &  1.71 & 1.39 & - & - &  - &  - & - & - &  - &  - \\
2014.5 &   44.52  &  -12.49  &  2.98 & 2.83 & -189.51 &  -9.45 &12.5 & 12.5  &-351.13 &-172.47 & 2.85   & 3.12\\      
2015.4 &   51.87 &  -31.21  &  2.05 & 3.00   & - & - &  - &  -                     &-348.02 &-163.56 & 10.15  & 4.12\\  
2016.5 &   41.30 &  -86.89  &  2.07 & 3.68 & - & - &  - &  -                     &-323.35  &-155.70   & 2.80    & 1.92\\
2017.5 &   37.48 & -112.20    &  2.65 & 2.5  & -177.72 &  -7.68 &12.5 & 12.5 &-331.28 &-133.11 & 29.07 & 4.43\\ 
2018.4 &   39.13 & -140.07  &  7.28 & 3.63 & -110.58 &  12.91 &12.5 & 12.5 &-315.61 &-134.58 & 3.05   & 3.85\\
2019.4 &   15.06  & -161.88   &  1.62  & 2.07 & - & - &  - &  - & - & - &  - &  -\\
\hline \hline
\end{tabular}
\caption{Positions of G2/DSO, OS1, and OS2. The values are in mas, the uncertainty is adapted from the Gaussian fit of the position. \normalfont{The epoch refers here to decimal years.}}
\label{tab:fit_positions}
\end{table*}

\begin{figure*}[htbp!]
	\centering
	\includegraphics[width=1.0\textwidth]{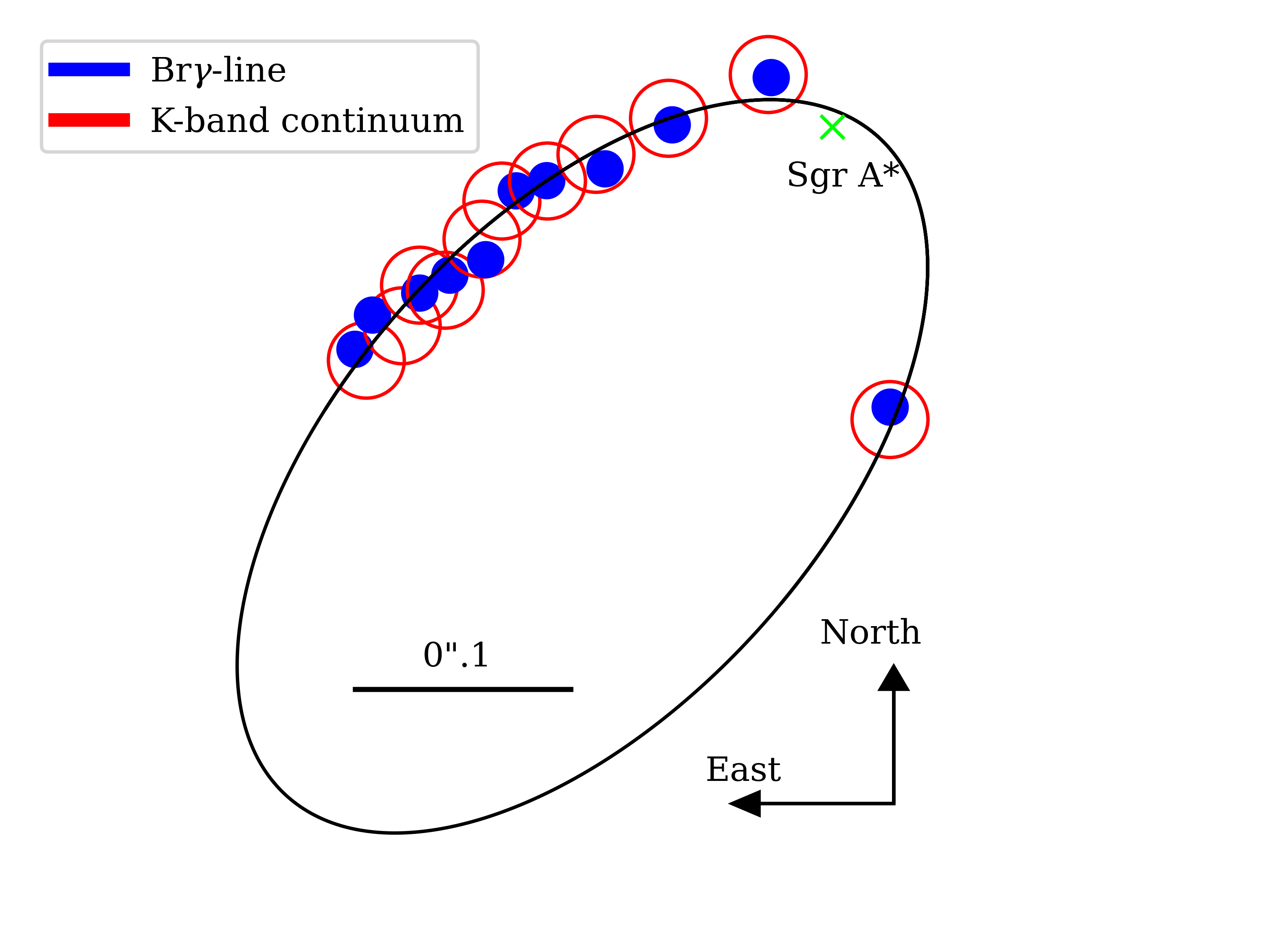}
	\caption{K-band continuum positions (red circle) derived from the high-pass filtered detection of G2/DSO as presented in Fig. \ref{fig:lr_results}. \normalfont{From Fig. \ref{fig:dso_line_evo}, we include the Br$\gamma$ line detection (blue points) for comparison.} The lime colored $\times$ marks the position of Sgr~A*. In 2015--2017 and 2019, the K-band counterpart of G2/DSO is confused with S23 and S31.}
\label{fig:kband_orbit}
\end{figure*}

\section{MCMC results of the orbital elements of \normalfont{G2/DSO and} the OS sources}
\label{sec:mcmc_results_app}
\begin{figure*}[htbp!]
	\centering
	\includegraphics[width=1.\textwidth]{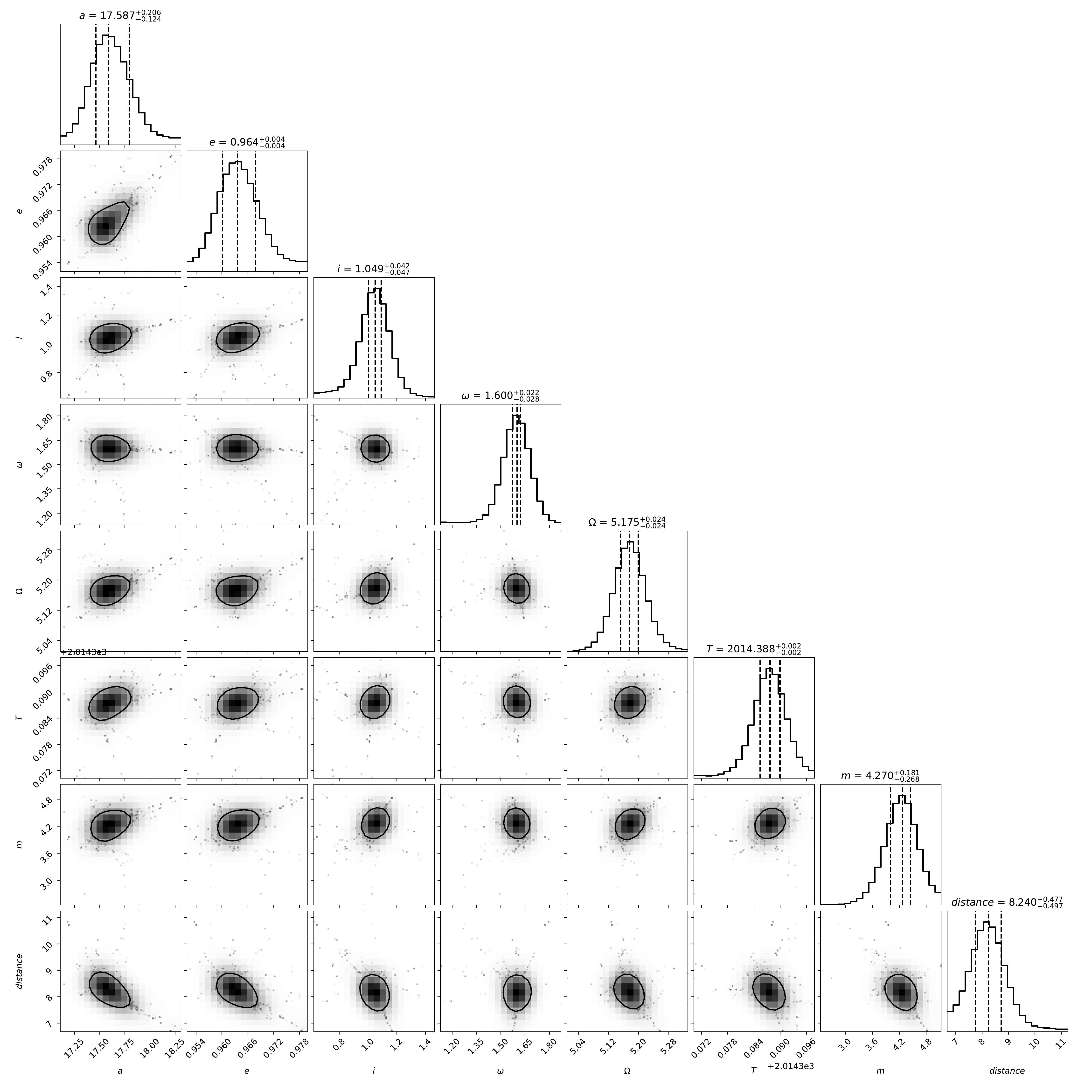}
	\caption{MCMC simulations of the Keplerian fit parameters of G2/DSO. The mean of the posterior distribution is in agreement with the input parameters which underlines the robustness of the Keplerian fit.}
\label{fig:dso_orbit_3}
\end{figure*}

\begin{figure*}[htbp!]
	\centering
	\includegraphics[width=1.0\textwidth]{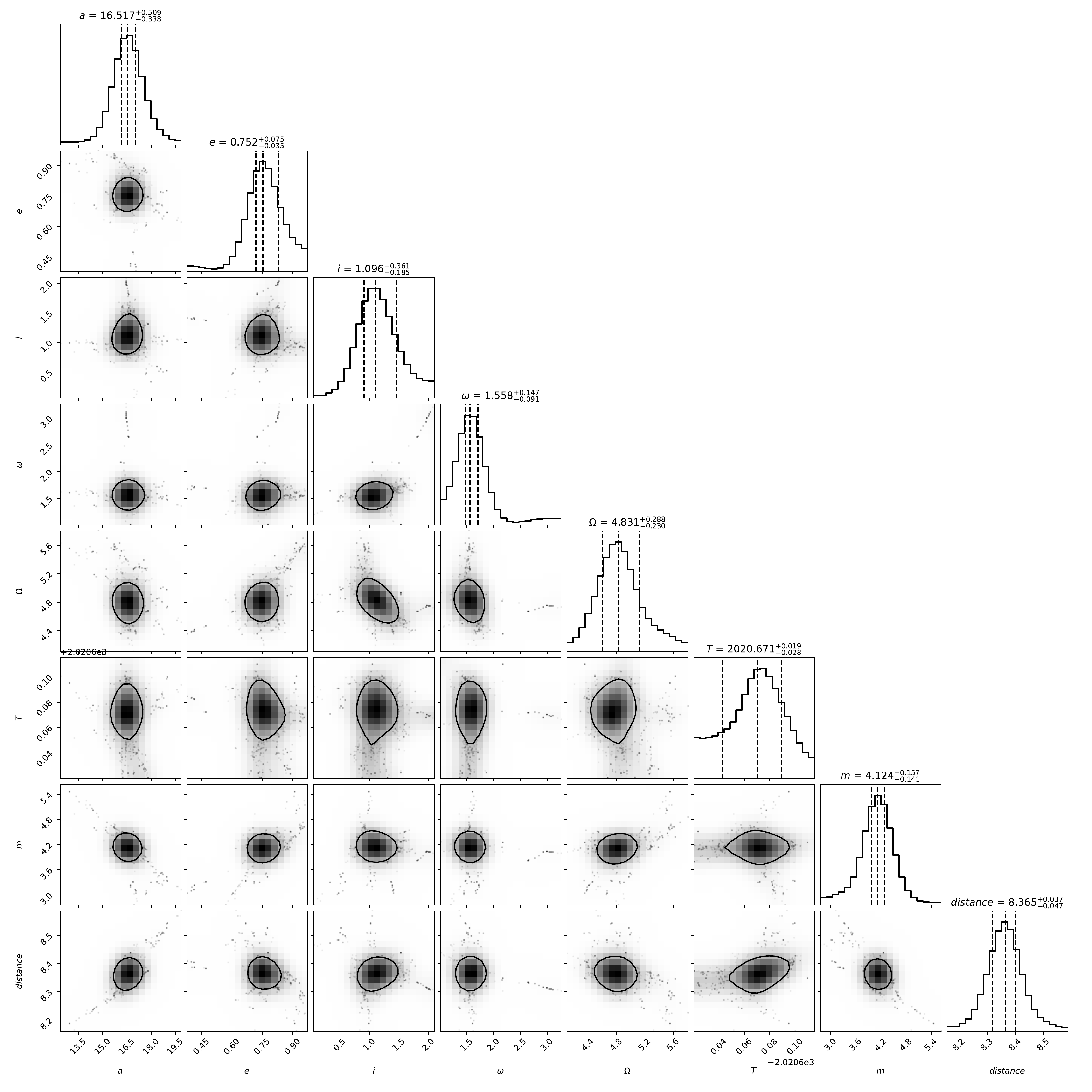}
	\caption{MCMC simulations for the orbital elements of OS1.}
\label{fig:os1_mcmc}
\end{figure*}

\begin{figure*}[htbp!]
	\centering
	\includegraphics[width=1.0\textwidth]{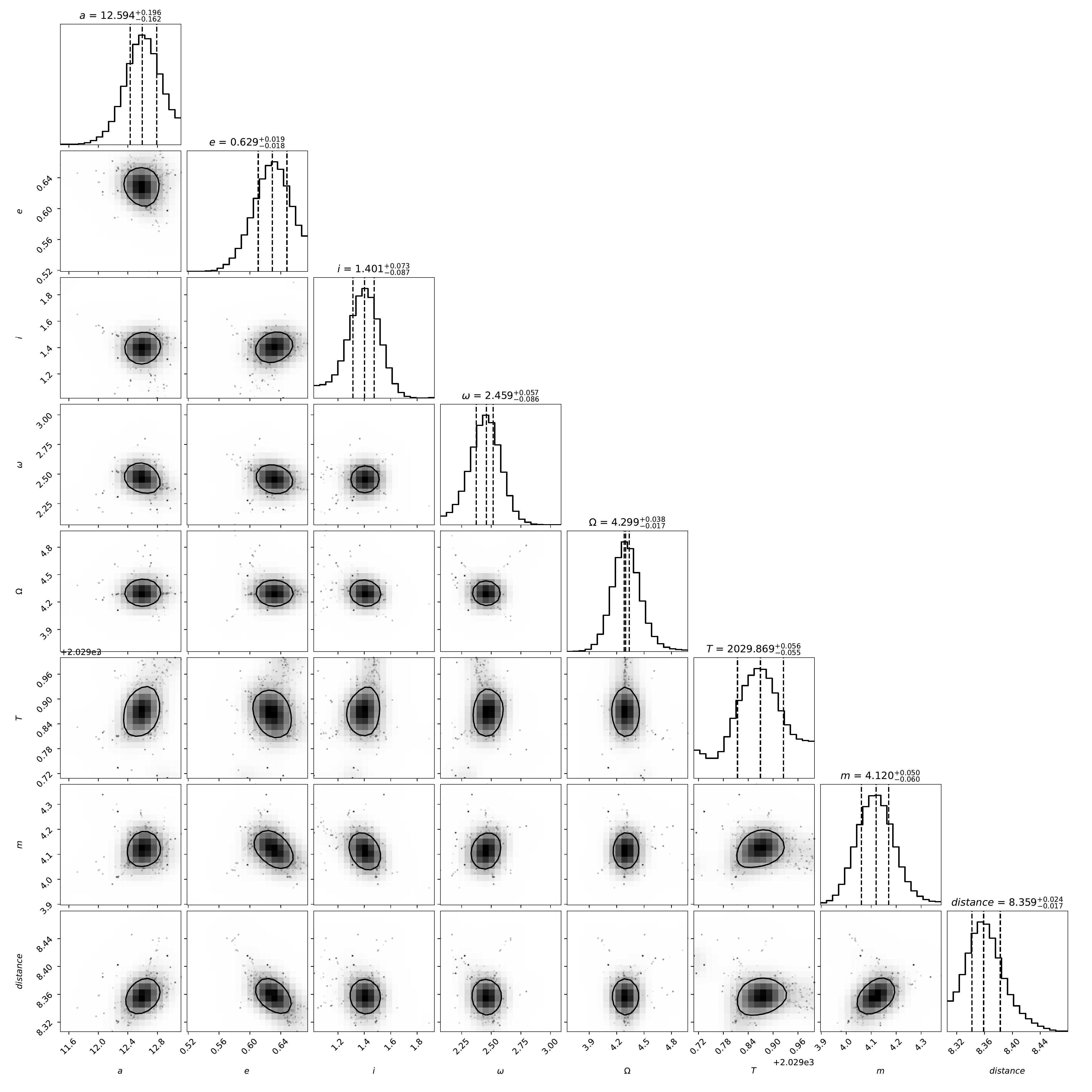}
	\caption{MCMC simulations for the orbital elements of OS1.}
\label{fig:os2_mcmc}
\end{figure*}

Here we present the results of our MCMC simulations. As for the G2/DSO, the likelihood for the Keplerian fit is minimized. Hence, the MCMC simulations confirm our initial parameters to a satisfying degree (see Fig. \ref{fig:dso_orbit_3}, Fig. \ref{fig:os1_mcmc}, and Fig. \ref{fig:os2_mcmc}). The quality of the derived Keplerian parameters is underlined by the compactness of the possible randomized value distribution.

\end{document}